\documentclass[fleqn,usenatbib]{mnras}

\usepackage{newtxtext,newtxmath}

\usepackage[T1]{fontenc}

\DeclareRobustCommand{\VAN}[3]{#2}
\let\VANthebibliography\thebibliography
\def\thebibliography{\DeclareRobustCommand{\VAN}[3]{##3}\VANthebibliography}

\usepackage{graphicx}	
\usepackage{amsmath}	
\usepackage{braket}
\usepackage{mathtools}

\usepackage{multirow}
\usepackage{rotating}
\usepackage{tabularx}
\usepackage{subcaption}
\usepackage{outlines}
\usepackage{booktabs}
\usepackage{natbib}
\newcolumntype{s}{>{\hsize=.3\hsize}X}
\usepackage{orcidlink}

\title[Planet characterization of TOI$-$757 b]{TOI$-$757 b: an eccentric transiting mini$-$Neptune on a 17.5$-$d orbit\thanks{Based on observations made with the ESO$-$3.6\,m telescope at La Silla Observatory under program 106.21TJ.001. This study uses \textit{CHEOPS} data observed as part of the Guaranteed Time Observation (GTO) programme CH\_PR100024. This paper includes data gathered with the 6.5 meter Magellan Telescopes located at Las Campanas Observatory, Chile.}}

\author[A. Alqasim et al.]{
\parbox{\textwidth}{
\large
A.~Alqasim\textsuperscript{\hyperlink{inst:1}{1}}\thanks{E-mail: \href{mailto:ahlam.alqasim.17@ucl.ac.uk}{ahlam.alqasim.17@ucl.ac.uk}.}\orcidlink{0000-0001-5102-5505}, 
N.~Grieves\textsuperscript{\hyperlink{inst:2}{2}}\orcidlink{0000-0002-4265-047X},
N. M. Ros\'{a}rio\textsuperscript{\hyperlink{inst:3}{3}, \hyperlink{inst:4}{4}}\orcidlink{0000-0002-8588-6730},
D.~Gandolfi\textsuperscript{\hyperlink{inst:5}{5}}\orcidlink{0000-0001-8627-9628},
J.~H.~Livingston\textsuperscript{\hyperlink{inst:6}{6}}\orcidlink{0000-0002-4881-3620},
S.~Sousa\textsuperscript{\hyperlink{inst:3}{3}}\orcidlink{0000-0001-9047-2965},
K.~A.~Collins\textsuperscript{\hyperlink{inst:7}{7}}\orcidlink{0000-0001-6588-9574},
J.~K.~Teske\textsuperscript{\hyperlink{inst:8}{8}}\orcidlink{0009-0008-2801-5040},
M.~Fridlund\textsuperscript{\hyperlink{inst:9}{9},\hyperlink{inst:10}{10}}\orcidlink{0000-0002-0855-8426},
J.~A.~Egger\textsuperscript{\hyperlink{inst:11}{11}}\orcidlink{0000-0003-1628-4231}
% ...
}
~\\
~\\
Affiliations are listed at the end of the paper in Appendix \ref{sec:affiliations}.
}

\date{Accepted XXX. Received YYY; in original form ZZZ}

\pubyear{2023}

\begin{document}
\label{firstpage}
\pagerange{\pageref{firstpage}--\pageref{lastpage}}
\maketitle

\begin{abstract}
We report the spectroscopic confirmation and fundamental properties of TOI$-$757\,b, a mini$-$Neptune on a 17.5$-$day orbit transiting a bright star ($V\,=\,9.7$ mag) discovered by the \textit{TESS} mission. 
We acquired high$-$precision radial velocity measurements with the HARPS, ESPRESSO, and PFS spectrographs to confirm the planet detection and determine its mass.
We also acquired space$-$borne transit photometry with the \textit{CHEOPS} space telescope to place stronger constraints on the planet radius, supported with ground$-$based LCOGT photometry.
WASP and KELT photometry were used to help constrain the stellar rotation period.
We also determined the fundamental parameters of the host star.
We find that TOI$-$757\,b has a radius of $R_{\mathrm{p}} = 2.5 \pm 0.1 R_{\oplus}$ and a mass of $M_{\mathrm{p}} = 10.5^{+2.2}_{-2.1} M_{\oplus}$, implying a bulk density of $\rho_{\text{p}} = 3.6 \pm 0.8$ g cm$^{-3}$.
Our internal composition modeling was unable to constrain the composition of TOI$-$757 b, highlighting the importance of atmospheric observations for the system.
We also find the planet to be highly eccentric with $e$ = 0.39$^{+0.08}_{-0.07}$, making it one of the very few highly eccentric planets among precisely characterized mini$-$Neptunes.
Based on comparisons to other similar eccentric systems, we find a likely scenario for TOI$-$757 b's formation to be high eccentricity migration due to a distant outer companion.
We additionally propose the possibility of a more intrinsic explanation for the high eccentricity due to star$-$star interactions during the earlier epoch of the Galactic disk formation, given the low metallicity and older age of TOI$-$757. 
\end{abstract}




\section{Introduction}\label{sec:intro}

There are two main populations of small planets, super$-$Earths and mini$-$Neptunes, separated by a gap known as the radius valley, centered around $R_{\mathrm{p}} \approx 1.8~R_{\oplus}$ \citep{owenwu2013photoevap, owenwu2017rvalley, fulton2017rvalley, ho2023rvalley}. 
The location of the radius valley is predicted to be dependent on the internal composition of the planet, and the valley is expected to be shifted to a larger planetary radius ($R_{\mathrm{p}}\,\approx\,2.3~R_{\oplus}$) if the internal composition of these planets is icy rather than rocky \citep{jin2018rvalleyice, mordasini2020rvalleyice}. 
Potential explanations of the radius gap focus on atmospheric mass loss mechanisms, which predict that the radius gap separates the super$-$Earth and mini$-$Neptune populations based on whether or not planets were able to retain their primordial hydrogen/helium atmosphere (H/He envelope). 
A possible explanation of the radius gap comes from thermally$-$driven mass loss mechanisms (also known as photoevaporation) induced by exposure to XUV irradiation from the host star \citep{owenwu2013photoevap, lopez2013photoevap}. 
Photoevaporation models can reproduce the position of the radius valley by assuming that super$-$Earth and mini$-$Neptune planets all have a rocky composition, with their different radii being a consequence of whether or not they lost their atmosphere. 
Another explanation proposed for the existence of the radius gap focuses on core$-$powered mass loss mechanisms, which results from the internal heating of the planet \citep{ginzburg2018corepower, gupta2019corepower, gupta2020corepower}. 
In the core$-$powered mass loss scernario, energy escaping from the planet's core heats up the H/He envelope and causes the atmospheric gases to escape, turning the planet into a stripped core.

We present the discovery, the confirmation, and the fundamental properties of TOI$-$757\,b, a mini$-$Neptune planet located above the radius valley.
The abundance of mini$-$Neptune planets that fall close to the radius valley is generally well$-$explained within the photoevaporation framework, where these planets are assumed to have rocky cores with a thin atmosphere \citep{bean2021subneptunes}. 
However, simply looking at the radius distribution of planets close to the radius valley does not give us clear insights about their densities or core compositions.
For that, mass measurements and (in many cases) atmospheric observations are necessary to break the degeneracies of what the core compositions of the mini$-$Neptune population is comprised of \citep{bitsch2019rockywater}.
In particular, studies of their mass measurements are very important in helping constrain their densities, which could give us clues about their internal structure \citep{benneke2013cloudy}.
In many cases however, constraints on the atmospheric chemical composition from transit spectroscopy are necessary in order to try to break the degeneracy of their internal composition [eg. mini-Neptune vs. water world].

Eccentric systems further complicate this scenario because they could suggest a different evolution path.
Such planets could have become eccentric due to gravitational interactions with other planets and/or migration \citep{barnes2009ecc}. 
They may therefore have formed further out, or undergone a different migration history compared to other planets, and as such it would be interesting to see if that suggests a different photo$-$evaporation history.
To make strides towards a better understanding of the mini$-$Neptune population, it is important to follow them up and gain a more precise picture of their formation mechanism. 

In this paper, we make use of multiple radial velocity (RV) monitoring campaigns to obtain a precise and accurate mass measurement of the planet, TOI$-$757 b.
We also make use of space$-$based photometric follow$-$up observations of the system, allowing us to derive a more precise radius measurement of the planet, as well as additional ground$-$based photometric data of TOI$-$757.
By studying the planet using both photometric and RV data, we were able to obtain a precise and accurate planet density and explore the implications of its composition. 
Determining the planet's composition allows us to interpret whether TOI$-$757 b indeed fits within the scenario presented by atmospheric mass loss models.

In Section \ref{sec:observations}, we describe the photometric data acquired from space$-$based telescopes, as well as the RV data acquired from ground$-$based telescopes.
In Section \ref{sec:starcharac}, we discuss the stellar characterization of TOI$-$757 and present the newly$-$derived stellar properties from this work.
In Section \ref{sec:sysconfirmation}, we describe how we validate the system using spaced$-$based astrometry as well as ground$-$based high$-$contrast imaging.
In Section \ref{sec:transitrvanalysis}, we detail the methods for modeling and fitting the photometric and RV data sets.
Finally, we present the discussions and conclusions in Section~\ref{sec:discussion}, with a summary of the work in Section~\ref{sec:summary}.

\section{Observations}\label{sec:observations}

Photometric data for TOI$-$757 were acquired through a combination of space$-$based and ground$-$based observations.
We make use of 4 transits from the \textit{TESS} space mission, spanning 3 separate sectors with 120s cadence lightcurves.
Additionally, we acquired follow$-$up photometric observations of the system.
We make use of 2 transits from the \textit{CHEOPS} space telescope with 60s cadence lightcurves, which allowed us to place stronger constraints on the planet radius, supported by ground$-$based LCOGT photometry.
We also acquired WASP and KELT ground$-$based photometry to aid in constraining the stellar rotation period.

High$-$precision RV data for the exoplanetary system were collected from multiple instruments, including the HARPS, ESPRESSO, and PFS spectrographs. 
We acquired a total of 73 high$-$precision RV measurements between 2021 and 2022, to confirm the planetary nature of the transit signal detected by \textit{TESS} and determine its mass measurement.

	\subsection{\textit{TESS}}\label{subsec:obstess}  

        The Transiting Exoplanet Survey Satellite (\textit{TESS}) is a NASA mission aiming to discover transiting exoplanets by an all$-$sky survey \citep{ricker2015tess}.
        The star TIC 130924120 was observed by \textit{TESS} at a 2$-$min cadence in Sectors 10, 37, and 64, resulting in a total of 4 transit events.    
        The raw images were reduced by the Science Processing Operations Center (SPOC) team at NASA Ames Research Center using the SPOC pipeline \citep{jenkins2016tess} and the photometry for all the 2$-$min target stars in the \textit{TESS} field of view were extracted, resulting in lightcurves ready to be searched for transit signals. 
        
        The SPOC conducted a transit search for Sector 10 with a noise$-$compensating matched filter \citep{jenkins2002solarvar, jenkins2010kepler, jenkins2020keplerhandbook} and detected the transit signature of TOI$-$757 b. 
        An initial limb$-$darkened model fit was performed \citep{li2019keplervalid} and TOI$-$757 b was found to have a period of 17.4655 $\pm$ 0.0029 days, a radius ratio of 0.03$\pm$ 0.01, and a transit depth of 1064 $\pm$ 68 ppm, transiting the star over a duration of 0.16 $\pm$ 0.02 days.
        A suite of diagnostic tests were conducted to help make or break the planetary nature of the signal \citep{twicken2018keplervalid}. 
        The candidate transit signature passed all these diagnostic tests including the difference image centroiding test, which constrained the location of the host star to within 2.5 $\pm$ 2.6 arcsec of the transit source. 
        The TESS Science Office (TSO) reviewed the vetting information and issued an alert for TOI$-$757.01 on 20 June 2019 \citep{guerrero2021toi}.
        In this work, we use the Presearch Data Conditioning$-$corrected Simple Aperture Photometry (PDCSAP) lightcurves of the system \citep{smith2012kepler, stumpe2012kepler, stumpe2014multiscale} for the transit modeling and analyses. 
        These lightcurves are publicly available at the Mikulski Archive for Space Telescopes (MAST\footnote{\protect\url{https://mast.stsci.edu/portal/Mashup/Clients/Mast/Portal.html}}).

        We independently analyzed the \textit{TESS} data using the Détection Spécial$-$isée de Transits (DST) algorithm~\citep{2012A&A...548A..44C} to confirm the presence of the candidate and ruled out any significant transit timing variations in the data. 
        Our analysis shows that we can confidently detect planetary signals in the range of 250 ppm to 450 ppm with orbital periods up to 20 days. 
        This would correspond, for central transits, to planets in the range of 1.3 to 1.8 Earth radii.

        \subsection{Light Curve Follow$-$up Observations}\label{subsec:obsfollowup} 
        
        The \textit{TESS} pixel scale is approximately $21\arcsec$ pixel$^{-1}$ and photometric apertures typically extend out to roughly 1 arcminute, generally causing multiple stars to blend in the \textit{TESS} photometric aperture. 
        To determine the true source of the \textit{TESS} detection, we acquired ground$-$based LCOGT time$-$series follow$-$up photometry of the field around TOI$-$757 as part of the \textit{TESS} Follow$-$up Observing Program \citep[TFOP;][]{collins:2019}\footnote{\url{https://tess.mit.edu/followup}}.
        The on$-$target follow$-$up light curves are also used to place constraints on the transit depth and the \textit{TESS} ephemeris. 
        We used the {\tt \textit{TESS} Transit Finder}, which is a customized version of the {\tt Tapir} software package \citep{Jensen:2013}, to schedule our transit observations.
        We also made use of WASP and KELT photometric data to search for activity signals and aid in constraining the stellar rotation period.

    	\subsubsection{WASP}\label{subsubsec:obswasp}  
     
            The Wide Angle Search for Planets (WASP) transit$-$search survey \citep{pollaco2006superwasp} is located at the Sutherland site of the South African Astronomical Observatory.
            The data was reduced using the pipeline described in \citep{pollaco2006superwasp}, with exposures that spanned 30 seconds (when observing with 200mm lenses) and 20 seconds (when observing with 85mm lenses).
            For data taken with the 200mm lenses, we used a broadband 400 to 700$-$nm filter; with the 85mm lenses, we used an SDSS$-$r filter.

            The field of TOI$-$757 was observed by WASP in 6 different years, first in 2007, 2008, 2011, and 2012, when WASP$-$South was equipped with 200$-$mm lenses, and then in 2013 and 2014, equipped with 85$-$mm lenses. 
            In each year the observations spanned 150 to 180 days. 
            A total of 66\,000 data points were obtained, with TOI$-$757 being by far the brightest star in the photometric extraction aperture. 
            The TOI$-$757 b transits are too shallow to be detected in the WASP photometry. 
            As such, the WASP data was used to check for stellar variability (see Sect. \ref{subsec:starvariability}).

    	\subsubsection{KELT}\label{subsubsec:obskelt}  

            TOI$-$757 was observed by the Kilodegree Extremely Little Telescope (KELT) survey in field S24 by the southern telescope during the observing seasons of 2010, 2011, 2012, and 2013. 
            The telescope includes an Apogee Alta U16M camera with a 4096$\times$4096, 9$\mu$m pixel Kodak KAF$-$16803 front illuminated CCD. 
            The optics include a Mamiya 645 80mm f/1.9 lens (42~mm effective aperture) and a Kodak Wratten \#8 red$-$pass filter. 
            The pixel scale of the detector is $\sim23$\arcsec pix$^{-1}$ leading to a total field of view (hereafter FoV) of $26^{\circ}\times26^{\circ}$ remarkably similar to the FoV of a single \textit{TESS} camera. 
            The KELT telescopes typically observe fields in two orientations (east and west) and the exposures are kept to 150~s. 
            Any given KELT field can be expected to be observed between 5$-$50 times a night given the position of the Moon \citep{Pepper2007, Pepper2012}.

            Following the work of \citet{oelkers2018kelt} the east and west light curves were combined into a single light curve (with a total of 4,540 exposures) and post$-$processed using the Trend$-$Filtering Algorithm \citep{Kovacs2005} to remove common detector systematics. 
            In practice, this algorithm has been found to remove large amplitude variability (such as stellar pulsations) and red$-$noise systematics while leaving most small amplitude variability (such as exoplanet transits and rotation signals) untouched.

            \subsubsection{LCOGT}\label{subsubsec:obslcogt}
            
            We observed an egress window of TOI$-$757 b on UTC 2019 June 04 simultaneously in Pan$-$STARRS $z-$short (henceforth, $z$s) band ($\lambda_{\rm c} = 8700$\,\AA, ${\rm Width} =1040$\,\AA) and Pan$-$STARRS Y$-$band ($\lambda_{\rm c} = 10040$\,\AA, ${\rm Width} =1120$\,\AA), using the Las Cumbres Observatory Global Telescope \citep[LCOGT;][]{Brown:2013} 1\,m network node at South Africa Astronomical Observatory (SAAO). 
            The $z$s$-$band and Y$-$band observations were observed simultaneously on two different LCO 1\,m telescopes at SAAO.
            We also observed an ingress window in $z$s$-$band on UTC 2021 January 20 from the same LCOGT 1.0\,m network node. 
            The images were calibrated by the standard LCOGT {\tt BANZAI} pipeline \citep{McCully:2018} and differential photometric data were extracted using {\tt AstroImageJ} \citep{Collins:2017}. 
            The egress lightcurve in Y$-$band rules out NEBs within $2\farcm5$ of TOI$-$757 to more than $\pm3\sigma$ in depth and more than $\pm6\sigma$ in ephemeris.
            We used circular photometric apertures with radius $6\farcs6$ to $9\farcs7$ to extract the TOI$-$757 light curves. 
            This excluded all of the flux from the nearest known neighbor in the Gaia DR3 catalog that is bright enough in \textit{TESS}$-$band to be capable of causing the \textit{TESS} detection (\textit{Gaia} DR3 6157753406083082112), which is $\sim16\arcsec$ east of TOI$-$757. 

            The 2019 $z$s$-$band observation was detrended using a linear fit to time, while the 2019 Y-band and 2021 $z$s$-$band observations were detrended using a linear fit to the total ensemble star counts. 
            The detrend parameter for each lightcurve was selected using a joint detrend plus transit model fit. 
            The detrend parameter selected minimized the BIC of the joint fit and was justified by a delta BIC > 2 compared to a transit model-only fit.
            The three light curves are included in the global photometric modelling described in section \ref{subsec:transitfit}.

    	\subsubsection{\textit{CHEOPS}}\label{subsubsec:obscheops}  
     
            The CHaracterizing ExOPlanet Satellite (\textit{CHEOPS}) is an ESA mission aiming to characterize known transiting exoplanets orbiting bright stars~\citep{2021ExA....51..109B}.
            The TOI$-$757 system was visited by \textit{CHEOPS} in two different occasions during the Guaranteed Time Observation (GTO) programme 24, first on March 31st 2021 and then on May 23rd 2021. 
            The target was observed for a total of 1.72 days. 
            We obtained a clear transit of TOI$-$757 b in the first visit, but since it was not centred with the observation window, we used the first \textit{CHEOPS} transit together with the \textit{TESS} transits to update the ephemerides and correct the time of the second visit. 
            The observation log with additional details of each visit is provided in Table~\ref{table:cheopslog}.
            
            We used the \textit{CHEOPS} data reduction pipeline  \citep[DRP;][]{hoyer2020expected} to reduce the raw lightcurves. 
            The pipeline automatically processes the data and removes instrumental variability and environmental effects such as background and smearing. 
            The DRP produces a file containing all the extracted lightcurves and additional data, including quality flags and the roll$-$angle and background time series that can be used for further corrections. 
            The signal is extracted in four different apertures and for this analysis we chose to work with the OPTIMAL set, which calculates the optimal aperture radius based on the best signal$-$to$-$noise ratio.
    
            We use the package \texttt{pycheops} \citep{maxted2022pycheops}, which is publicly available\footnote{\url{https://github.com/pmaxted/pycheops}}, to correct and de$-$correlate the \textit{CHEOPS} light curves. 
            We use \texttt{pycheops} to download the light curves from the Data \& Analysis Center for Exoplanets (DACE\footnote{\protect\url{https://dace.unige.ch}}) website, which is used to store \textit{CHEOPS} observation data. 
            DACE is a facility dedicated to extrasolar planets data visualisation, exchange, and analysis  hosted by the University of Geneva.
            We first subtract the contamination of nearby sources and perform an initial outlier clipping at $5\sigma$. 
            The de$-$correlation parameters are selected by calculated the Bayesian Information Criterion (BIC) value of the model for each combination of parameters, assuming that the lowest BIC corresponds to the combination that best describes the data. 
            We de$-$correlate against time, position of the centroid, contamination and smearing as well as the roll angle of the telescope. 
            Some \textit{CHEOPS} observations can be contaminated by observation$-$specific sources, such as the Moon. 
            Therefore, we use \texttt{pycheops} to correct for this `glint' effect as well. 
            Then, we perform a preliminary least squares fit of the light curves, without any de$-$correlation and calculate the root mean square (RMS) error. 
            We select Gaussian distributions centred at zero with a standard deviation of two times the RMS error from the previous fit, to use as priors for the de$-$correlation parameters. 
            Finally, we perform an MCMC fit of the \textit{CHEOPS} light curves within \texttt{pycheops}, including a Gaussian Process (GP) regression method with a SHOTerm kernel to fit for the chosen de$-$correlation parameters. 
            We then retrieve the light curves corrected with the GP and save them in a separate file, to be included in the photometric fit.
    
            \begin{table*}
            \caption{\textit{CHEOPS} observation log for TOI$-$757. The first column has the observation ID and the remaining columns show the unique file key of each \textit{CHEOPS} visit, the start date of the observations, the duration, the exposure time, the efficiency and the number of non$-$flagged data points on each visit.}
            \label{table:cheopslog}      
            \centering  
                \setlength{\extrarowheight}{3pt}
                \resizebox{\textwidth}{!}{
                \begin{tabular}{c c c c c c c}       
                \hline
                ID & File Key & Start Date & Duration (h) & Exp. Time (s) & Efficiency & No. Points\\
                \hline
                1 & CH\_PR100024\_TG013901\_V0200 & 2021$-$03$-$31T15:26:09 & 28.06 & 60 & 76.8\% & 1294\\
                2 & CH\_PR100024\_TG014401\_V0200 & 2021$-$05$-$23T03:32:08 & 13.41 & 60 & 55.9\% & 450\\
                \hline
                \end{tabular}}
            \end{table*}

	\subsection{HARPS}\label{subsec:obsharps}  
 
        We acquired 27 high$-$resolution (R$\approx$115,000) spectra of TOI$-$757 with the High Accuracy Radial velocity Planet Searcher \citep[HARPS;][]{Mayor2003} spectrograph mounted at the ESO$-$3.6m telescope at the European Southern Observatory (ESO), La Silla, Chile. 
        The observations were carried out between 22 January and 30 May 2021 (UT), as part of our HARPS large program for the RV follow$-$up of \textit{TESS} transiting planet candidates (ID: 106.21TJ.001, PI: Gandolfi). 
        We set the exposure time to 1500$-$1800 s leading to a median signal$-$to$-$noise (S/N) ratio per pixel at 550\,nm of $\sim$87 on the extracted spectra and a median RV uncertainty of  0.98\,m\,s$^{-1}$. 
        We reduced the HARPS data using the Data Reduction Software \citep[DRS;][]{Lovis2007} available at the telescope and extracted the RV measurements by cross$-$correlating each Echelle spectrum against a G2 numerical mask \citep{Baranne1996,Pepe2002}. 
        We also used the DRS to extract the Ca II H \& K lines activity indicator logR$\prime_\mathrm{HK}$, and the profile activity diagnostics of the cross$-$correlation function (CCF), namely, the full$-$width at half maximum (FWHM), the bisector inverse slope (BIS), and the contrast.
        We finally extracted relative RV measurements and additional activity indicators $--$ such as the H$\alpha$ and Na D lines indicators $--$ using the Template Enhanced Radial velocity Re$-$analysis Application \citep[TERRA;][]{Anglada2012}.
        Table~\ref{table:rvdata} lists the time stamps of our HARPS observations in Barycentric Julian Dates in Barycentric Dynamical Time (BJD$_\mathrm{TDB} - 2457000$), along with the DRS and TERRA RV measurements and activity indicators.

	\subsection{ESPRESSO}\label{subsec:obsespresso}  
 
        We obtained 22 spectra of TOI$-$757 with the Echelle SPectrograph for Rocky Exoplanets and Stable Spectroscopic Observations \citep[ESPRESSO;][]{pepe2021} at the 8.2 m ESO Very Large Telescope (VLT) array, at the Paranal Observatory in Chile. 
        The observations were obtained from 2021 April 04 to 2021 August 06 for an ESPRESSO program designed to explore the densities of sub$-$Neptunes (Progam ID: 105.20P7.001, PI: Bouchy). 
        We observed TOI$-$757 in HR mode (1 UT, R$\sim$140,000) over a spectral range from $\sim$380 to $\sim$780 nm. 
        ESPRESSO is contained in a temperature and pressure controlled vacuum vessel to avoid spectral drifts. 
        ESPRESSO observations can be carried out with simultaneous calibration using a Fabry$-$P\'erot etalon on a second calibration fiber, which was used for our observations. 
        
        The ESPRESSO RVs were processed using DACE.
        The RV and stellar activity indicators provided in DACE are computed by cross$-$correlating the calibrated spectra with stellar templates for the specific spectral type, using the latest version (3.0.0) of the publicly$-$available ESPRESSO Data Reduction Software\footnote{\protect\url{https://www.eso.org/sci/software/pipelines/espresso/espresso-pipe-recipes.html}}. 
        Stellar activity indicators such as the FWHM and the BIS are derived from the CCF, while other activity indicators (e.g. S-index, Na-index, etc.) are extracted from the spectra.
        The 22 ESPRESSO observations of TOI$-$757, listed in Table \ref{table:rvdata}, had an exposure time of 600 s and a median RV uncertainty of 0.54\,m\,s$^{-1}$.

	\subsection{PFS}\label{subsec:obspfs}  
 
        We obtained a total of 26 RV data points (15 epochs) using the Carnegie Planet Finder Spectrograph \citep[PFS;][]{crane2006, crane2008, crane2010} on the 6.5-meter Magellan II Clay telescope. 
        PFS is an optical high-resolution echelle spectrograph covering 391$-$784 nm, and uses an iodine cell for wavelength calibration.    
        The observations presented here were taken with the $0.3\arcsec\times 2.5\arcsec$ slit, resulting in a resolving power of $R \sim 130,000$ and a median RV uncertainty of  0.89\,m\,s$^{-1}$.
        Individual exposure times for TOI$-$757 ranged from 600$-$1500 s and were often taken $\times$2 per night separated by a few hours (to average over granulation) and binned on a nightly basis. 
        The time stamps were converted from JD to BJD in Barycentric Dynamical Time using the tools provided by \cite{eastman2010timeconversion}.
        The extraction of stellar spectra and the measurement of radial velocities were conducted with an IDL pipeline customized for PFS \citep{butler1996}. 
        The PFS data presented here (see Table \ref{table:rvdata}) were collected as part of the Magellan \textit{TESS} Survey (MTS), described in detail in \cite{Teske2021}, although they were not included/published in that paper. 
        An update to MTS with this and additional TOI$-$757 data is forthcoming.

\section{Stellar Characterization of TOI$-$757}\label{sec:starcharac}

We characterize the stellar properties of TOI$-$757 (TIC 130924120) using the HARPS and ESPRESSO spectra. 
A summary of the stellar parameters, including the \textit{Gaia} astrometry, are listed in Table \ref{table:starparams}. 
The HARPS$-$derived stellar parameters give consistent results with their ESPRESSO counterparts.
Hence, we decided to use the ESPRESSO stellar parameters in our final analysis, given that they are slightly more precise.
The reported values of our work in Table \ref{table:starparams} thus correspond to the ESPRESSO$-$derived stellar parameters.

Based on the derived $T_{\mathrm{eff}}$ for TOI$-$757, we find that it is most likely a K0 V type star \citep{mamajek2013intrinsic}. 
The following subsections describe the different properties of the system and how we derived its basic parameters listed in Table \ref{table:starparams}.

\begin{table}
        \caption{Identifiers, equatorial coordinates, parallax, distance, optical and infrared magnitudes, and stellar fundamental parameters for TOI$-$757. The parameters reported for this work were derived using the ESPRESSO spectra.}
        \centering
        \setlength{\extrarowheight}{3pt}
        \begin{tabularx}{\linewidth}{X X s}
        	\hline
                                   Parameter          &               Value     & Reference \\
                \addlinespace[3pt]
                \hline
                \multicolumn{3}{l}{ \textit{Identifiers} }                                  \\
                                          CD          &         $-$34 8236      &           \\                    
                                         TIC          &           130924120     &           \\
                                       2MASS          &   J12315870$-$3533177   &           \\
                                       APASS          &            18444726     &           \\
                           \textit{Gaia} DR3          & 6157753406084087168     &           \\
                                         TOI          &                 757     &           \\
                                         TYC          &    7242$-$01598$-$1     &           \\
                                       UCAC4          &        273$-$064695     &           \\
                                        WISE          & J123158.74$-$353316.5   &           \\
                \hline
                \multicolumn{3}{l}{ \textit{Equatorial Coordinates} }                       \\
                                    RA (J2000) [deg]  & 187.99                  &         1 \\
                                   Dec (J2000) [deg]  & $-$35.55                &         1 \\
                              Parallax [mas]          &  16.58 $\pm$ 0.02       &         1 \\
                               Distance [pc]          & $60.4_{-0.2}^{+0.3}$    &         1 \\
                                        RUWE          &               0.97      &         1 \\
                \hline
                \multicolumn{3}{l}{ \textit{Apparent Magnitudes} }                          \\
                         \textit{Gaia} G [mag]        & 9.5615 $\pm$ 0.0001     &         1 \\
                         \textit{Gaia} BP [mag]       & 9.9566 $\pm$ 0.0005     &         1 \\
                         \textit{Gaia} RP [mag]       & 8.9937 $\pm$ 0.0003     &         1 \\
                         \textit{TESS} T [mag]        &  9.048 $\pm$ 0.006      &         2 \\
                                     B [mag]          & 10.56 $\pm$ 0.06        &         2 \\
                                     V [mag]          &  9.743 $\pm$ 0.004      &         2 \\
                                     J [mag]          &  8.35 $\pm$ 0.03        &         4 \\
                                     H [mag]          &  7.98 $\pm$ 0.04        &         4 \\
                                    Ks [mag]          &  7.87 $\pm$ 0.02        &         4 \\
                       WISE 3.4 $\mu$m [mag]          &  7.81 $\pm$ 0.03        &         5 \\
                       WISE 4.6 $\mu$m [mag]          &  7.88 $\pm$ 0.02        &         5 \\
                        WISE 12 $\mu$m [mag]          &  7.83 $\pm$ 0.02        &         5 \\
                        WISE 22 $\mu$m [mag]          &  7.67 $\pm$ 0.13        &         5 \\
                \hline
                \multicolumn{3}{l}{ \textit{Stellar Parameters} }                           \\
                   $R_{\star}$ [$R_{\odot}$]          &    0.773 $\pm$ 0.005    &         6 \\
                   $M_{\star}$ [$M_{\odot}$]          &      0.80 $\pm$ 0.04    &         6 \\
                $\rho_{\star}$ [g cm$^{-3}$]          &     2.4 $\pm$ 0.1       &         6 \\
                      $T_{\mathrm{eff}}$ [K]          &       5278 $\pm$ 63     &         6 \\
                log $g$ (\textit{Gaia}) [cm s$^{-2}$] &     4.54 $\pm$ 0.03     &         6 \\
                log $g$ (spec) [cm s$^{-2}$]          &     4.5 $\pm$ 0.1       &         6 \\
                   $L_{\star}$ [$L_{\odot}$]          &     0.42 $\pm$ 0.02     &         6 \\
                                   Age [Gyr]          &       7.5 $\pm$ 3.0     &         6 \\
                   $v \sin(i)$ [km s$^{-1}$]          &      2.4 $\pm$ 0.4      &         6 \\
                                     ~[Fe/H]          &    $-$0.27 $\pm$ 0.04   &         6 \\
                                     ~[M/H]           &     $-$0.21 $\pm$ 0.05  &         6 \\
                                     ~[Mg/H]          &     $-$0.2 $\pm$ 0.03   &         6 \\
                                     ~[Si/H]          &    $-$0.23 $\pm$ 0.04   &         6 \\
                                     ~[Ti/H]          &    $-$0.14 $\pm$ 0.05   &         6 \\
                               ~[$\alpha$/Fe]         &    0.08 $\pm$ 0.05      &         6 \\
                \hline
                \addlinespace[4pt]
                \multicolumn{3}{l}{\multirow{2}{*}{\parbox{8cm}{ \textbf{References:} (1) \cite{gaia2023dr3}; (2) \cite{paegert2022tess}; (3) \cite{hog2000tycho}; (4) \cite{cutri2003sloan}; (5) \cite{cutri2014wise}; 
                (6) This work. }}} \\
            \end{tabularx}
        \label{table:starparams}
\end{table}

        \subsection{Stellar atmospheric properties and abundances}\label{subsec:staratmos}
        
        The spectroscopic stellar parameters ($T_{\mathrm{eff}}$, $\log g$, microturbulence, [Fe/H]) were estimated using the ARES+MOOG methodology. 
        The methodology is described in detail in \citet[][]{Sousa-21, Sousa-14, Santos-13}. 
        This was done by using the latest version of ARES \footnote{The last version, ARES v2, can be downloaded at \url{https://github.com/sousasag/ARES}.} \citep{Sousa-07, Sousa-15} to consistently measure the equivalent widths (EW) of selected iron lines on the combined ESPRESSO spectrum. 
        We used the list of iron lines presented in \citet[][]{Sousa-08}. 
        A minimization process is then used in this analysis to find the ionization and excitation equilibrium and converge for the best set of spectroscopic parameters. 
        This process makes use of a grid of Kurucz model atmospheres \citep{Kurucz-93} and the radiative transfer code MOOG \citep{Sneden-73}. 
        We also derived a more accurate trigonometric surface gravity using recent \textit{Gaia} data following the same procedure as described in \citet[][]{Sousa-21} which provided a very consistent value when compared with the spectroscopic surface gravity.
        
        Abundances of Mg, Si, and Ti were derived using the same tools and models as for stellar parameter determination. 
        Although the EWs of the spectral lines were automatically measured with ARES, we performed careful visual inspection of the EWs measurements. 
        For the derivation of abundances, we closely followed the methods described in \citet{Adibekyan-12, Adibekyan-15}. 
        Considering the mean of the abundances of Mg, Si, and Ti as [$\alpha$/H], we calculated the $\alpha$ enhancement of the star relative to iron and its metallicity ($[\mathrm{M/H}]$) following \citet{Yi-01}. 
        These parameters are presented in Table~\ref{table:starparams}.

        \subsection{Stellar radius, mass, and age}\label{subsec:starproperties}  
        
        We determined the stellar radius of TOI$-$757 via a MCMC modified infrared flux method \citep{Blackwell1977,Schanche2020} to build spectral energy distributions (SEDs) using stellar atmospheric models from two catalogues \citep{Kurucz1993,Castelli2003} with priors coming from our spectral analysis described above. 
        We compute synthetic photometry from these SEDs that are compared to the observed broadband photometry in the following bandpasses: \textit{Gaia} $G$, $G_\mathrm{BP}$, and $G_\mathrm{RP}$, 2MASS $J$, $H$, and $K$, and \textit{WISE} $W1$ and $W2$ \citep{Skrutskie2006,Wright2010,GaiaCollaboration2022} to derive the stellar bolometric flux of TOI$-$757. 
        Using known physical relations, we convert this flux into effective temperature and angular diameter that is subsequently translated into stellar radius using the offset$-$corrected \textit{Gaia} parallax \citep{Lindegren2021}. 
        To correct for model uncertainties between the two atmospheric catalogues we used a radius posterior distribution Bayesian modelling averaging and report the weighted average in Table~\ref{table:starparams} \citep{fragoso2015bma}.
        Thus, we obtain $R_{\star}$ = 0.773 $\pm$ 0.005 $R_{\odot}$.
        
        We finally derived both the stellar mass $M_{\star}$ and age $t_{\star}$ according to stellar evolutionary models. 
        In detail, we inputted $T_{\mathrm{eff}}$, [M/H], and $R_{\star}$ in the isochrone placement algorithm \citep{bonfanti2015,bonfanti2016}, which is designed to interpolate the input set of values along with their uncertainties within pre$-$computed grids of PARSEC\footnote{\textsl{PA}dova and T\textsl{R}ieste \textsl{S}tellar \textsl{E}volutionary \textsl{C}ode: \url{http://stev.oapd.inaf.it/cgi-bin/cmd}.} v1.2S \citep{marigo2017} isochrones and tracks. 
        We inflate the mass and age uncertainties by 4\% and 20\% (respectively) by adding them in quadrature to their respective internal uncertainties.
        This is done to account for isochrone systematics \citep[see][]{bonfanti2021}. 
        After inflating the internal uncertainties, we obtained $M_{\star}=0.80\pm0.04\,M_{\odot}$ and $t_{\star}=7.5\pm3.0$ Gyr.

        \subsection{Stellar variability and rotation period}\label{subsec:starvariability}  

        To measure the rotational period of the star, we searched our WASP light curves for any rotational modulation using the methods outlined in \citet{maxted2011wasp41b}, analyzing both individual years of data and combined years.  
        We found no significant periodicity over the range 1 d to 100 d, with a 95\%$-$confidence upper limit on the amplitude of 2 mmag.

\begin{figure}
            \includegraphics[width=\linewidth, trim={2cm 0.7cm 11.5cm 23cm}, clip]{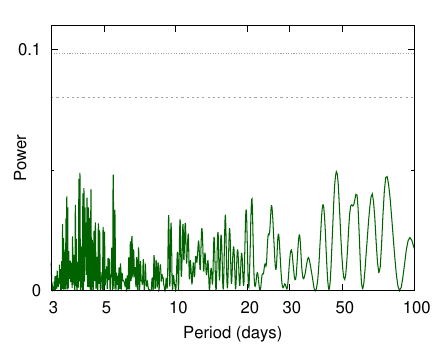}
            \includegraphics[width=\linewidth, trim={2cm 0.7cm 11.5cm 23cm}, clip]{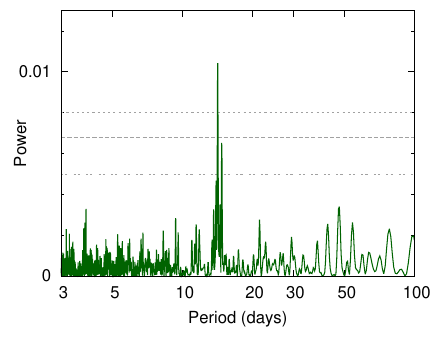}
        \caption{\textit{Top:} A sample periodogram of the WASP$-$South photometric time series data on TOI$-$757, being from 2011 and 2012 combined, showing no significant periodicities over the range 1 d to 100 d, with a 95\%$-$confidence upper limit on the amplitude of 1 mmag. The dashed horizontal lines are the estimated 1\% and 10\% false alarm probability levels. \textit{Bottom:} Periodogram of the KELT data for TOI$-$757, accumulated from 2010 to 2013, showing a 14.2$-$d periodicity. The horizontal lines mark the false$-$alarm levels at 10\%, 1\%\ \&\ 0.1\%\ likelihood, estimated using methods from \protect\cite{maxted2011wasp41b}. The amplitude of the KELT modulation (around 1 mmag) is sufficiently consistent with the WASP upper limit, which is also about that level.}
        \label{fig:keltwasp}
        \end{figure}

        Using KELT time$-$series photometry, which spans 5$-$year and 9$-$year baselines (depending on the instrument used), \citet{oelkers2018kelt} found a significant rotation period signal of $\sim$14.2\,days for TOI$-$757, above a FAP level of 0.1\% (see Fig.~\ref{fig:keltwasp}).  
        The amplitude of the KELT modulation (around 1 mmag) is consistent with the WASP upper limit. 
        The WASP data show no modulation stronger than 2 mmag (at the 95\% confidence level), whereas the KELT data exhibit a 1 mmag modulation.  
        Thus, the two datasets are consistent.
        To further verify the stellar signal we see from the periodogram analysis of the KELT photometry, we fit a GP to the KELT light curves, which have the trend$-$filtering$-$algorithm (TFA) applied to it. 
        These were the same TFA light curves used in the \cite{oelkers2018kelt} paper to calculate the rotation period of TOI$-$757. 
        We use the \texttt{celerite2} RotationTerm kernel \citep{exoplanet:foremanmackey17} for the KELT GP, and additionally fit for the instrumental jitter term (which is added in quadrature to the errors of the data) and the systemic mean magnitude.
        We use wide, Uniform priors for all parameters, with the rotation term $P_{\text{GP}}$ prior being $\mathcal{U}$[1, 40].
        We find a well$-$constrained $P_{\text{GP}}$ with a best$-$fit mean value of 14.18 $\pm$ 0.04 days, which is consistent with the results reported in the \cite{oelkers2018kelt} paper.

\begin{figure*}
        \centering
            \begin{subfigure}{0.5\textwidth}
                \centering
                \includegraphics[width=\linewidth]{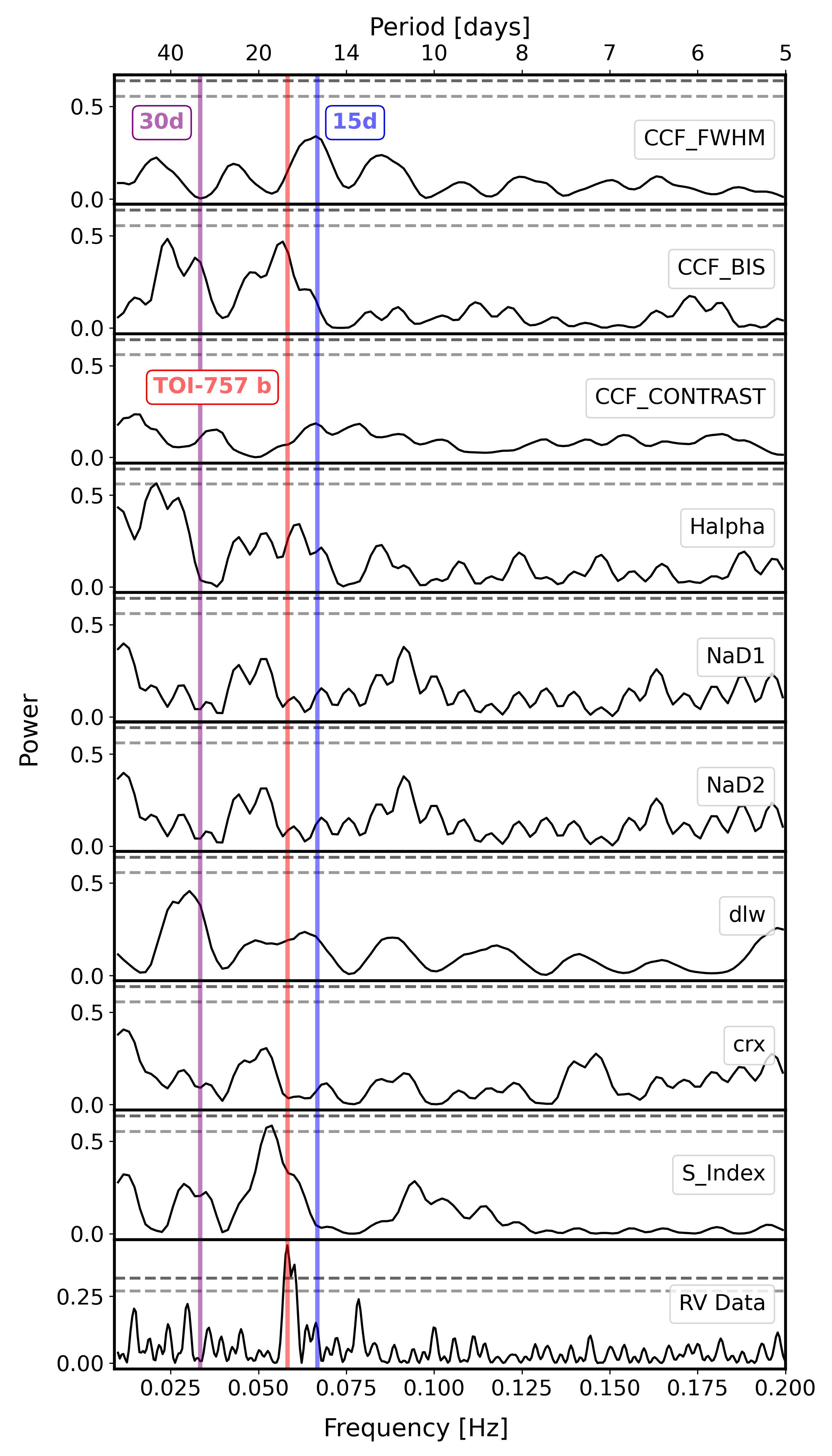}
                \caption{HARPS}
                \label{fig:HARPS_periodogram_lombsc_astropy}
            \end{subfigure}%
            \begin{subfigure}{0.5\textwidth}
                \centering
                \includegraphics[width=\linewidth]{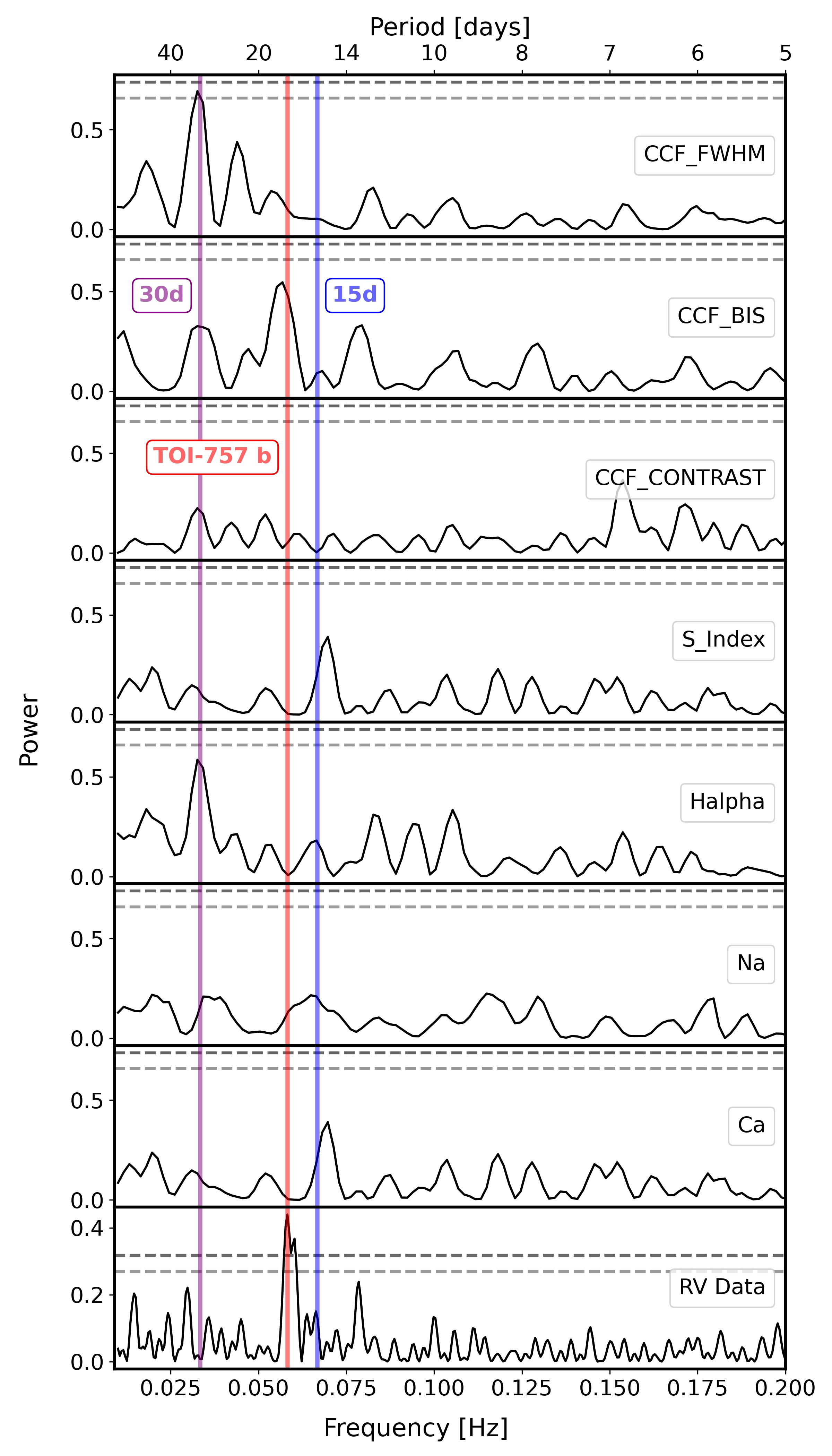}
                \caption{ESPRESSO}
                \label{fig:ESPRESSO_periodogram_lombsc_astropy}
            \end{subfigure}%
        \caption{GLS periodograms of the HARPS and ESPRESSO RV activity indicators. The last panel at the bottom of each figure is the periodogram of the combined HARPS, ESPRESSO and PFS RV data. The solid vertical lines correspond to the 15$-$d (blue) and 30$-$d (purple) rotation signals of the star, overplotted across the entire figure for comparison. The planet signal of TOI$-$757 b from the RV periodogram is shown in red solid lines across both figures. The dashed gray lines represent the FAP significance at the 0.1$\%$ (dark gray) and 1$\%$ (light gray) levels. \textit{Left:} HARPS periodograms of the activity indicators. Starting from the top, we show the CCF$_{\text{FWHM}}$, CCF$_{\text{BIS}}$, CCF$_{\text{CONTRAST}}$, H$_{\alpha}$ Index, Na D$_1$ Index, Na D$_2$ Index, dlw Index, crx index, and S$-$Index activity indicators. \textit{Right:} ESPRESSO periodograms of the activity indicators. Starting from the top, we show the CCF$_{\text{FWHM}}$, CCF$_{\text{BIS}}$, CCF$_{\text{CONTRAST}}$, S$-$Index, H$_{\alpha}$ Index, Na Index, and Ca Index activity indicators.}
        \label{fig:lombsc_periodograms}
        \end{figure*}

        We also computed the Generalized Lomb$-$Scargle (GLS) periodograms \citep{astropy:2013, astropy:2018, astropy:2022}\footnote{We note that the \texttt{astropy} package defines the LombScargle class under the timeseries module as a "LombScargle Periodogram". However, the application of this is the same as the Generalized Lomb$-$Scargle Periodogram when using the \texttt{fit\_mean=True} argument, which is what we did for all our periodograms.} of the HARPS and ESPRESSO activity indicators (Sect.~\ref{sec:observations}).
        We found repeated peaks across different activity indicators for $\sim$15 days and $\sim$30 days (Figure \ref{fig:lombsc_periodograms}).
        To check the significance of these peaks, we computed their False Alarm Probability (FAP) using the bootstrap method \citep[see, e.g., ][]{Kuester1997,Hatzes2019}.
        Briefly, we computed the LS periodogram of 10,000 simulated time$-$series obtained  by randomly shuffling the measurements and their error bars, while keeping the observation time$-$stamps fixed, and defined the FAP as the fraction of those periodograms whose highest power exceeds the power of the signal found in the original data at any frequency. 
        We considered a peak to be significant if it has a FAP\,>\,0.1\%, and found that the majority of the peaks are not significant at the 0.1\% level.

        We also used the chromospheric activity index $\mathrm{\log\,R ^\prime_\mathrm{HK}}$ of the Ca {\sc ii} H\,\&\,K lines to estimate the rotation period of the star. We found a mean $\mathrm{\log\,R ^\prime_\mathrm{HK}}$ value of about $-$4.82 and $-$5.04 for the HARPS and ESPRESSO spectra, respectively.
        The $P_{\star}-\log\,R ^\prime_\mathrm{HK}$ empirical relation from Figure 13 in \cite{suarez2015} suggests a rotation period $P_{\star}$ of $\sim$30 days for TOI$-$757.
        If this is the case, the 15$-$d period found in the KELT photometry, as well as in the periodograms of the spectroscopic activity indicators, would be the first harmonic of the rotation period.
        
        We also estimated the stellar rotation period using the projected equatorial rotational velocity $v\sin(i)$ as derived from the HARPS and ESPRESSO spectra (Sect.~\ref{subsec:staratmos}), and found a value of $v\sin(i)$ = 2.4 $\pm$ 0.4 km s$^{-1}$. 
        Using the stellar radius (Sect.~\ref{subsec:starproperties}) and assuming the star is seen equator$-$on, we found a rotation period of $P_{\star} = 16.3_{-2.3}^{+1.2}$ days, which is in agreement with the 14.2$-$d period found by \citet{oelkers2018kelt} as well as the 15$-$d signal seen in the activity indicators.
        However, due to the well known degeneracy between $v\sin(i)$ and the macroturbulent velocity $v_{\text{mac}}$ for slowly rotating stars \citep[see, e.g.,][]{Sundqvist2013,Simon2014}, it is possible that the $v_{\text{mac}}$ has been underestimated and, consequently, the $v\sin(i)$ overestimated.
        
        We finally fitted the RV time$-$series using GP regression to account for the stellar activity, as described in detail in Sect. \ref{subsec:rvfit}. 
        Using a wide Uniform prior of $\mathcal{U}$[10, 50] for $P_{\text{RV,GP}}$, the solution finds a bimodal distribution in the posteriors with a primary peak at 30$-$d and a secondary peak at 15$-$d.
        To test how this affects our analysis, we constrained the $P_{\text{RV,GP}}$ prior to be centered around 15$-$d ($\mathcal{U}$[1, 22]) and 30$-$d ($\mathcal{U}$[20, 50]), and we found no significant difference within 1$-\sigma$ in the final planet parameters derived from the fits.

        It does not seem clear which of the 15$-$d or 30$-$d signals represent the true rotation period of the star from this analysis alone. 
        Since we could not verify the validity of the periodogram signals with certainty, we looked at evidence from other independent measurements using photometric data (WASP, KELT), the projected equatorial rotational velocity of the star $v\sin(i)$, and the chromospheric activity index $\mathrm{\log\,R ^\prime_\mathrm{HK}}$.
        However, these avenues also did not provide a conclusive picture of the rotation period.
        The photometric variability amplitude in the WASP and KELT data is also found to be low, providing no definitive clues.
        One possible explanation that could account for these two stellar signals is that the star rotates at $\sim$30$-$d and has two active regions at opposite longitudes, producing power at the first harmonic at $\sim$15$-$d.
        The ambiguity of the stellar rotation does not affect the main results and conclusions of the paper.
        We verified this by adopting tighter priors for the GP \texttt{RotationTerm} kernel in our RV fit centered around 15$-$d and 30$-$d, and both yielded consistent results within 1$-\sigma$ for our planet parameters.
        Given that it causes no significant difference in our final planet parameters, we decided to adopt a prior of $\mathcal{U}$[20, 50] for the $P_{\text{RV,GP}}$ in our RV fitting (see Sect. \ref{subsec:rvfit}).

\section{Planet Confirmation}\label{sec:sysconfirmation}

In this section, we describe how we confirmed the planetary nature of our transiting system and ruled out the presence of nearby stellar companions.
We performed periodogram analyses of the RV data, and used precise \textit{Gaia} astrometric measurements, as well as high$-$contrast speckle imaging. 
The system was also previously analyzed by \cite{giacalone2021vetting} using TRICERATOPS, a vetting and validation tool that is capable of modeling transits for \textit{TESS} planet candidates from the target star as well as nearby contaminant stars.
The study reported TOI$-$757 b as a "\emph{Likely Planet}".

        \subsection{Periodogram Analysis}\label{subsec:rvperiodogorams}  
        
        To verify the transiting planet candidate, we first computed the GLS periodograms of the full RV dataset, which independently found the planet signal at $P$\,$\approx$\,17.2\,days (see the bottom panel of Fig. \ref{fig:lombsc_periodograms}, which shows the periodogram of the combined HARPS, ESPRESSO and PFS RV data).
        To test the significance of the signal, we estimated its FAP using the bootstrap method described in Sect.~\ref{subsec:starvariability}, and found that the peak is significant at the 0.1$\%$ level.
        This period agrees with the period found from the photometric transit fitting ($P$\,$\approx$\,17.5 days), described in Section \ref{subsec:transitfit} and with no counterpart of statistical significance in any of the periodograms of the activity indicators (Fig.~\ref{fig:lombsc_periodograms}).
        The clear presence of the Doppler reflex motion of the planet in the power spectrum of the radial velocities spectroscopically confirms the planetary nature of the transit signal found by \textit{TESS}.

        \subsection{\textit{Gaia} Astrometry}\label{subsec:gaia}  

        We checked the \textit{Gaia} archive to determine whether TOI$-$757 is a single star or whether it might be part of a binary system.
        The Renormalised Unit Weight Error (RUWE) calculated for each \textit{Gaia} source is a good indicator of the multiplicity of the system \citep{lindegren2018ruwe, stassun2021ruwe}.
        \textit{Gaia} defines stars being likely well$-$defined, single systems if the RUWE parameter is less than 1.4. 
        Additionally, if the single$-$star model is well$-$fitted to the astrometric observations of a source, the RUWE parameter is expected to be close to 1.0.
        We checked the \textit{Gaia} astrometric measurements of TOI$-$757, which we found to be precise and well$-$fitted, and we also found the errors of the parallax measurement to be very small (see Table \ref{table:starparams}).
        The RUWE value of TOI$-$757 is 0.968, which is also very close to 1.0. 
        To check for a wide$-$separation binary companion, we also searched through the \cite{elbadry2021gaiabinaries} binary candidate catalogue, and we did not find anything reported for TOI$-$757. 
        This, along with the precise astrometric measurements of the system, indicate that our star is not likely to be part of a binary system with a long$-$period stellar companion.

	\subsection{Speckle Imaging}\label{subsec:speckle}  

        To confirm whether TOI$-$757 is a single$-$star system, we also made use of speckle imaging observations, to check for the presence of nearby stars that could either potentially be the sources of false positive scenarios, or contaminate the transit signal.
        
        On the night of July 14, 2019 (UTC), TOI$-$757 was observed with the High Resolution Camera (HRCam) instrument on the Southern Astrophysical Research (SOAR) 4.1$-$m telescope \citep{sebring2003soar, tokovinin2008soar}. 
        HRCam has a 15$^{\prime\prime}$ field of view and observations are taken using an exposure time of 0.2 s. 
        A minimum of two data cubes (each consisting of 400 frames) are recorded and processed independently for every target, and the final results are averaged across the cubes \citep{tokovinin2018soar}. 
        We process the data as described in \cite{tokovinin2010soar} by computing the power spectrum, which relates to the Fourier transform of the per$-$frame intensity distribution, averaged over all cube frames.

\begin{figure}
        \centering
            \begin{subfigure}{\linewidth}
                \centering
                \includegraphics[width=\linewidth]{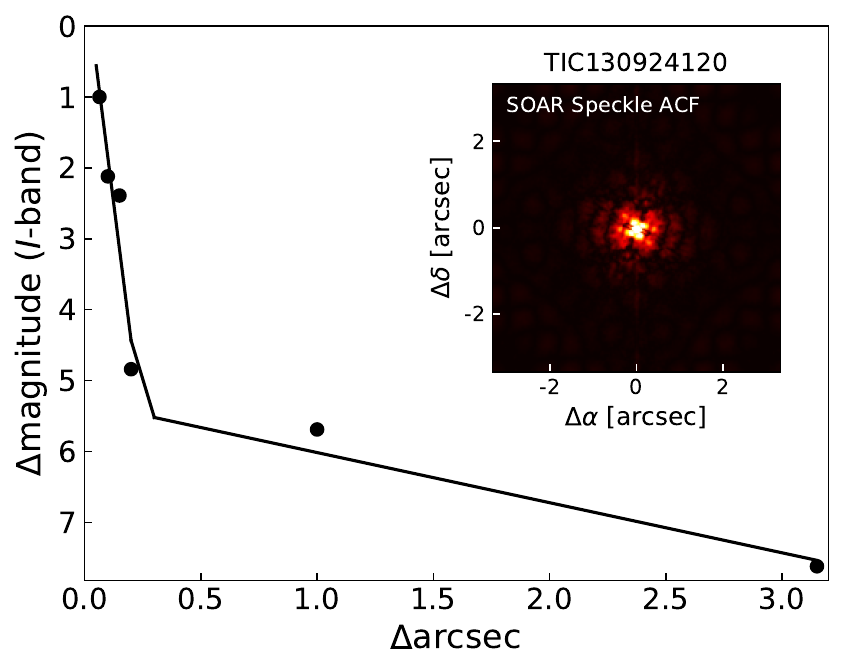}
                \label{fig:speckle_soar_toi757}
            \end{subfigure}%
            
            \begin{subfigure}{\linewidth}
                \centering
                \includegraphics[width=\linewidth, trim={0.4cm 0.3cm 0.2cm 0.8cm}, clip]{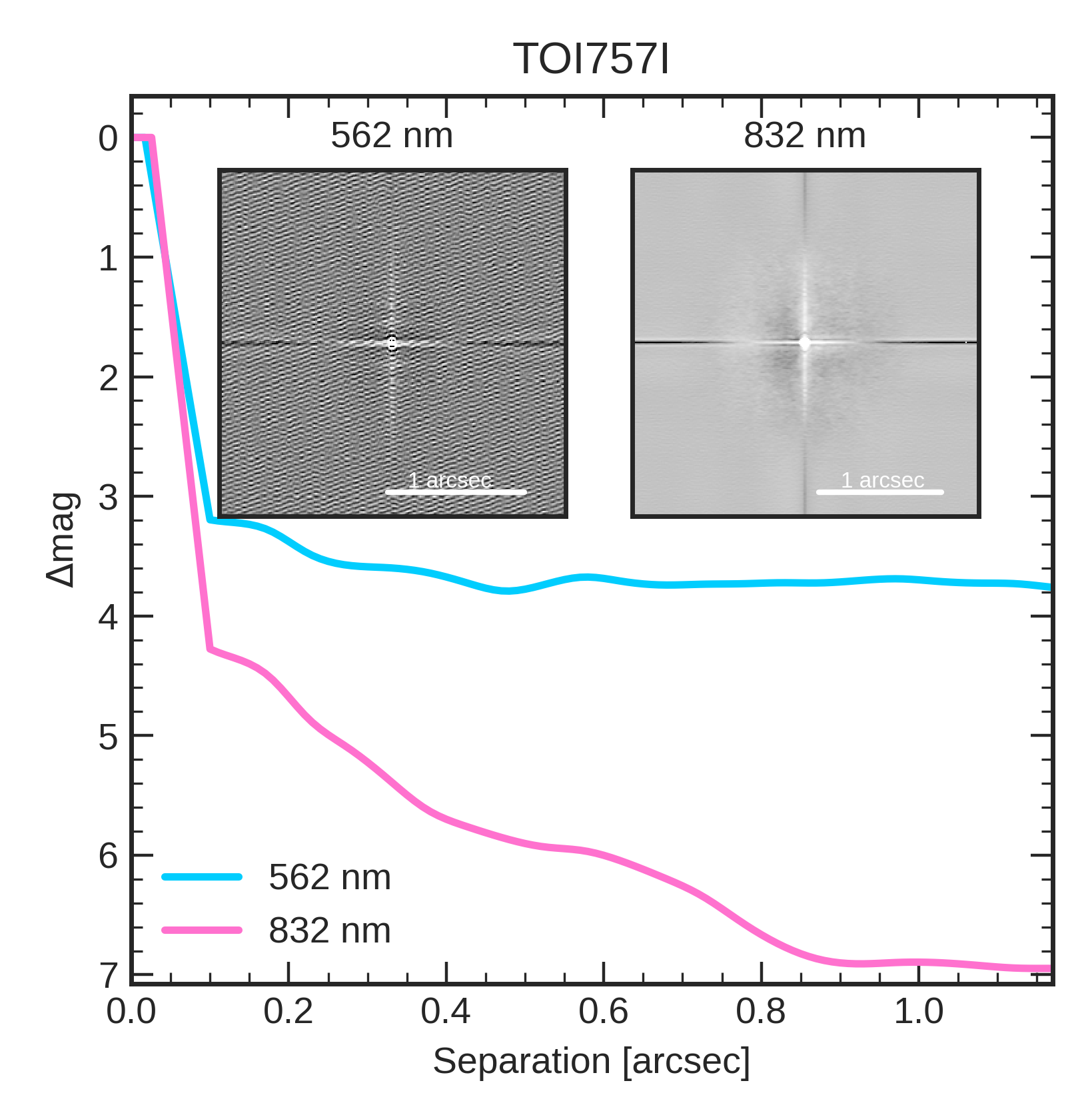}
                \label{fig:speckle_gemini_toi757}
            \end{subfigure}%
    
        \caption{Speckle imaging sensitivity curves resulting from the High$-$contrast imaging instruments listed in Section \ref{subsec:speckle}. \textit{Top:} SOAR/HRCam reconstructed images, showing no signs of stellar companions out to 3.0$^{\prime\prime}$ and within $\Delta$mag~7. \textit{Bottom:} Gemini North/’Alopeke reconstructed images, showing no signs of stellar companions out to 1.2$^{\prime\prime}$ and within $\Delta$mag~7 in the 832 nm filter.}
        \label{fig:speckleimages}
        \end{figure}

        On the night of February 17, 2020 (UT), TOI$-$757 was observed with the 'Alopeke speckle imager \citep{scott2021gemini}, mounted on the 8.1\,m Gemini North telescope on Mauna Kea. 'Alopeke simultaneously acquires data in two bands centered at 562\,nm and 832\,nm using high speed electron$-$multiplying CCDs (EMCCDs). We collected and reduced the data following the procedures described in \citet{howell2011speckle}. The resulting reconstructed image achieved a contrast of $\Delta\mathrm{mag}=7$ at a separation of 1\arcsec in the 832\,nm band (see Fig.~\ref{fig:speckleimages}).

        From our imaging observations, we do not find evidence of nearby stellar companions that might be the source of false positives, or that could be diluting the transit signal.

\section{Photometric and Radial Velocity Analysis}\label{sec:transitrvanalysis}

We make use of the \texttt{exoplanet} toolkit \citep{exoplanet:joss,
exoplanet:zenodo} and its dependencies \citep{exoplanet:foremanmackey17,
exoplanet:foremanmackey18, exoplanet:agol20, exoplanet:arviz, astropy:2013,
astropy:2018, exoplanet:kipping13, exoplanet:luger18, exoplanet:pymc3, exoplanet:theano} in the analyses described in this section.
We perform separate Markov Chain Monte Carlo (MCMC) fits for the photometric and RV data to derive independent radius and mass measurements of the planet.
We describe the fitting method for each case in the subsections below.
For each case, prior to running the MCMC, we initially run the defined global model through a custom non$-$linear optimizer using the \texttt{pymc3-ext} library. 
This enables us to optimize certain parameters and check that the model looks reasonable before running the MCMC sampler.
We then explore the parameter space using the gradient$-$based Hamiltonian Monte Carlo (HMC) sampler from \texttt{PyMC3}, which is more efficient than the traditional Metropolis–Hastings MCMC and also makes use of the No U$-$Turn Sampling (NUTS) algorithm described in \cite{hoffman2014nuts}.

We use Gaussian priors the stellar parameters $R_{\star}$ and $M_{\star}$ in all fits described below.
By extension, this applies a Gaussian prior to $\rho_{\star}$, and particularly helps with constraining the transit model.
For the MCMC fits, we inflate the errors on the stellar mass ($\sim$5\%) and stellar radius ($\sim$4\%) as described in \cite{tayar2022realistic} to achieve realistic uncertainties for the purpose of fitting.
For clarity, the final planet parameters reported in the results, such as planet radius and planet mass, are not extracted using the inflated errors and are calculated using the precise stellar errors derived from the spectra.
All time stamps for the photometric and RV data are given in \textit{TESS} BJD time (BTJD), which is defined as BJD $-$ 2457000.

The complete list of best$-$fit results from the transit and RV analyses can be found in Table \ref{table:priors}, including the priors used as well as the posterior results from the MCMC fit.
The reported best$-$fit posterior values represent the 50$^{\text{th}}$ percentile from the MCMC fit, and the upper and lower uncertainties of the results reflect the 16$^{\text{th}}$ and 84$^{\text{th}}$ percentiles, respectively.
We also performed a joint fit of the RV and transit data using \texttt{pyaneti} \citep{barragan2019pyaneti}, and we found that the results yielded consistent planet parameters with the separate fits done using \texttt{exoplanet}. 
Consequently, performing a joint or separate fit made no difference in the final results and conclusions of the paper.

\begin{table*}
\centering
\caption{Summary of fitted parameters, priors and best$-$fit results from the posteriors for the photometric and RV fitting. Parameters that are left blank for the distribution and prior entries are either derived using correlated parameterization (eg. $e$ and $\omega_{\text{p}}$) or calculated using the results of the orbit model. The upper and lower uncertainties of the best$-$fit posterior values reflect the 16$^{\text{th}}$ and 84$^{\text{th}}$ percentiles from the MCMC fit, respectively, while the reported best$-$fit value is the 50$^{\text{th}}$ percentile. $\mathcal{U}$ refers to a Uniform distribution, $\mathcal{N}$ a Normal, $\mathcal{N}[\mathcal{U}]$ a Bounded Normal, and $\mathcal{C}$ is a fixed constant.}
\label{table:priors}
\resizebox*{!}{0.95\textheight}{
    \begin{tabular}{llll}
    \toprule
    Parameter                                                         & Distribution               & Prior                    & Posterior                         \\
    \midrule
    \textbf{Stellar Parameters}                                       &                            &                          &                                   \\
    \midrule
    $M_{\star} [M_{\odot}]$                                           & $\mathcal{N}[\mathcal{U}]$ & [0.797, 0.05], [0, 3]    &            0.80 $\pm$ 0.05 \\         
    $R_{\star} [R_{\odot}]$                                           & $\mathcal{N}[\mathcal{U}]$ & [0.773, 0.04], [0, 3]    &            0.78 $\pm$ 0.03 \\         
    $\rho_{\star}$ [g cm$^{-3}$]                                      & $--$                         & $--$                       &            2.4 $\pm$ 0.3 \\         
    $u_{1,\text{\textit{TESS}}}$                                      & $--$                         & $--$                       &            0.6 $\pm$ 0.3 \\         
    $u_{2,\text{\textit{TESS}}}$                                      & $--$                         & $--$                       &           $-$0.07$^{+0.40}_{-0.26}$ \\         
    $u_{1,\text{\textit{CHEOPS}}}$                                    & $--$                         & $--$                       &            0.5$^{+0.2}_{-0.3}$ \\         
    $u_{2,\text{\textit{CHEOPS}}}$                                    & $--$                         & $--$                       &            0.1$^{+0.4}_{-0.3}$ \\         
    $u_{1,\text{LCO-Y}}$                                              & $--$                         & $--$                       &            0.4$^{+0.5}_{-0.3}$ \\             
    $u_{2,\text{LCO-Y}}$                                              & $--$                         & $--$                       &            0.08$^{+0.41}_{-0.36}$ \\             
    $u_{1,\text{LCO-zs}}$                                             & $--$                         & $--$                       &            0.7 $\pm$ 0.5 \\             
    $u_{2,\text{LCO-zs}}$                                             & $--$                         & $--$                       &           $-$0.02$^{+0.49}_{-0.43}$ \\             
    \midrule
    \textbf{Planet Parameters}                                        &                            &                          &                                   \\
    \midrule
    \addlinespace[3pt]
    \multicolumn{4}{l}{ \textit{Transit} } 
    \\
    \addlinespace[3pt]
    $P$ [days]                                                        & $\mathcal{U}$              & [16.97, 17.97]           &  17.46819 $\pm$ 0.00002 \\
    $t_0$ [BJD $-$ 2457000]                                             & $\mathcal{U}$              & [2304.53, 2306.53]       &   2305.5322 $\pm$ 0.0006 \\
    $b$                                                               & $\mathcal{U}$              & [0, 1.03]                &            0.3 $\pm$ 0.2 \\     
    $R_{\text{p}}/R_{\star}$                                          & $\mathcal{U}$              & [0.024, 0.034]           &      0.0299$^{+0.0009}_{-0.0006}$ \\
    $R_{\text{p}} [R_{\oplus}]$                                       & $--$                         & $--$                       &                   2.5 $\pm$ 0.1 \\  
    $a$ [AU]                                                          & $--$                         & $--$                       &         0.122 $\pm$ 0.002 \\ 
    $i$ [$^{\circ}$]                                                  & $--$                         & $--$                       &           89.5$^{+0.4}_{-0.3}$ \\
    $t_{\text{dur}}$ [days]                                           & $--$                         & $--$                       &         0.155 $\pm$ 0.002 \\    
    \addlinespace[3pt]
    \multicolumn{4}{l}{ \textit{Radial Velocity} }
    \\
    \addlinespace[3pt]
    $K$ [m s$^{-1}$]                                                  & $\mathcal{N}$              & [2.07, 3]                &            3.3 $\pm$ 0.6 \\
    $M_{\text{p}} [M_{\oplus}]$                                       & $--$                         & $--$                       &           10.5$^{+2.2}_{-2.1}$ \\
    $e$                                                               & $--$                         & $--$                       &            0.39$^{+0.08}_{-0.07}$ \\
    $\omega_{\text{p}}$ [$^{\circ}$]                                  & $--$                         & $--$                       &        $-$36$^{+14}_{-10}$ \\ 
    \addlinespace[3pt]
    \multicolumn{4}{l}{ \textit{Derived Properties} }
    \\
    \addlinespace[3pt]
    $T_{\text{eq}}$ [K]                                               & $--$                         & $--$                       &                $641 \pm 10$ \\
    $\rho_{\text{p}}$ [g cm$^{-3}$]                                   & $--$                         & $--$                       &                   $3.6 \pm 0.8$ \\
    \midrule
    \textbf{Instrument Parameters}                                    &                            &                          &                                   \\
    \midrule
    \addlinespace[3pt]
    \multicolumn{4}{l}{ \textit{Transit} } 
    \\
    \addlinespace[3pt]
    Systemic Mean Flux $\mu_{\text{\textit{TESS}}}$ [ppt]             & $\mathcal{N}$              & [0, 10]                  &         0.019 $\pm$ 0.009 \\
    Systemic Mean Flux $\mu_{\text{\textit{CHEOPS}}}$ [ppt]           & $\mathcal{N}$              & [0, 10]                  &         0.002 $\pm$ 0.009 \\
    Systemic Mean Flux $\mu_{\text{LCO-Y}}$ [ppt]                     & $\mathcal{N}$              & [0, 10]                  &           $-$0.25 $\pm$ 0.1 \\
    Systemic Mean Flux $\mu_{\text{LCO-zs}}$ [ppt]                    & $\mathcal{N}$              & [0, 10]                  &           $-$0.1 $\pm$ 0.1 \\
    Jitter $\sigma_{\text{\textit{TESS}}}$ [ppt]                      & $\mathcal{N}$              & [0.62, 10]               &         0.616$^{+0.002}_{-0.002}$ \\
    Jitter $\sigma_{\text{\textit{CHEOPS}}}$ [ppt]                    & $\mathcal{N}$              & [0.5, 10]                &         0.228 $\pm$ 0.009 \\
    Jitter $\sigma_{\text{LCO-Y}}$ [ppt]                              & $\mathcal{N}$              & [1.62, 10]               &         0.003$^{+0.270}_{-0.003}$ \\
    Jitter $\sigma_{\text{LCO-zs}}$ [ppt]                             & $\mathcal{N}$              & [2, 10]                  &            1.69 $\pm$ 0.07 \\
    \addlinespace[3pt]
    \multicolumn{4}{l}{ \textit{Radial Velocity} }
    \\
    \addlinespace[3pt]
    Systemic Mean Velocity $\mu_{\text{HARPS}}$ [m s$^{-1}$]          & $\mathcal{N}$              & [1.04, 10]               &            0.8 $\pm$ 0.6 \\
    Systemic Mean Velocity $\mu_{\text{HARPS\_FWHM}}$ [m s$^{-1}$]    & $\mathcal{N}$              & [6591.91, 2.01]          &         6592.4$^{+1.8}_{-1.9}$ \\
    Systemic Mean Velocity $\mu_{\text{ESPRESSO}}$ [m s$^{-1}$]       & $\mathcal{N}$              & [1.65e-13, 10]           &            0.20 $\pm$ 0.5 \\
    Systemic Mean Velocity $\mu_{\text{ESPRESSO\_FWHM}}$ [m s$^{-1}$] & $\mathcal{N}$              & [6886.15, 1.22]          &         6887.0 $\pm$ 1.1 \\
    Systemic Mean Velocity $\mu_{\text{PFS}}$ [m s$^{-1}$]            & $\mathcal{N}$              & [1.83, 10]               &           $-$0.5$^{+1.4}_{-1.7}$ \\
    Jitter $\sigma_{\text{HARPS}}$ [m s$^{-1}$]                       & $\mathcal{U}$              & [0, 5]                   &            0.4$^{+0.4}_{-0.3}$ \\
    Jitter $\sigma_{\text{HARPS\_FWHM}}$ [m s$^{-1}$]                 & $\mathcal{U}$              & [0, 5]                   &            3.9$^{+0.8}_{-1.4}$ \\
    Jitter $\sigma_{\text{ESPRESSO}}$ [m s$^{-1}$]                    & $\mathcal{U}$              & [0, 5]                   &            1.2$^{+0.6}_{-0.7}$ \\
    Jitter $\sigma_{\text{ESPRESSO\_FWHM}}$ [m s$^{-1}$]              & $\mathcal{U}$              & [0, 5]                   &            4.0$^{+0.8}_{-1.3}$ \\
    Jitter $\sigma_{\text{PFS}}$ [m s$^{-1}$]                         & $\mathcal{U}$              & [0, 5]                   &            0.8 $\pm$ 0.5 \\
    \midrule
    \textbf{GP Parameters}                                            &                            &                          &                                   \\
    \midrule
    \addlinespace[3pt]
    \multicolumn{4}{l}{ \textit{Transit} } 
    \\
    \addlinespace[3pt]
    $P_{\text{PHOT,GP}}$ [days]                                       & $\mathcal{U}$              & [1, 50]                 &            2.0 $\pm$ 0.3 \\
    $Q_{\text{PHOT,GP}}$                                              & $\mathcal{C}$              & $1 / \sqrt{2}$           &                               $--$  \\
    $\sigma_{\text{\textit{TESS},GP}}$ [ppt]                          & $\mathcal{U}$              & [0, 10]                  &         0.072$^{+0.006}_{-0.005}$ \\
    \addlinespace[3pt]
    \multicolumn{4}{l}{ \textit{Radial Velocity} }
    \\
    \addlinespace[3pt]
    $P_{\text{RV,GP}}$ [days]                                         & $\mathcal{U}$              & [20, 50]                 &           31.0$^{+1.4}_{-1.7}$ \\
    $Q_{0~\text{RV,GP}}$                                              & $\mathcal{U}$              & [$\pi$, 10000]           &            4.13$^{+2.03}_{-0.76}$ \\
    $\delta Q_{\text{RV,GP}}$                                         & $\mathcal{U}$              & [0.001, 10000]           &          1.9$^{+295.4}_{-1.9}$ \\
    $f_{\text{RV,GP}}$                                                & $\mathcal{U}$              & [0, 1]                   &            0.7$^{+0.2}_{-0.3}$ \\
    $\sigma_{\text{HARPS,GP}}$ [m s$^{-1}$]                           & $\mathcal{U}$              & [0, 10]                  &            2.8$^{+1.0}_{-0.7}$ \\
    $\sigma_{\text{HARPS\_FWHM,GP}}$ [m s$^{-1}$]                     & $\mathcal{U}$              & [0, 100]                 &          30.4$^{+12.5}_{-8.9}$ \\
    $\sigma_{\text{ESPRESSO,GP}}$ [m s$^{-1}$]                        & $\mathcal{U}$              & [0, 10]                  &            1.9$^{+1.0}_{-0.9}$ \\
    $\sigma_{\text{ESPRESSO\_FWHM,GP}}$ [m s$^{-1}$]                  & $\mathcal{U}$              & [0, 100]                 &           14.2$^{+4.5}_{-3.3}$ \\
    $\sigma_{\text{PFS,GP}}$ [m s$^{-1}$]                             & $\mathcal{U}$              & [0, 10]                  &            4.0$^{+2.0}_{-1.3}$ \\
    \bottomrule
    \end{tabular}
    }
\end{table*}

	\subsection{Photometric Fit}\label{subsec:transitfit}  

        In this section, we describe how we perform the global photometric transit fitting of space$-$based (\textit{TESS}, \textit{CHEOPS}) and ground$-$based (LCOGT) data acquired for TOI$-$757 b. 
        We treat LCOGT data taken with different filters (Y-band and $z$s$-$band) separately when fitting the lightcurves, henceforth referred to as LCO$-$Y and LCO$-z$s respectively.

    	\subsubsection{Transit model and MCMC Fit}\label{subsubsec:transitmethods}  

            We define a global transit model in \texttt{exoplanet} to fit over stellar, planetary and telescope$-$dependent parameters, as described below.
            We use the full, unbinned \textit{TESS} lightcurves with the Out$-$of$-$Transit (OOT) data, which enables us to simultaneously detrend the lightcurves using a GP component while performing the transit fitting.
            The \textit{CHEOPS} and LCOGT lightcurves are pre$-$processed and detrended as described in Sect. \ref{subsubsec:obscheops} and Sect. \ref{subsubsec:obslcogt}, respectively.
            As such, no GPs were used on these lightcurves while performing the fitting.
            
            For the stellar and planetary parameters, we use the built$-$in Keplerian orbit model and fit for the stellar radius $R_{\star}$, the stellar mass $M_{\star}$, the planet period $P$, the transit center time $t_{0}$, the planet eccentricity $e$, the argument of periastron of the planetary orbit $\omega_{\text{p}}$, the impact parameter $b$, and the radius ratio $R_{\text{p}}/R_{\star}$. 
            The $R_{\text{p}}/R_{\star}$ parameter is used to set the bounds on $b$ via the \texttt{ImpactParameter} distribution from \texttt{exoplanet} based on physical considerations, since the impact parameter cannot exceed the sum of the planet and star radii, which places a Uniform distribution prior in the range [0, 1 + $R_{\text{p}}/R_{\star}$].
            Since $e$ and $\omega_{\text{p}}$ are correlated, we also parameterize them using $\sqrt{e} \sin(\omega_{\text{p}})$ and $\sqrt{e} \cos(\omega_{\text{p}})$ when performing the fitting \citep{eastman2013exofast, anderson2011wasp30b}. 
            Additionally, we fit over the Limb$-$Darkening coefficients $u_{1}$ and $u_{2}$.            
            We fit for this separately for each telescope (due to instrumentation and observing differences between them) using the \cite{kipping2013quadlimb} parameterization for quadratic limb darkening. 
            This follows the form described in Eq. \ref{eq:quadlimb}, using Uniform priors in the range [0,1], and constrained by the boundary conditions on $u_{1}$ and $u_{2}$.
            We fit over $u_{1}$ and $u_{2}$ directly, which are constrained by the parametrization of $q_{1}$ and $q_{2}$.

                \begin{equation}\label{eq:quadlimb}
                \begin{dcases}
                    \text{Parametrization}                            & \text{Boundary Conditions} \\
                    q_{1} = (u_{1} + u_{2})^{2}                       & u_{1} + u_{2} < 1 \\
                    q_{2} = 0.5 ~ u_{1} ~ (u_{1} + u_{2})^{-1}        & u_{1} > 0 \\
                                                                      & u_{1} + 2 ~ u_{2} > 0 \\
                \end{dcases}     
                \end{equation}

            From the orbit model, we track other planetary parameters such as the implied stellar density $\rho_{\star}$, the semi$-$major axis $a/R_{\star}$ and the inclination angle $i$.
            To calculate the transit duration time $t_{\text{dur}}$, we approximate it using the following equation (see \cite{winn2010transitdur} for more details):

                \begin{equation}\label{eq:duration}
                t_{\text{dur}} = \frac{P}{\pi} \times ~ \arcsin \Bigg( \frac{R_{\star}}{a} \frac{\sqrt{(1 - R_{\text{p}}/R_{\star})^{2} - b^{2}}}{\sin(i)} \Bigg) \times \Bigg( \frac{\sqrt{1 - e^{2}}}{1 + e \sin(\omega)} \Bigg)
                \end{equation}

            This describes the full transit duration, which is defined as the time between the second and third contact (when the planetary disk is completely in front of the stellar disk), also commonly referred to as $t_{\text{23}}$.
            
            For each of the \textit{TESS}, \textit{CHEOPS}, LCO$-$Y and LCO$-z$s lightcurves, we define sets of telescope$-$dependent parameters: the systemic mean flux $\mu$ and a jitter term $\sigma$. 
            For the \textit{TESS} data, we use an additional GP component for the data residuals to account for any additional instrumental noise and stellar activity signals.
            The kernel used for the TESS GP component is the \texttt{celerite2} SHOTerm, a stochastically$-$driven, damped harmonic oscillator \citep{exoplanet:foremanmackey17}.
            This kernel includes three free hyperparameters that we fit: the amplitude of the noise model $\sigma$, a rotation term $\rho$ representing the star's periodicity, and the complexity term $Q$, which is related to the decay timescale of stellar activity signals.
            The planet parameters are shared across all the photometric data when fitting, but the stellar and telescope$-$dependent parameters are fit separately for each instrument.
            We use loose, uninformative priors when fitting all our parameters.
            
            When running the MCMC, we used 2 chains with 8000 tuning steps and 8000 draws, resulting in a total of 16,000 independent samples for each parameter.
            For convergence diagnostics, we checked the Gelman$-$Rubin statistic for our resulting MCMC chains, and we found them all to be <1.01 for all our sampled parameters, indicating that there were no major issues during sampling.
            We also checked the number of effective samples for each parameter, and all were well above 1000 samples per chain, indicating that a sufficient number of samples were independently drawn for every parameter.

    	\subsubsection{Transit$-$only Fit Results}\label{subsubsec:transitresults}  
            
            We find that TOI$-$757 b has an orbital period of $P$ = 17.46819$\pm$ 0.00002 days, with a best$-$fit planet$-$to$-$star radius ratio of $R_{\text{p}}/R_{\star}$ = 0.0299$^{+0.0009}_{-0.0006}$, resulting in a planetary radius of $R_{\text{p}}$ = 2.5 $\pm$ 0.1 $R_{\oplus}$.
            The full transit parameters are listed in Table \ref{table:priors}.
            Since the measurements for $e$ and $\omega$ are not well$-$constrained from transit modeling, we rely on the RV fitting for a more robust result of the planet's eccentricity, described in the next section.
    
\begin{figure*}
            \centering
            \begin{subfigure}{0.5\linewidth}
                \centering
                \includegraphics[width=\linewidth, trim={0.9cm 0.5cm 2.5cm 2cm}, clip]{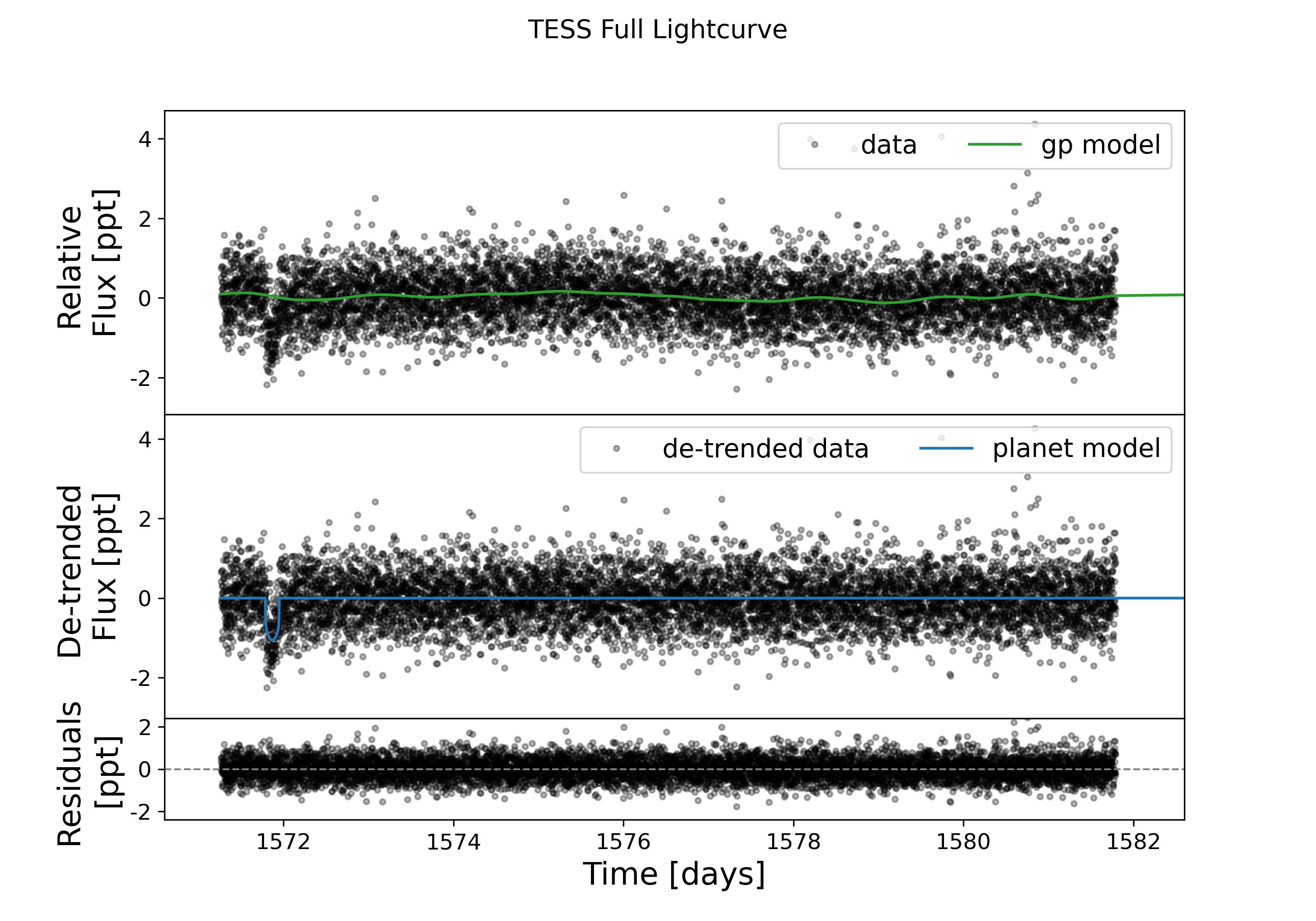}
                \label{fig:tesslctransit1}
            \end{subfigure}%
            \begin{subfigure}{0.5\linewidth}
                \centering
                \includegraphics[width=\linewidth, trim={0.9cm 0.5cm 2.5cm 2cm}, clip]{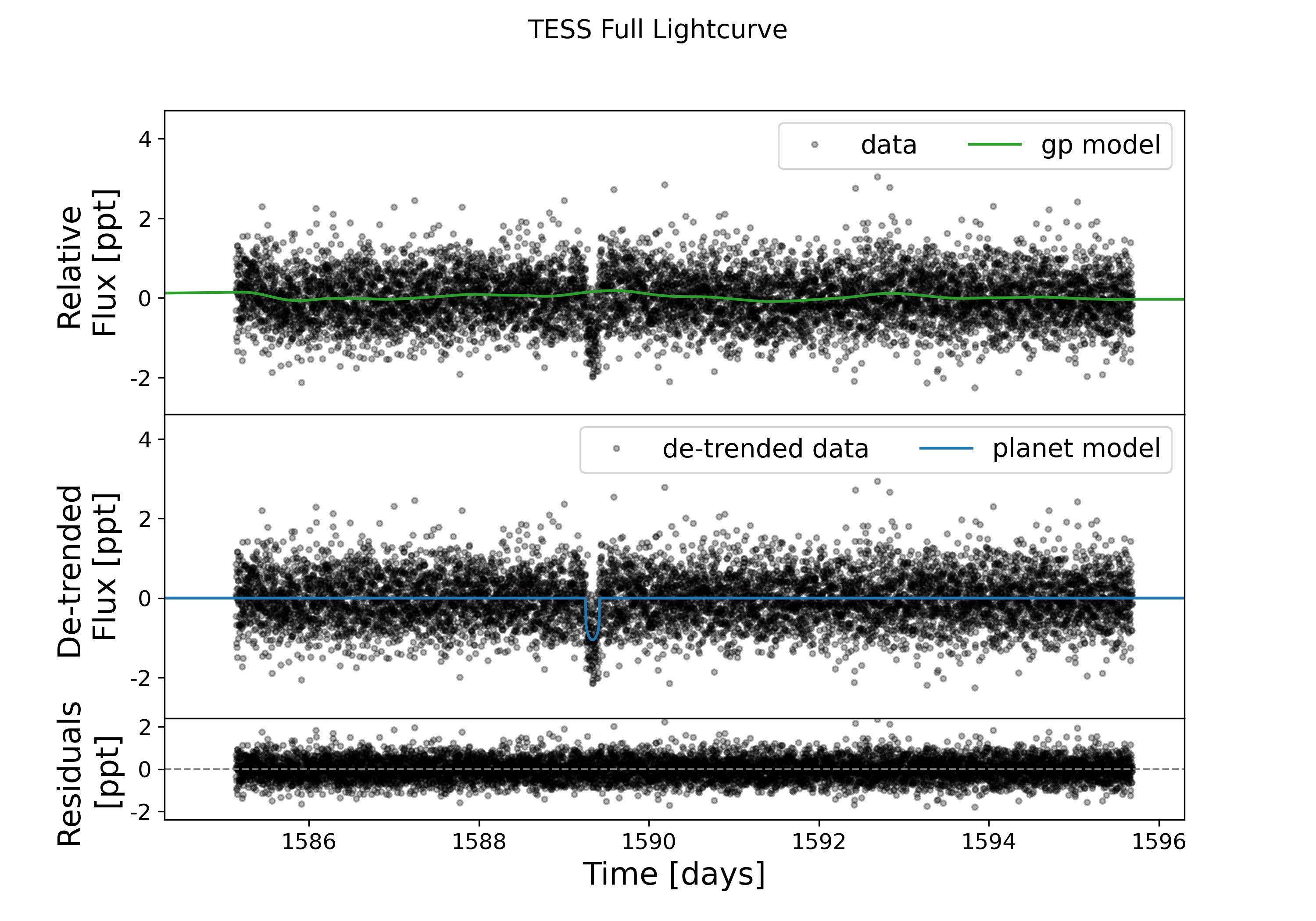}
                \label{fig:tesslctransit2}
            \end{subfigure}%
         
            \begin{subfigure}{0.5\linewidth}
                \centering
                \includegraphics[width=\linewidth, trim={0.9cm 0.5cm 2.5cm 2cm}, clip]{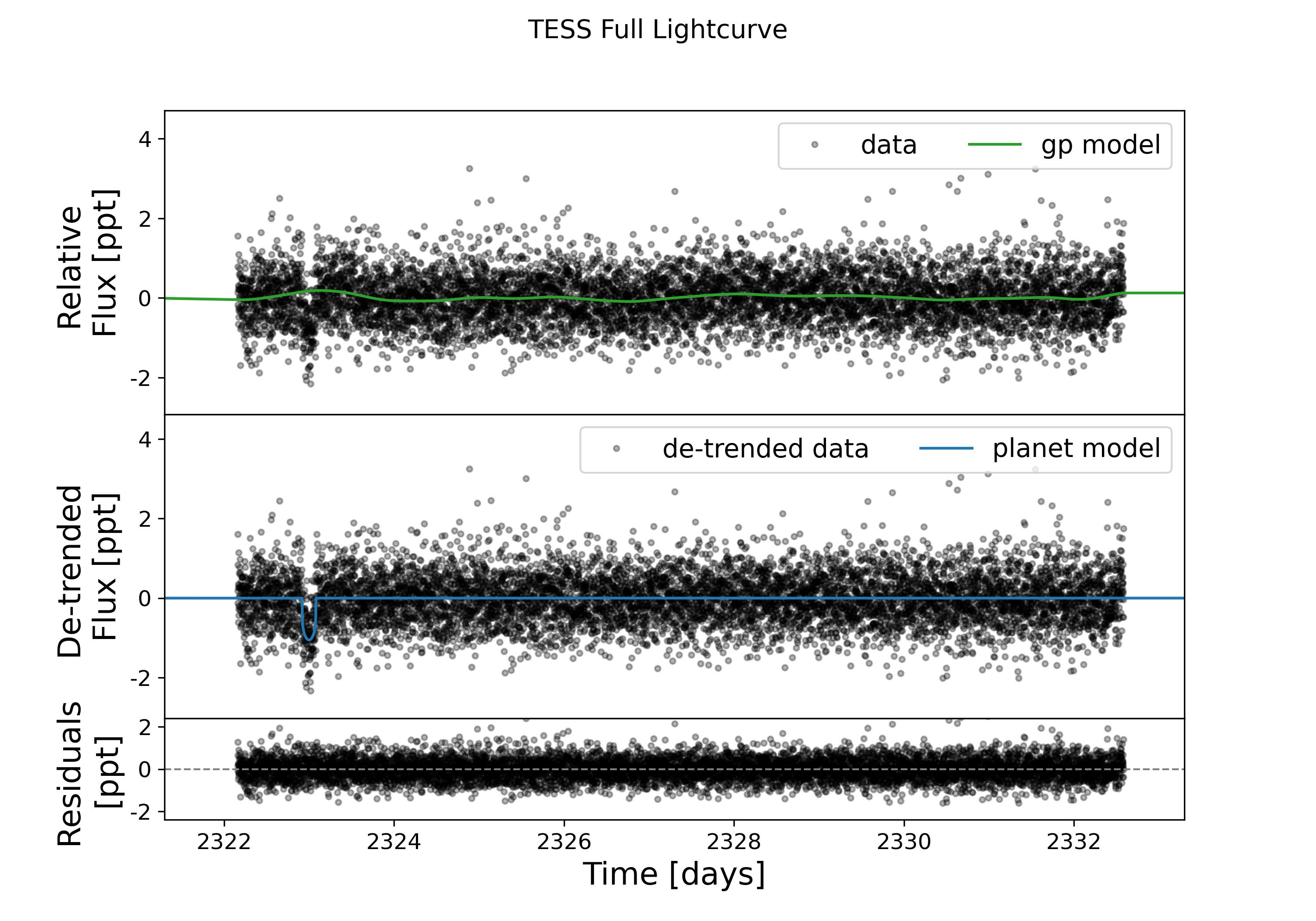}
                \label{fig:tesslctransit3}
            \end{subfigure}%
            \begin{subfigure}{0.5\linewidth}
                \centering
                \includegraphics[width=\linewidth, trim={0.9cm 0.5cm 2.5cm 2cm}, clip]{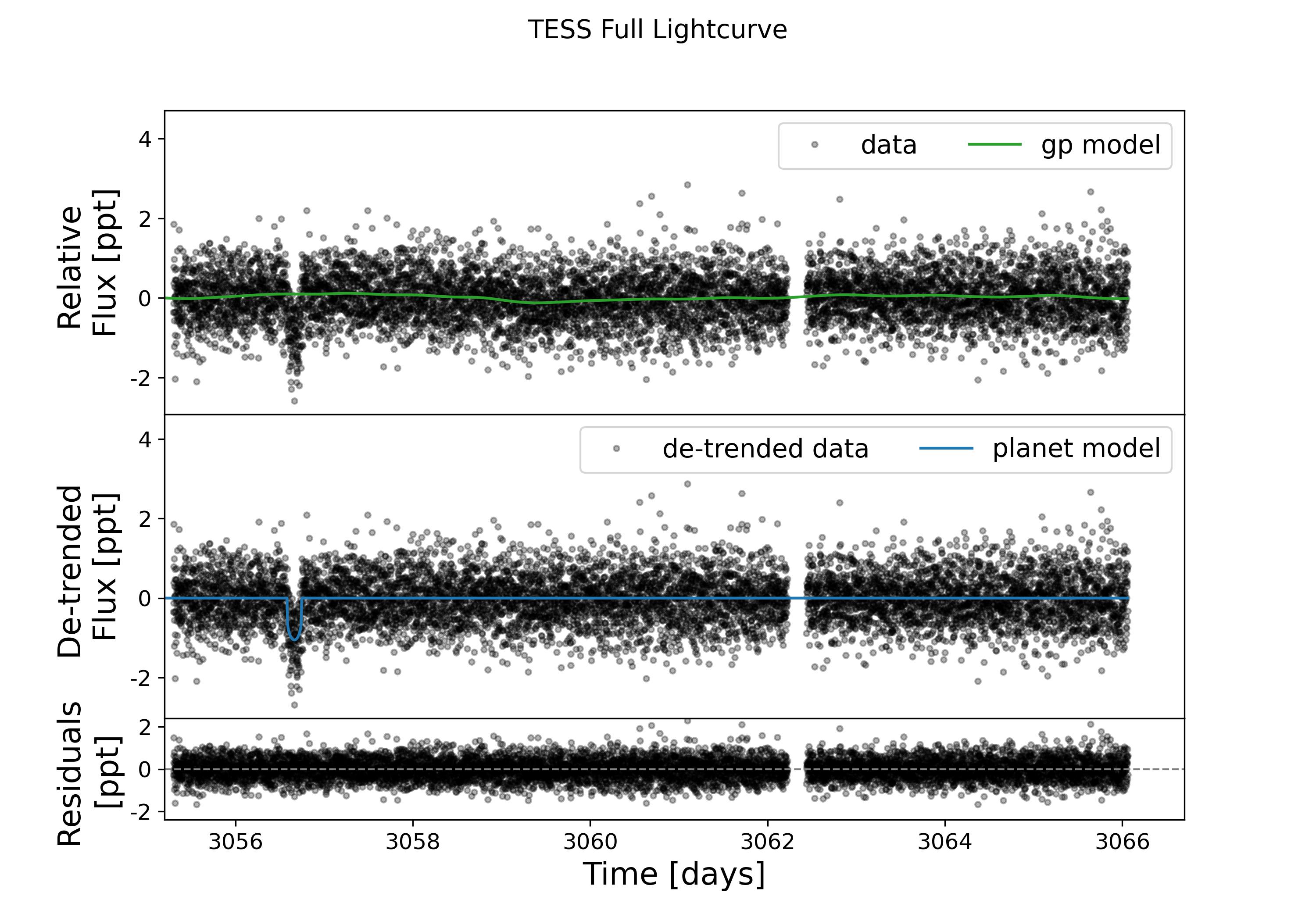}
                \label{fig:tesslctransit4}
            \end{subfigure}%
            \caption{\textit{TESS} SPOC$-$PDCSAP lightcurve data of TOI$-$757 (see Section \ref{subsec:obstess}) taken from Sectors 10 (top two figures), 37 (bottom left figure) and 64 (bottom right figure).
            In the top panel of each figure, we show the raw lightcurves in black dotted points along with the best$-$fit GP model plotted on top in a solid green line.
            The middle panel of the figures shows the detrended data in black dotted points (where the GP model as well as the systemic mean flux $\mu$ of the lightcurve were subtracted out), along with the best$-$fit planet model plotted on top in a solid blue line (see Section \ref{subsec:transitfit}). 
            The bottom panel of the figures shows the residuals of the detrended data after subtracting the planet model.}
            \label{fig:tesslcgp}
            \end{figure*}

\begin{figure*}
            \centering
            \begin{subfigure}{0.5\linewidth}
                \centering
                \includegraphics[width=\linewidth, trim={1.4cm 0.5cm 2.5cm 2cm}, clip]{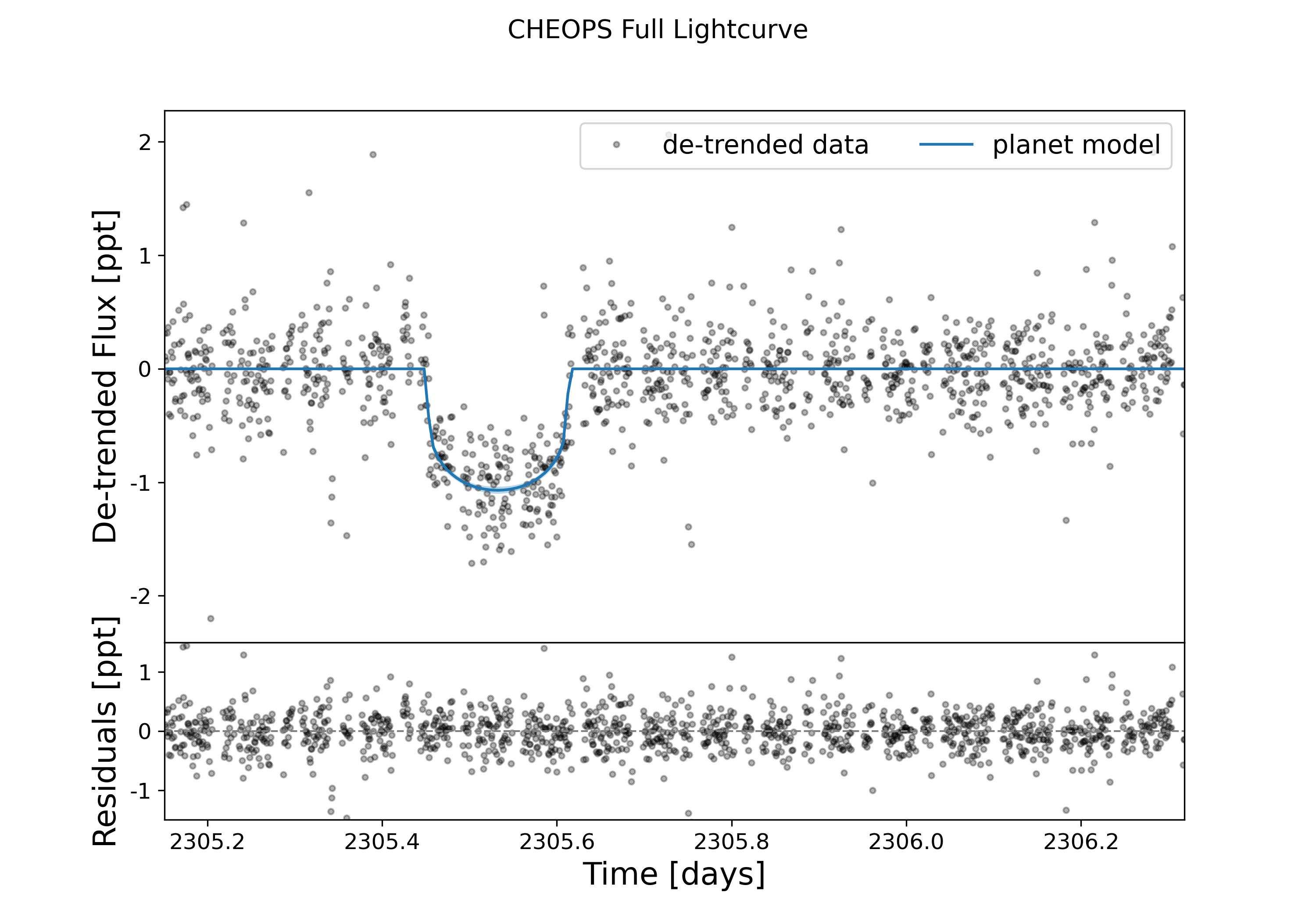}
                \label{fig:cheopslctransit4}
            \end{subfigure}%
            \begin{subfigure}{0.5\linewidth}
                \centering
                \includegraphics[width=\linewidth, trim={1.4cm 0.5cm 2.5cm 2cm}, clip]{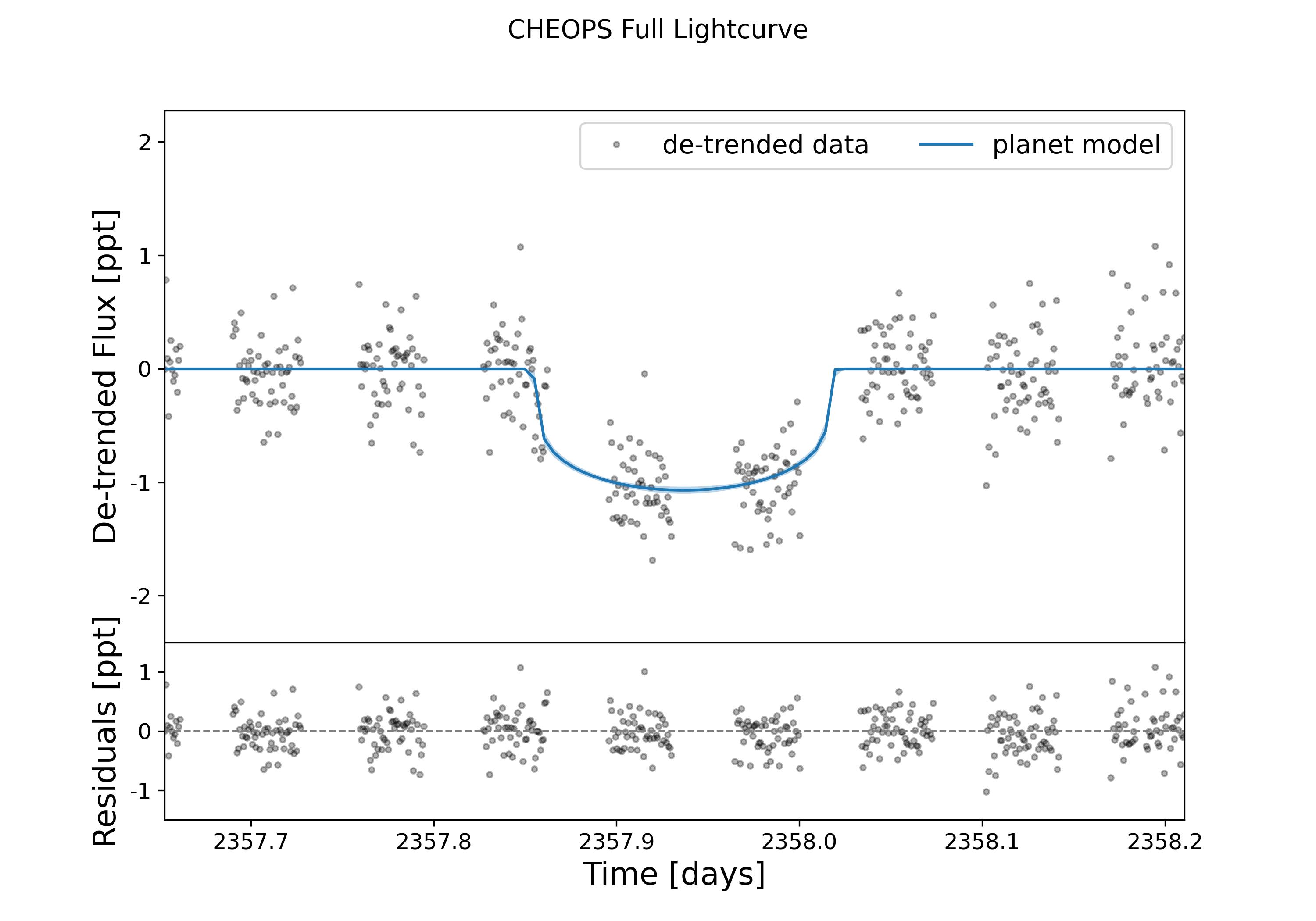}
                \label{fig:cheopslctransit5}
            \end{subfigure}%
            \caption{\textit{CHEOPS} lightcurve data of TOI$-$757 (see Section \ref{subsubsec:obscheops}).
            In the top panel of each figure, we show the detrended data in black dotted points, along with the best$-$fit planet model plotted on top in a solid blue line (see Section \ref{subsec:transitfit}). 
            The bottom panel of the figures shows the residuals of the detrended data after subtracting the planet model.}
            \label{fig:cheopslcgp}
            \end{figure*}

\begin{figure*}
            \centering
            \begin{subfigure}{0.5\linewidth}
                \centering
                \includegraphics[width=\linewidth, trim={0.5cm 0.5cm 2cm 1.1cm}, clip]{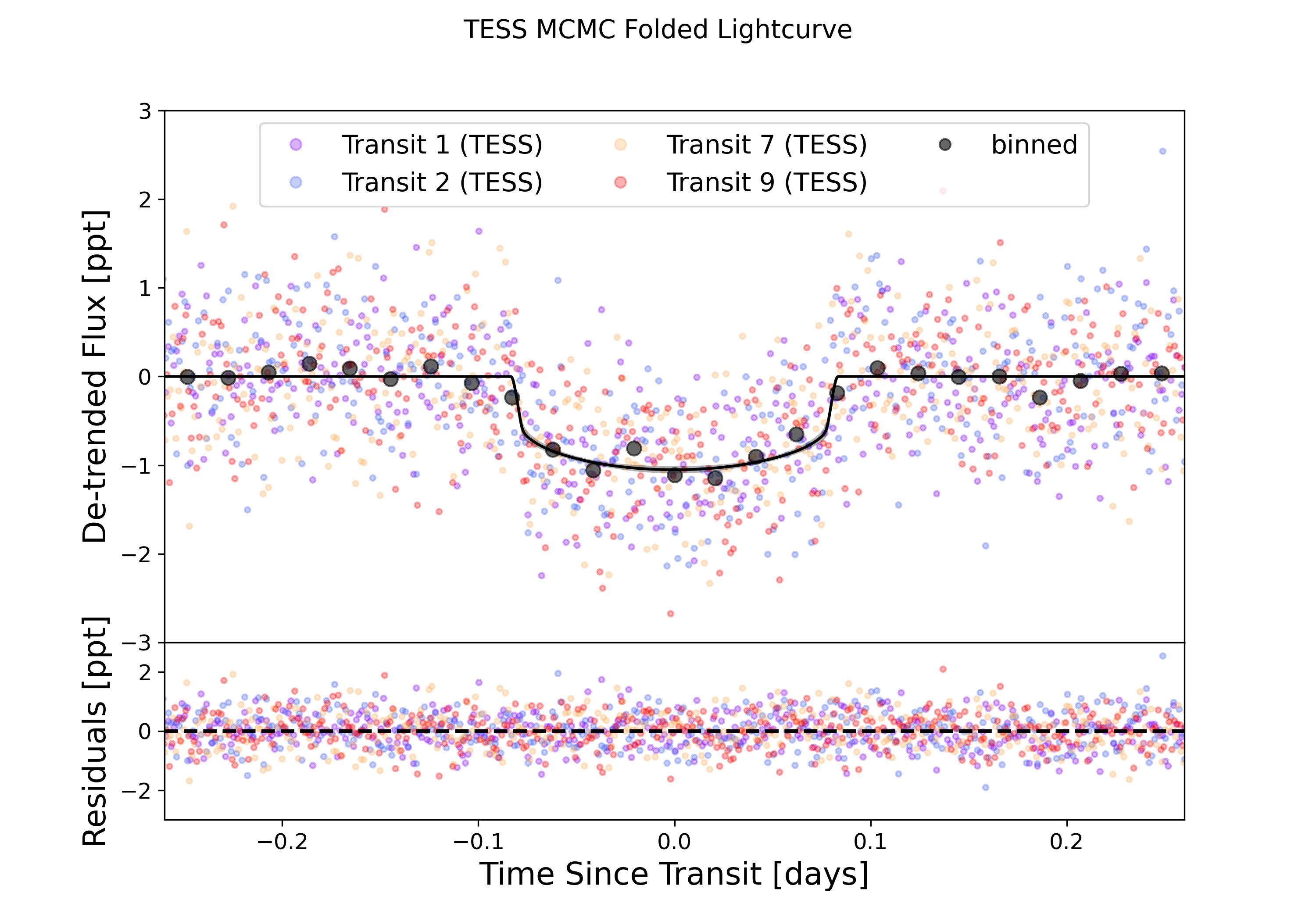}
                \label{fig:tess_lc_trace_folded}
            \end{subfigure}%
            \begin{subfigure}{0.5\linewidth}
                \centering
                \includegraphics[width=\linewidth, trim={0.5cm 0.5cm 2cm 1.1cm}, clip]{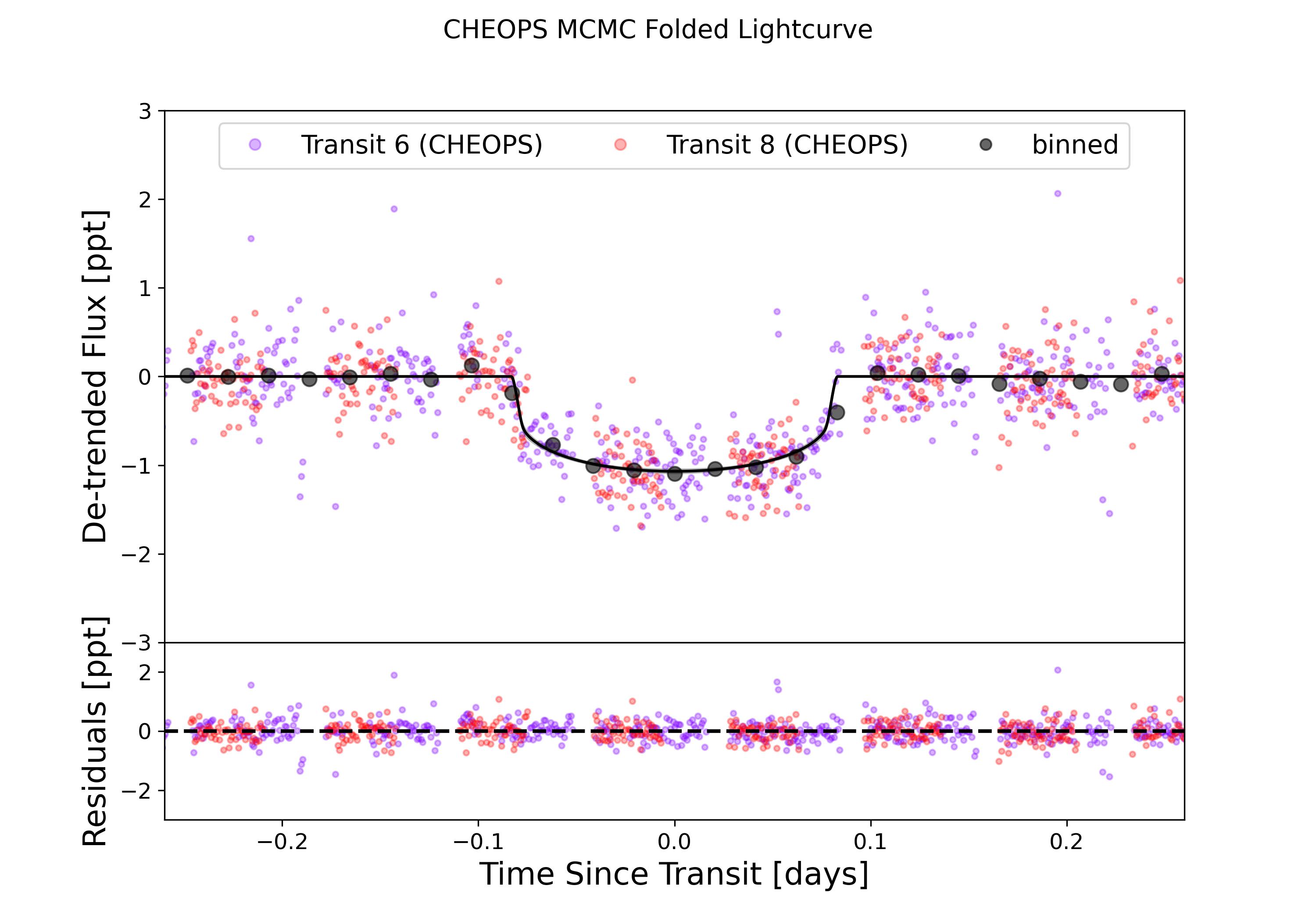}
                \label{fig:cheops_lc_trace_folded}
            \end{subfigure}%
    
            \begin{subfigure}{0.5\linewidth}
                \centering
                \includegraphics[width=\linewidth, trim={0.5cm 0.5cm 2cm 1.1cm}, clip]{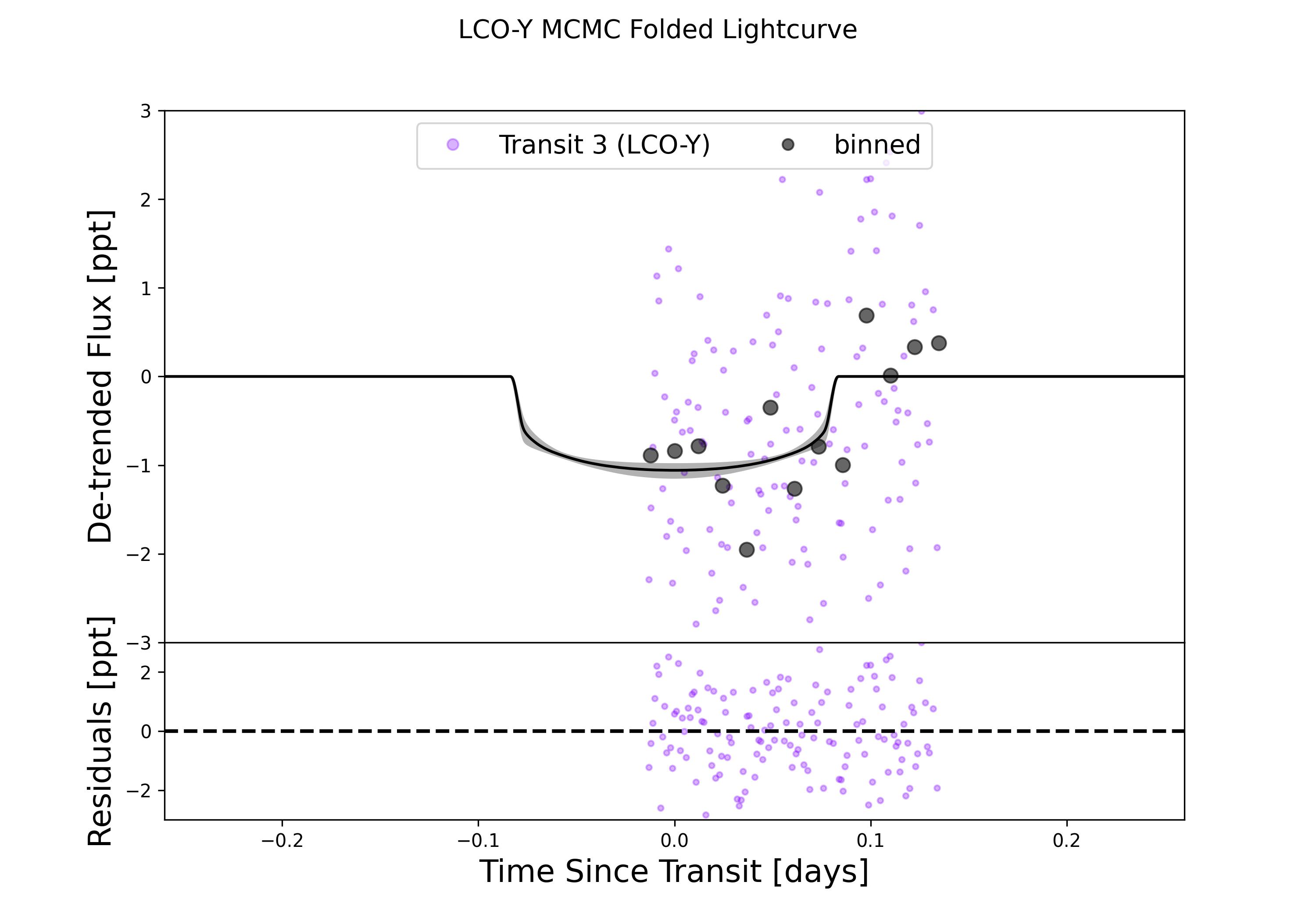}
                \label{fig:lco_y_lc_trace_folded}
            \end{subfigure}%
            \begin{subfigure}{0.5\linewidth}
                \centering
                \includegraphics[width=\linewidth, trim={0.5cm 0.5cm 2cm 1.1cm}, clip]{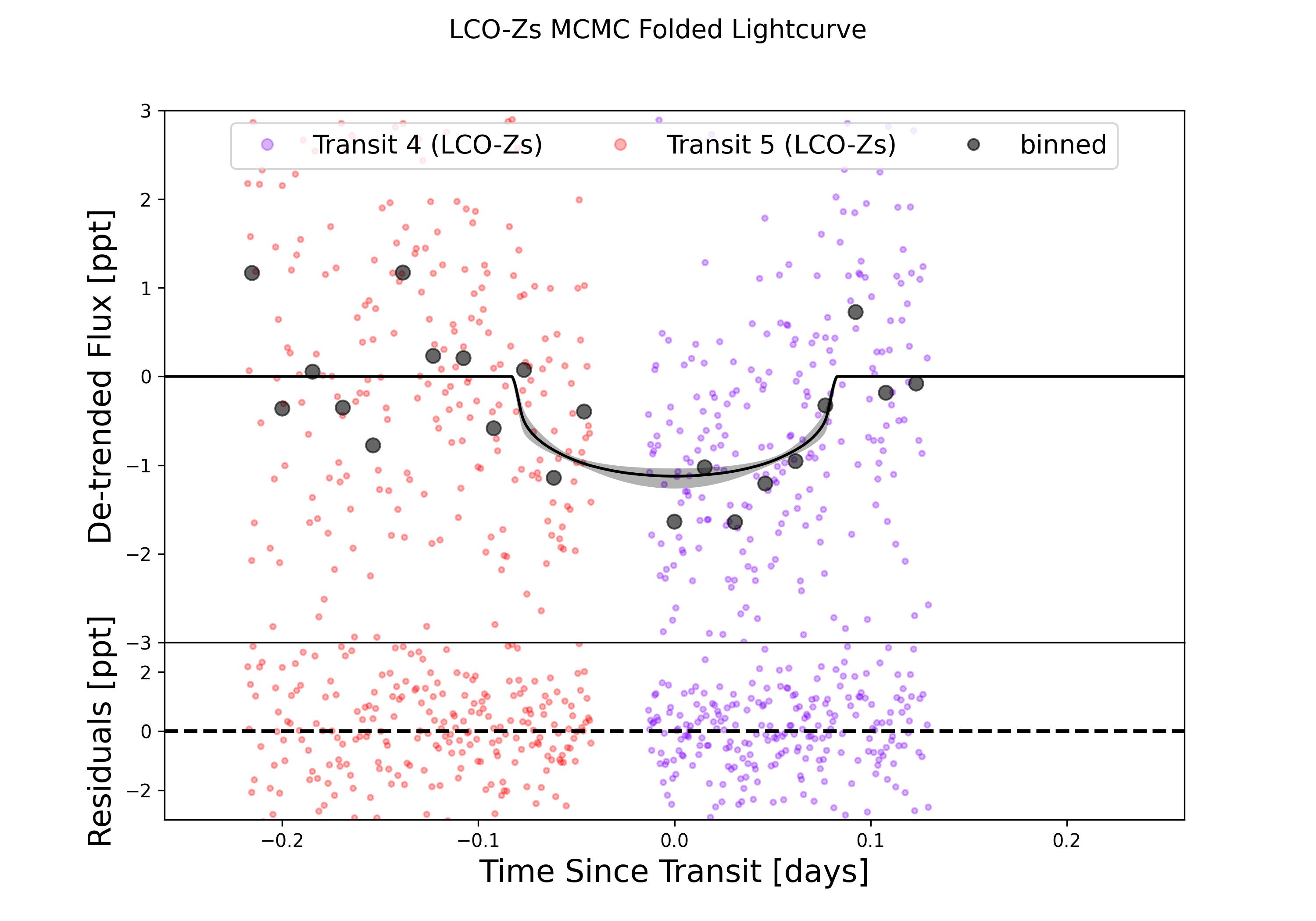}
                \label{fig:lco_zs_lc_trace_folded}
            \end{subfigure}%
            \caption{Detrended photometric lightcurves, folded over the best$-$fit orbital period $P$, along with the best$-$fit planet model and the posterior constraints corresponding to the 16$^{\text{th}}$ and 84$^{\text{th}}$ percentiles (see Section \ref{subsec:transitfit}). The dotted data points are color$-$coded to show the individual transits for each telescope. For clarity, we also overplot the binned detrended data in black dots. \textit{Top Panel:} \textit{TESS} (\textit{left}) and \textit{CHEOPS} (\textit{right}) folded lightcurves. \textit{Bottom Panel:} LCO$-$Y (\textit{left}) and LCO$-z$s (\textit{right}) folded lightcurves.}
            \label{fig:lc_trace_folded}
            \end{figure*}

            In the top panel of Fig.~\ref{fig:tesslcgp}, we show the original PDCSAP \textit{TESS} light curves with the best$-$fit GP model overplotted.
            The middle panel of the figures show the detrended photometric data by subtracting the GP model as well as the systemic mean flux $\mu$ of the light curve, with the best$-$fit planet model overplotted.
            The bottom panel of the plot shows the residuals of the detrended data after subtracting the planet model.
            Fig. \ref{fig:cheopslcgp} shows a similar plot for the \textit{CHEOPS} data, but without any GP component, since the \textit{CHEOPS} data were already pre$-$processed and detrended prior to fitting.
            Fig. \ref{fig:lc_trace_folded} displays the same detrended \textit{TESS} and \textit{CHEOPS} data, as well as the detrended LCO$-$Y and LCO$-z$s lightcurves, folded over the best$-$fit orbital period. 
            We overplot the best$-$fit planet model for each folded lightcurve along with the posterior constraints corresponding to the 16$^{\text{th}}$ and 84$^{\text{th}}$ percentiles.

	\subsection{Radial Velocity Fit}\label{subsec:rvfit}  

        In this section, we describe how we perform the RV fitting of the HARPS, ESPRESSO and PFS data. 
        The RV data for all instruments can be found in Table \ref{table:rvdata}.

    	\subsubsection{RV model and MCMC Fit}\label{subsubsec:rvmethods}  

            We define a global RV model in \texttt{exoplanet} to fit over stellar, planetary and instrument$-$dependent parameters, as described below.
            The stellar and planet parameters are shared between the HARPS, ESPRESSO and PFS data when fitting, but the instrument$-$dependent parameters are fit separately.
            
            For the stellar and planetary parameters, we use the built$-$in Keplerian orbit model and fit for the stellar radius $R_{\star}$, the stellar mass $M_{\star}$, the planet period $P$, the transit center time $t_{0}$, the planet eccentricity $e$, the argument of periastron of the planetary orbit $\omega_{\text{p}}$, the inclination $i$, and the planet mass $M_{\text{p}}$. 
            We use the best$-$fit MCMC values from the results of the photometric transit fit in Section \ref{subsec:transitfit} to set priors on $P$, $i$ and $t_{0}$ when performing the RV fit. 
            Since $e$ and $\omega_{\text{p}}$ are correlated, we also parametrize them using $\sqrt{e} \sin(\omega_{\text{p}})$ and $\sqrt{e} \cos(\omega_{\text{p}})$ when performing the fitting. 
            From the orbit model, we track other planetary parameters such as the implied stellar density $\rho_{\star}$.
    
            For each of the HARPS, ESPRESSO and PFS instruments, we define sets of instrument$-$dependent parameters, including the systemic mean velocity $\mu$, a jitter term $\sigma$, and a GP component for the data residuals to account for any additional instrumental noise and stellar activity signals.
            We also fit GPs to the HARPS and ESPRESSO CCF$_{\text{FWHM}}$ activity indicators to simultaneously model the stellar activity along with the RV data. 
            The kernel used for the RV GP components described in this section is the \texttt{celerite2} RotationTerm, which is a combination of two SHOTerms that can be used to model stellar rotation using two modes in Fourier space, giving us SHO($P_{\star}$) + SHO($P_{\star}$ / 2).
            It includes 5 free hyperparameters that we fit for: the amplitude of the noise model $\sigma$, a rotation term $P_{\star}$ representing the star's periodicity, the complexity term $Q_{0}$ for the secondary oscillation, a term representing the difference between the quality factors ($Q_{1}$ and $Q_{2}$) of the two modes $\delta Q$, and the fractional amplitude of the second mode compared to the first $f$.
            Four of the hyperparameters ($P_{\star}$, $Q_{0}$, $\delta Q$ and $f$) are shared between the activity indicator GPs and the RV GPs, while the fifth hyperparameter $\sigma$ (corresponding to the amplitude of the noise model) is fit separately for each dataset. 
            This is to help constrain the RV data better using the stellar rotation period and any additional activity signals that could be contaminating the RV data.
            Apart from the priors placed on $P$, $i$ and $t_{0}$ from the photometric MCMC results, we use loose, uninformative priors on all other parameters.
            When running the MCMC, we used 2 chains with 8000 tuning steps and 8000 draws, resulting in a total of 16,000 independent samples for each parameter.
            For convergence diagnostics, we checked the Gelman$-$Rubin statistic for our resulting MCMC chains, and we found them all to be <1.01 for all our sampled parameters, indicating that there were no major issues during sampling.
            We also checked the number of effective samples for each parameter, and all were well above 1000 samples per chain, indicating that a sufficient number of samples were independently drawn for every parameter.

    	\subsubsection{Model Comparison}\label{subsubsec:modelcomp}  
            
            We initially tested running the fit for different sets of models to see which one is best suited to use for our analysis.
            Examples of tests include the inclusion or exclusion of GPs, whether to use any given set(s) of activity indicators during fitting, and adjusting the different GP kernels that could be used.
            We then performed a model comparison test by comparing the BIC value for each model case.
            We define $\Delta$BIC = BIC$_{\text{new}}$ $-$ BIC$_{\text{old}}$. 
            A smaller $\Delta$BIC signifies that the model is less complex and favored, while still providing a good fit to the data.
            A more complex model could still have a lower BIC if it fits the data sufficiently better than simpler models.
            We found that including GPs for all RV instruments, using only the CCF$_{\text{FWHM}}$ activity indicator from HARPS and ESPRESSO, and using the RotationTerm kernel for the GP analysis, was the favored model.
            Table \ref{table:bic} shows a list of the different model combinations we tested, along with their corresponding BIC values.
            The BIC value for the preferred model was 361.6, which we subtract from the BIC values of the other models and report under the column labeled $\Delta$BIC.

            \begin{table}
            \caption{A list of the different model combinations we tested for RV fitting, along with their corresponding BIC values. The BIC value for the preferred model was 361.6, which we subtract from the BIC values of the other models and report under the column labeled $\Delta$BIC.}
                \centering
                \setlength{\extrarowheight}{3pt}
                \begin{tabularx}{\linewidth}{XXXXX}
                \hline
                    GP            & Kernel (\texttt{celerite2}) & Activity Indicator & BIC Value & $\Delta$BIC \\
                \hline
                    All           & \texttt{Rotation Term}      & FWHM               & 361.60    & 0           \\
                    All           & \texttt{SHO Term}           & FWHM               & 385.68    & 24.08       \\
                    All           & \texttt{Matern34 Term}      & FWHM               & 397.20    & 35.6        \\
                    All           & \texttt{SHO Term}           & FWHM + S$-$Index     & 427.04    & 65.44       \\
                    HARPS only    & \texttt{SHO Term}           & FWHM + S$-$Index     & 455.04    & 93.44       \\
                    PFS only      & \texttt{SHO Term}           & FWHM + S$-$Index     & 490.03    & 128.43      \\
                    None          & None                        & None               & 508.90    & 147.3       \\
                    ESPRESSO only & \texttt{SHO Term}           & FWHM + S$-$Index     & 707.37    & 345.77      \\
                    \hline
                \end{tabularx}
                \label{table:bic}
            \end{table}

    	\subsubsection{RV$-$only Fit Results}\label{subsubsec:rvresults}  
            
            From the RV$-$only MCMC fit, we find that TOI$-$757 b is on a highly eccentric orbit with an eccentricity of $e$ = 0.39$^{+0.08}_{-0.07}$ with an argument of periastron for the planetary orbit of $\omega_{\text{p}}$ = $-$36.02$^{\circ~+13.69}_{-10.10}$. 
            The RV semi$-$amplitude is found to be $K$ = 3.27$^{+0.63}_{-0.61}$ m/s, resulting in a planetary mass of $M_{\text{p}}$ = 10.45$^{+2.18}_{-2.09}$ $M_{\oplus}$. 
            The full RV best$-$fit parameters are listed in Table \ref{table:priors}.

\begin{figure}
            \centering
            \begin{subfigure}{\linewidth}
                \centering
                \includegraphics[width=\linewidth, trim={1.5cm 0.6cm 3cm 2cm}, clip]{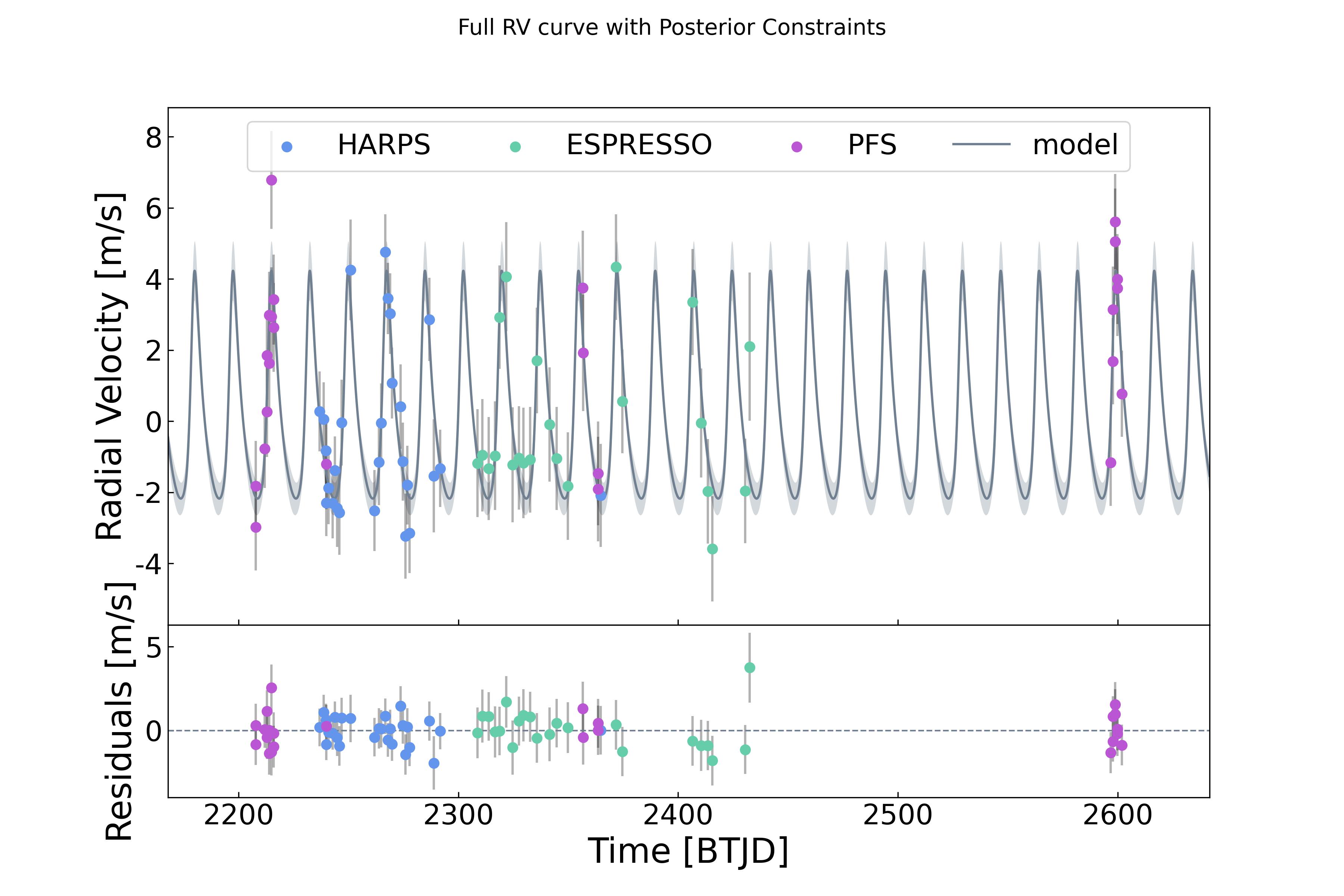}
                \label{fig:rv_trace_full}
            \end{subfigure}%
    
                \begin{subfigure}{\linewidth}
                \centering
                \includegraphics[width=\linewidth, trim={1.5cm 0.6cm 3cm 2cm}, clip]{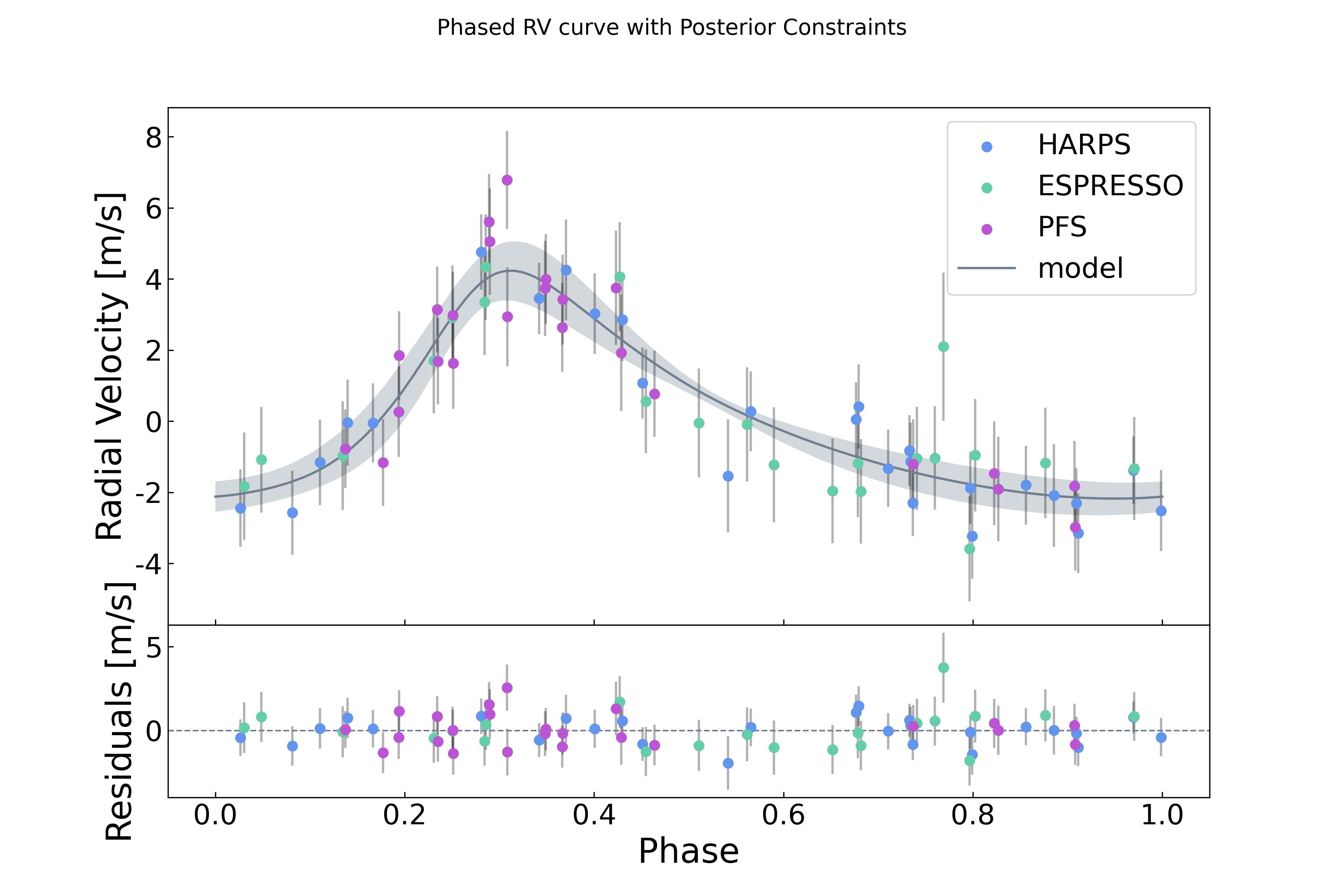}
                \label{fig:rv_trace_folded}
            \end{subfigure}%
            \caption{RV data of the HARPS (\textit{blue dots}), ESPRESSO (\textit{green dots}), and PFS (\textit{magenta dots}) instruments. 
            The error bars include the jitter component $\sigma$ for each instrument, which was added in quadrature to the RV error for each data point.
            The data were detrended by subtracting the GP model as well as the systemic mean velocity $\mu$ of the instrument (see Section \ref{subsec:rvfit}).
            The best$-$fit planet model is plotted on top in a gray solid line, along with the posterior constraints for the 16$^{\text{th}}$ and 84$^{\text{th}}$ percentiles in the gray shaded region. 
            The top panel shows the data across the full time baseline, while the lower panel displays the same detrended data folded over the best$-$fit orbital period $P$.}
            \label{fig:rv_trace}
            \end{figure}

\begin{figure*}
            \centering
              \includegraphics[width=\linewidth, trim={0.5cm 0 3.5cm 3cm}, clip]{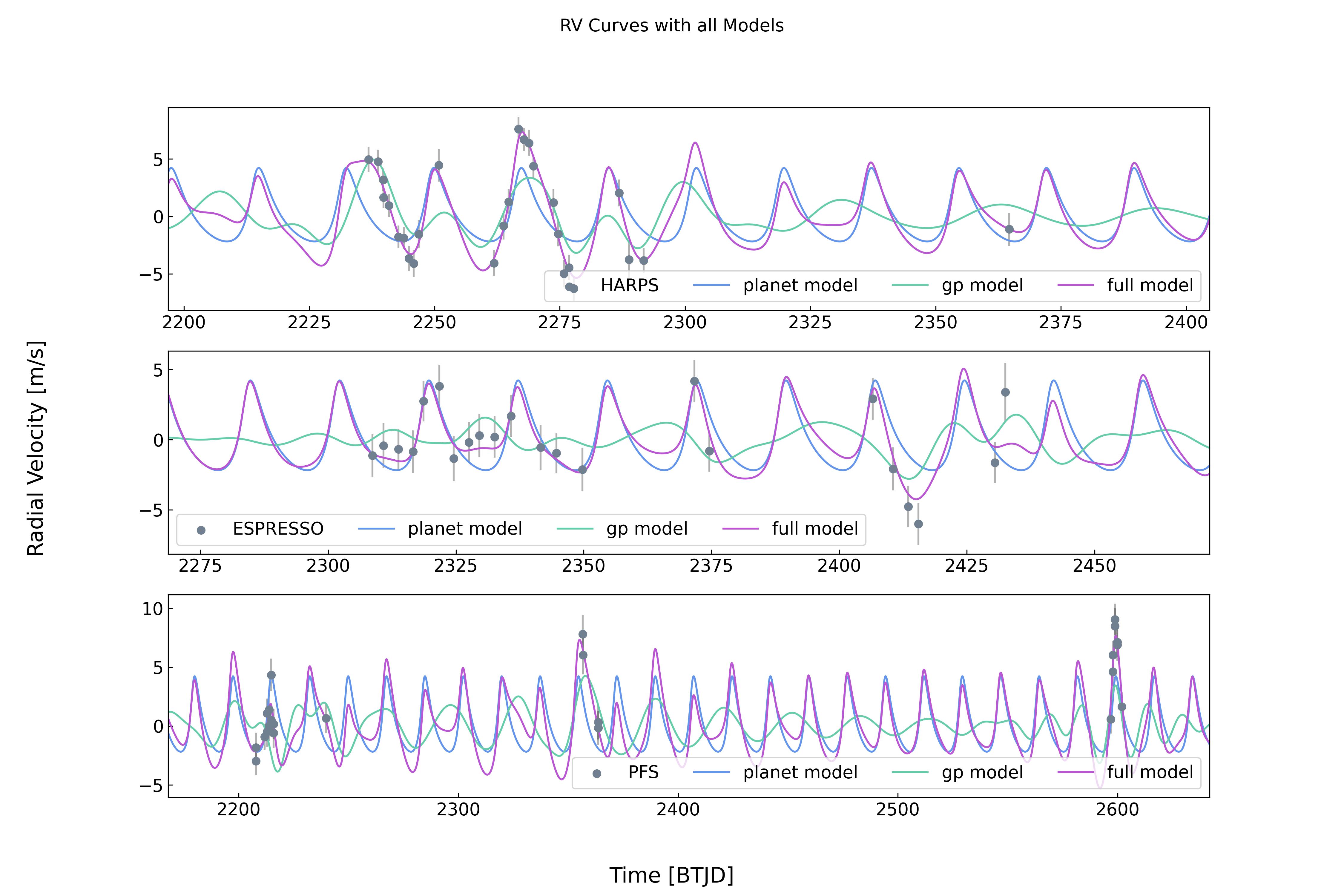}
            \caption{Mean$-$subtracted RV data plotted in gray dotted points for each of the HARPS (\textit{top panel}), ESPRESSO (\textit{middle panel}), and PFS (\textit{bottom panel}) instruments, along with separate model curves for the planet model (\textit{blue solid line}), the GP model (\textit{green solid line}) and the full planet + GP model (\textit{magenta solid line}). The error bars include the jitter component $\sigma$ for each instrument, which was added in quadrature to the RV error for each data point.}
            \label{fig:rv_models_gp}
            \end{figure*}

\begin{figure}
                \centering
                \includegraphics[width=\linewidth, trim={1cm 0.7cm 3cm 2cm}, clip]{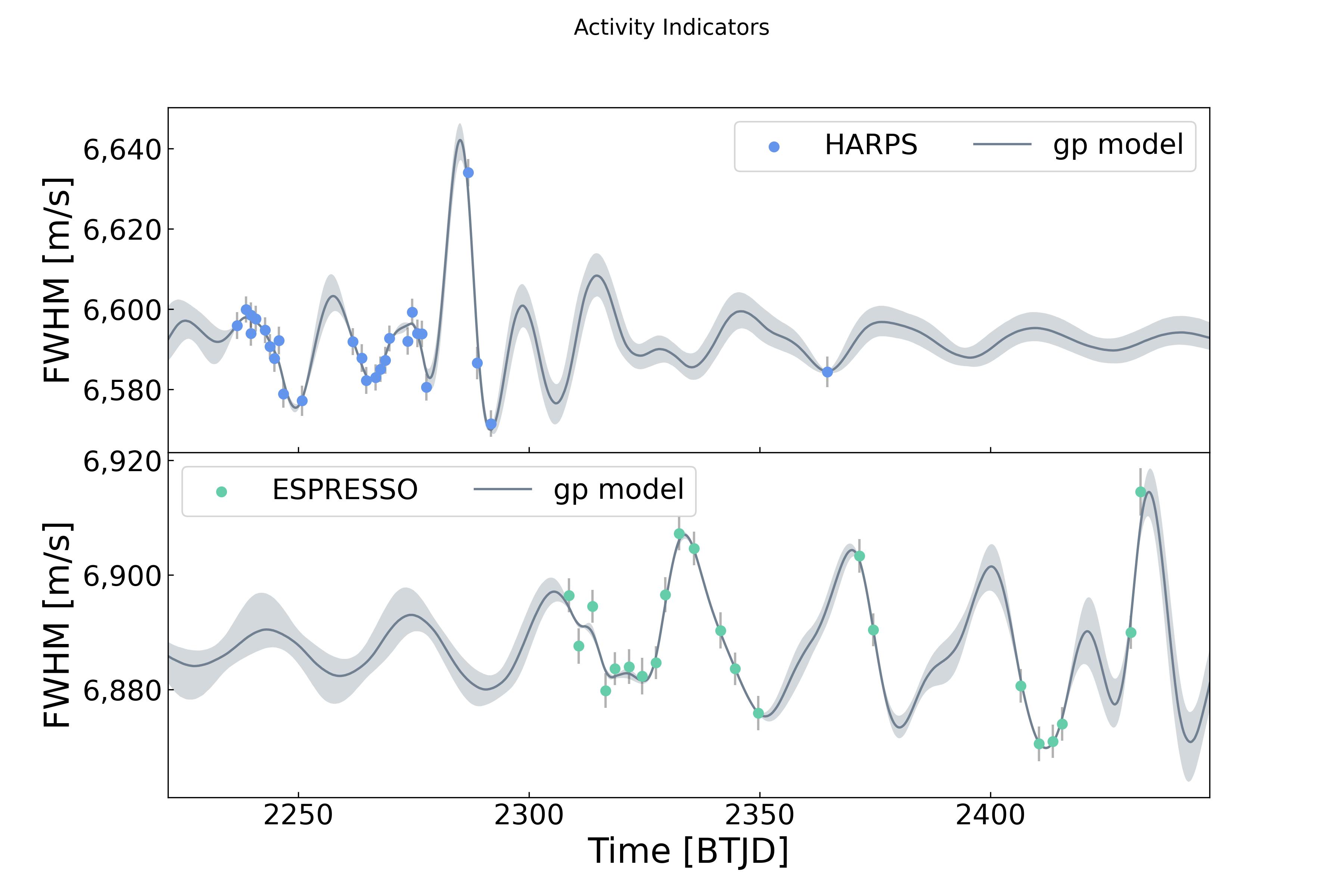}
            \caption{GP models, shown in gray solid lines, of the CCF$_{\text{FWHM}}$ activity indicators from HARPS (\textit{top panel}, represented by the blue dots) and ESPRESSO (\textit{bottom panel}, represented by the green dots). The shaded regions are the posterior constraints for the 16$^{\text{th}}$ and 84$^{\text{th}}$ percentiles of the GP models. The error bars include the jitter component $\sigma$ for each instrument, which was added in quadrature to the error for each data point.}
            \label{fig:activity_indicators_gp}
            \end{figure}

            We plot the detrended RV data by subtracting the GP model as well as the systemic mean velocity $\mu$ of the instrument in Fig. \ref{fig:rv_trace}.
            The detrended data is shown along with the best$-$fit planet model and the posterior constraints for the 16$^{\text{th}}$ and 84$^{\text{th}}$ percentiles in the top panel, while the lower panel displays the same data folded over the orbital period, along with the best$-$fit planet model and the posterior constraints.
            Fig. \ref{fig:rv_models_gp} shows the mean$-$subtracted RV data points for each instrument, along with separate curves for the planet model, the GP model and the full planet+GP model.
            In Fig. \ref{fig:activity_indicators_gp}, we show the GP models of the CCF$_{\text{FWHM}}$ activity indicators with the posterior constraints for the 16$^{\text{th}}$ and 84$^{\text{th}}$ percentiles, which were used to help constrain the GP hypermarameters of the RV data.

        \subsection{Results Comparison}\label{subsec:resultscomp}  

    	\subsubsection{Stellar Density Tests}\label{subsubsec:densitytests}  

            We checked what solution the MCMC finds when we placed wide uniform priors versus tighter Gaussian priors on the stellar parameters, $R_{\star}$ and $M_{\star}$ (and by extension, on the stellar density).
            For a proper comparison, we ran this test for all of our MCMC fits: the transit fit as well as the RV fit.
            We find that in the case of the transit$-$only fit, the eccentricities derived when we test the stellar density with ($e$ = $0.3 \pm 0.2$) and without ($e$ = $0.5 \pm 0.3$) informative stellar priors are both poorly constrained by the photometric data alone.
            When testing the case of the RV$-$only fit, we find a well$-$constrained eccentricity of $e$ = $0.4 \pm 0.1$, with a narrower posterior distribution compared with the transit$-$only fit, regardless of whether we applied a prior on the stellar density or not.
            As expected, the RV data alone are able to constrain the eccentricity much better than the photometric data, and independently predicts a significantly high eccentricity.

    	\subsubsection{GP Tests}\label{subsubsec:gptests}  
     
            To test whether our RV GPs are overfitting the planet signal or contributing to the high eccentricity, we placed a tighter prior on all the GP $\sigma$ terms (which correspond to the amplitude of the noise model).
            More specifically, we changed the $\sigma_{\text{GP}}$ priors from wide ($\mathcal{U}$[0, 10]) to narrow ($\mathcal{U}$[0, 1]) priors and re$-$ran the MCMC fit.
            We additionally tested fitting all the RVs using a single$-$GP (while still using wide priors for $\sigma_{\text{GP}}$), and we also do not fit over any activity indicators in this test case for simplicity.
            Finally, we tested fitting without any RV$-$related GP components, including the exclusion of activity indicators.
            A summary of the results can be found in Table \ref{table:sigmagpcomparison}.
            The planet parameters resulting from each test case did not change significantly compared to the results from the wide $\sigma_{\text{GP}}$ priors case (which is what is used in the final analysis of this paper), and the mean best$-$fit values are all within the 1$-\sigma$ error limits.
            Moreover, the eccentricity measurement does not appear to be affected by using narrow $\sigma_{\text{GP}}$ priors, using a single GP, nor by having no RV GP components at all.
            This shows that the high eccentricity of 0.4 from the RV fit is most likely not an artifact of the GP behavior. 
            Additionally, the fact that we do not find an argument of periastron $\omega_{\text{p}}$ near 90$^{\circ}$ or $-$90$^{\circ}$ provides further confidence to our eccentricity result.
            Obtaining a spurious eccentricity is more likely when $\omega_{\text{p}}$ is close to such values, given the shape (and similarity to a circular orbit) of the RV curve for these orientations.

            \begin{table}
                \centering
                \caption{Comparison of resulting planet parameters in the RV$-$only fit when testing different priors for $\sigma_{\text{GP}}$. For the wide$-$priors case, we use a Uniform distribution of $\mathcal{U}$[0, 10] for all the $\sigma_{\text{GP}}$ terms of the RV instrument GPs. For the narrow$-$priors case, we use a Uniform distribution of $\mathcal{U}$[0, 1]. For the single$-$GP case, we use wide priors for $\sigma_{\text{GP}}$ as done in the wide$-$priors case, and we also do not fit over any activity indicators for simplicity. For the no$-$GP case, we remove all RV$-$related GP components, including activity indicators. The reported values correspond to the mean best$-$fit value from the MCMC fit, with the standard deviation as the error.}
                \resizebox{\linewidth}{!}{
                \begin{tabular}{lllll}
                    \toprule
                    Parameter                        & Wide priors         & Narrow priors       & Single GP          & No GP               \\           
                    \midrule
                    \addlinespace[3pt]
                    $K$ [m s$^{-1}$]                 & $3.3 \pm 0.6$     & $3.3 \pm 0.4$     & $3.8 \pm 0.7$    & $3.1 \pm 0.5$     \\
                    $M_{\text{p}} [M_{\oplus}]$      & $10.5 \pm 2.2$    & $10.6 \pm 1.4$    & $12.3 \pm 2.4$   & $9.8 \pm 1.6$     \\
                    $e$                              & $0.39 \pm 0.08$     & $0.39 \pm 0.06$     & $0.36 \pm 0.07$    & $0.39 \pm 0.08$     \\
                    $\omega_{\text{p}}$ [$^{\circ}$] & $-34 \pm 14$  & $-36 \pm 13$  & $-28 \pm 15$ & $-26 \pm 24$  \\
                    \bottomrule
                \end{tabular}
                }
                \label{table:sigmagpcomparison}
            \end{table}

\section{Discussion}\label{sec:discussion}

From our analysis, we clearly detect TOI$-$757 b in the photometric data from \textit{TESS} and \textit{CHEOPS}, supported by the transits acquired from ground$-$based telescopes. 
We also spectroscopically confirm the transiting planet by significantly detecting the Doppler reflex motion induced by TOI$-$757 b in our HARPS, ESPRESSO and PFS RV data.
The photometric and RV fitting yield a planet radius of $R_{\text{p}}$ = 2.5 $\pm$ 0.1 $R_{\oplus}$, a mass of $M_{\text{p}}$ = 10.5$^{+2.2}_{-2.1}$ $M_{\oplus}$, and an eccentricity of $e$ = 0.39$^{+0.08}_{-0.07}$ for TOI$-$757 b.

	\subsection{Radius Valley}\label{subsec:radiusvalley}  
        
\begin{figure}
        \centering
          \includegraphics[width=\linewidth, trim={0.2cm 0.3cm 0 0}, clip]{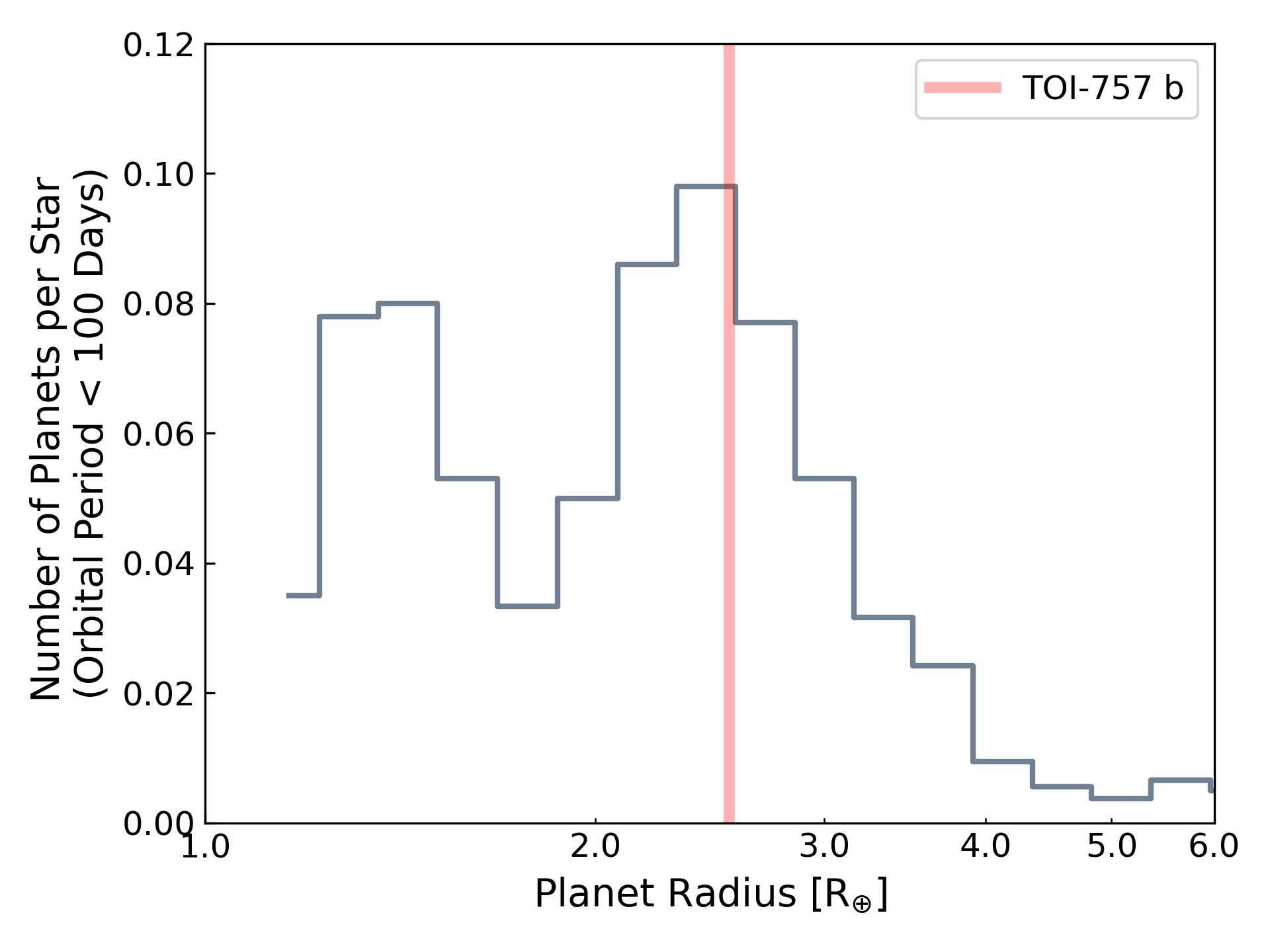}
        \caption{The observed planet radius distribution adapted from \protect\cite{fulton2017rvalley}. The well$-$known radius gap feature of the exoplanet population is apparent from 1.5 to 2.0 $R_{\oplus}$. The overplotted red line represents the radius of TOI$-$757 b, showing that the planet lies above the valley and is among the abundant population of mini$-$Neptunes.}
        \label{fig:fulton2017_rvalley}
        \end{figure}

        The composition of mini$-$Neptunes plays an important role in our understanding of the radius valley.
        Whether super$-$Earths and mini$-$Neptunes are the same population that evolved differently, or are two distinct populations that formed differently, is still widely debated.      
        In Fig. \ref{fig:fulton2017_rvalley}, we plot the observed planet radius distribution adapted from \cite{fulton2017rvalley}. 
        The well$-$known radius gap feature of the exoplanet population is apparent from 1.5 to 2.0 $R_{\oplus}$. 
        The overplotted red line represents the radius of TOI$-$757 b, showing that the planet lies to the right of the valley and is among the abundant population of mini$-$Neptunes.
        With our newly characterized mass measurement of TOI$-$757 b, we are able to derive a density measurement of the planet and understand its potential composition better, giving us clues as to whether it belongs to the mini$-$Neptune population or whether it could be a water$-$world.

	\subsection{Mass$-$Radius Relation}\label{subsec:massradius}  

        To understand the possible internal composition of TOI$-$757 b, we calculate the expected density of the planet based on the mass and radius measurements derived from MCMC fits, giving us an implied density of $\rho_{\text{p}} = 3.6 \pm 0.8$ g cm$^{-3}$.
        It is unclear whether the planet could harbour any water/ice components or whether it has a fraction of a H/He envelope.
        The composition of TOI$-$757 b might include a significant fraction of water/ice, since water$-$rich worlds form outside the snow line and later migrate inward, while rocky planets are expected to form within the snow line \citep{armitage2016snowline}.
        This raises questions about the formation mechanisms of the mini$-$Neptune population, and might allude to possible migration scenarios in the evolutionary history of these planets, particularly in the case of TOI$-$757 b.
        In such cases, atmospheric studies may be needed to distinguish between water$-$rich planets and H/He atmospheres.
        
\begin{figure}
        \centering
          \includegraphics[width=\linewidth, trim={0.4cm 0.5cm 1.5cm 2cm}, clip]{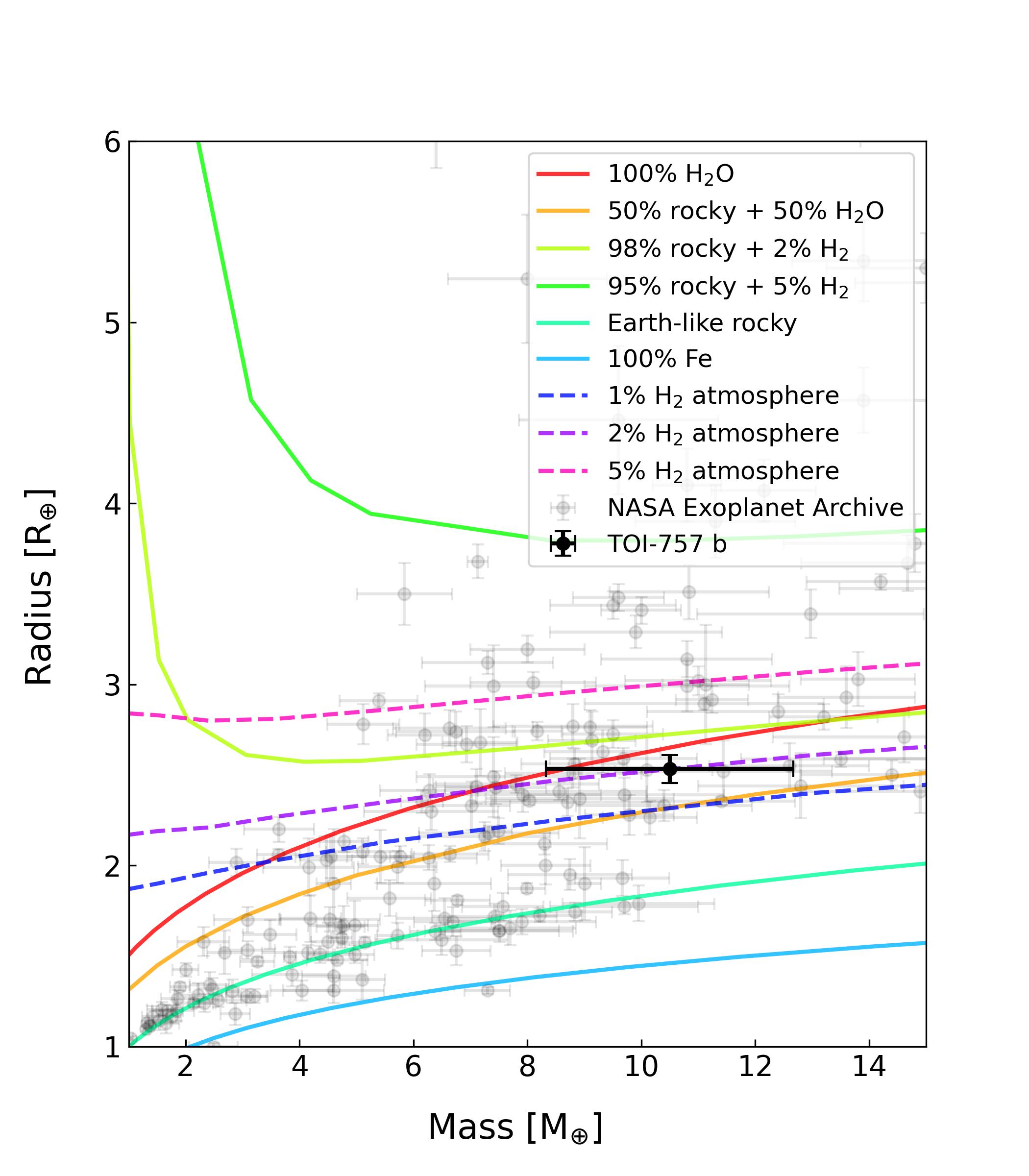}
        \caption{Mass$-$Radius diagram showing composition tracks from \protect\cite{zeng2019massradius} in solid lines, assuming a temperature of 500 K and a surface pressure level of 1 mbar. 
        We also show composition tracks from \protect\cite{lopez2014massradius} in dashed lines, assuming an insolation flux of 10 on the planet (relative to the solar constant) and a stellar age of 10 Gyr.
        The black dot represents TOI$-$757 b, while the gray dots show confirmed exoplanets with mass and radius uncertainties < 20$\%$, taken from the NASA Exoplanet Archive.}
        \label{fig:zeng_lopez_MR_diagram}
        \end{figure}

        To investigate this, we plot TOI$-$757 b on a mass$-$radius diagram in Fig.~\ref{fig:zeng_lopez_MR_diagram}, with composition tracks from \cite{zeng2019massradius} in solid lines, assuming a temperature of 500\,K and a surface pressure level of 1\,mbar.
        While the \cite{zeng2019massradius} composition tracks have been widely used in the exoplanet community, they assume a constant specific entropy. 
        This could pose issues when determining the inner composition of small planets, such as TOI$-$757 b.
        Thus, following the suggestion from \cite{rogers2023waterworlds}, we also show composition tracks from \cite{lopez2014massradius}, which assume a constant planet age.
        These tracks are shown in Fig.~\ref{fig:zeng_lopez_MR_diagram} using dashed lines, assuming an insolation flux of 10 on the planet (relative to the solar constant) and a stellar age of 10\,Gyr.
        The black dot represents TOI$-$757 b, while the gray dots show confirmed exoplanets with mass and radius uncertainties <\,20$\%$, taken from the NASA Exoplanet Archive\footnote{\url{https://exoplanetarchive.ipac.caltech.edu}} \citep{akeson2013nexi} as of 25 January 2024. 

        Normally, there is a clear degeneracy between composition models when trying to separate H/He envelopes and water$-$rich atmospheres for mini$-$Neptunes.
        This becomes particularly difficult to constrain for single$-$planet systems, even with well$-$characterized mass and radius measurements \citep{rogers2023waterworlds}.
        While the position of TOI$-$757 b on the mass-radius diagram falls with other known mini-Neptunes (see Fig. \ref{fig:zeng_lopez_MR_diagram}), the planet composition remains vague due to the aforementioned model degeneracy. 
        Looking at the \cite{zeng2019massradius} models, the composition of TOI$-$757 b appears to be consistent with a 100$\%$ water world.
        However, we note that some studies indicate that a planet entirely made of water is not considered to be physically realistic (see e.g. \cite{Marboeuf+2014}).
        The \cite{lopez2014massradius} models seem to suggest a composition of a rocky core with a 2$\%$ hydrogen atmosphere for TOI$-$757 b.
        To further investigate which composition is more likely for our system, we performed internal composition modeling described in Sect. \ref{subsec:internalstruc}.
        While a H/He gas layer was found to be well constrained for the system, we were unable to rule out the presence of a significant water fraction.
        
        To be able to more confidently characterize TOI$-$757 b's composition, atmospheric observations would be necessary in order to break the degeneracy between planet composition models for mini$-$Neptunes and water$-$rich worlds. 
        This could potentially make TOI$-$757 b an exciting target for James Webb Space Telescope (JWST) observations, and the host star is also bright, making it favorable for atmospheric follow up.
        To evaluate whether TOI$-$757 b is suitable for atmospheric characterization or not, we use the two metrics introduced by \cite{kempton2018esmtsm}, the Transmission Spectroscopy Metric (TSM) and the Emission Spectroscopy Metric (ESM).
        We find that TOI$-$757 b has a TSM of 44.75 and an ESM of 3.19.
        Considering the recommended metric threshdolds of TSM > 90 and ESM > 7.5 for small mini$-$Neptunes \citep{kempton2018esmtsm}, this makes TOI$-$757 b a challenging (albeit still interesting) target for atmospheric observations given these guidelines.

	\subsection{Internal Structure Modeling and Atmospheric Evolution}\label{subsec:internalstruc} 

        We used the method introduced in \citet{Leleu+2021} and based on \citet{Dorn+2017} to model the internal structure of TOI$-$757 b. 
        The planet is modelled as a fully spherically symmetric structure consisting of an inner iron core with up to 19\% sulphur \citep{Hakim+2018}, a silicate mantle made up of Si, Mg and Fe \citep{Sotin+2007}, a condensed water layer \citep{Haldemann+2020} and a fully separately modelled H/He envelope \citep{lopez2014massradius}. 
        We assume that the Si/Mg/Fe ratios of the planet match those of the star exactly \citep{Thiabaud+2015}.

        A Bayesian inference scheme is then used to infer posterior distributions for the composition and interior structure of the planet, with the radius ratio, period and RV semi$-$amplitude of the planet, along with the stellar parameters (radius, mass, age, effective temperature and Si, Mg and Fe abundances) as input parameters. 
        We chose priors that are uniform in log for the mass of the H/He envelope and uniform on the simplex for the mass fractions of the inner iron core, silicate mantle and water layer (all with respect to the solid part of the planet without the H/He envelope). 
        In agreement with \citet{Thiabaud+2014} and \citet{Marboeuf+2014}, we chose an upper limit of 50\% for the water mass fraction. 
        We do note that results of our analysis do depend to some extent on the chosen priors, as determining the internal structure of a planet is a highly degenerate problem.

\begin{figure}
            \centering
            \includegraphics[width=\linewidth]{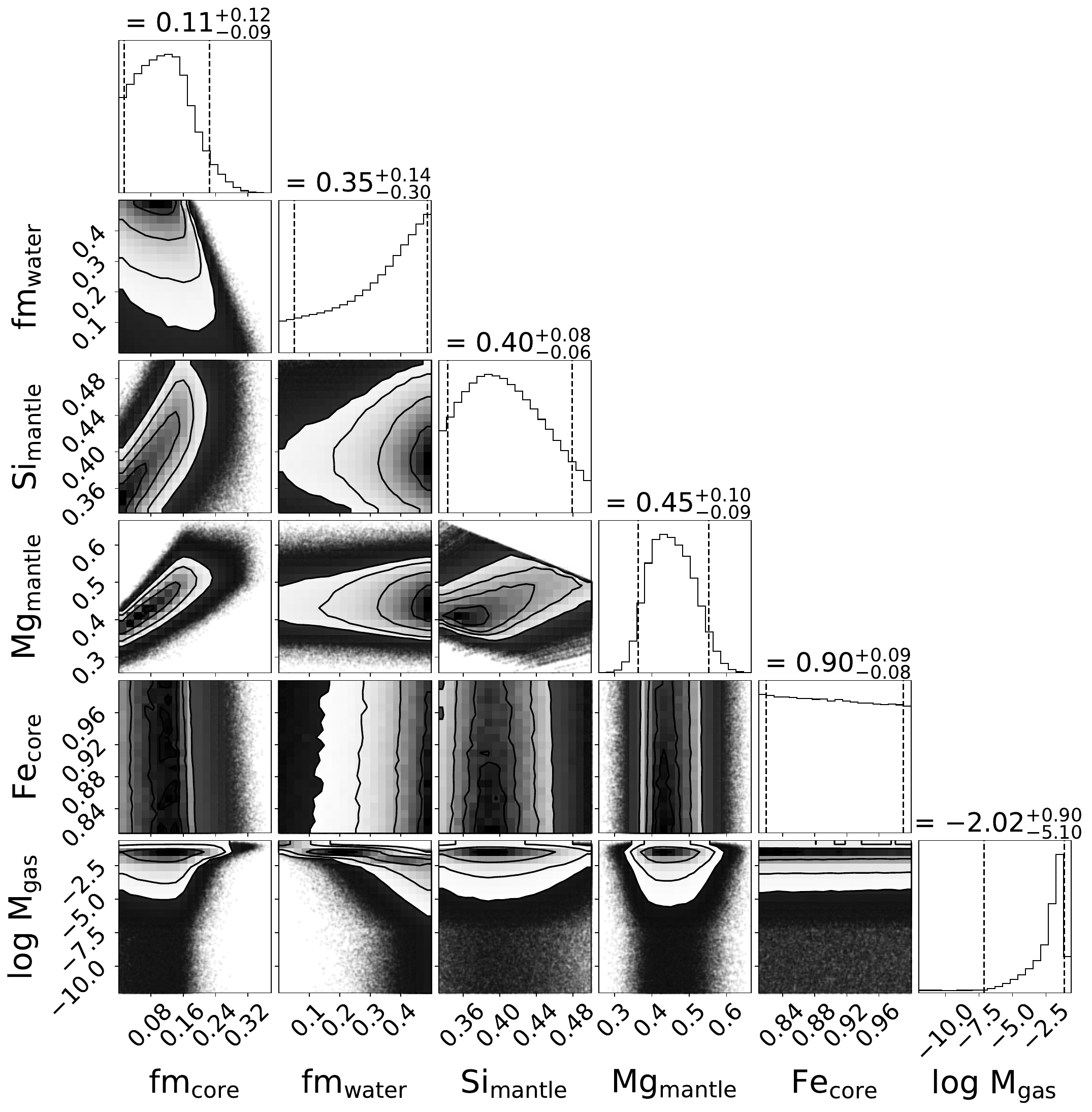}
            \caption{Posterior distributions of the most relevant internal structure parameters for TOI$-$757 b. The values above the histograms are the median and the 5 and 95th percentiles of the distributions, same as the dashed lines. The plotted internal structure parameters are the iron core and water mass fractions of the solid planet without gas, the molar fraction of Si and Mg within the mantle and of Fe in the core and the $\log_{10}$ of the absolute gas mass in Earth masses.}
            \label{fig:int_struct}
        \end{figure}

        Figure~\ref{fig:int_struct} shows the results of our analysis. While the presence of a water layer remains unconstrained with a tendency towards larger water mass fractions, the mass of the H/He layer of TOI$-$757 b is quite well constrained with a median value of $M_{\mathrm{gas}}=0.01^{+0.07}_{-0.01}$ $M_{\oplus}$, where the uncertainties correspond to the 5th and 95th percentiles of the posterior. This corresponds to a thickness of $R_{\mathrm{gas}}=0.28_{-0.27}^{+0.38}\,R_{\oplus}$ of the gas layer.

	\subsection{Tidal Damping of the Eccentricity and its Consequences}\label{subsec:tidaldis} 
        
        We estimate the current timescale for the decay of the eccentricity of TOI$-$757 b adopting the model by \citet{hut1981tidalevo}. 
        To apply that model, we express the product of their constant tidal time lag $\Delta t_{\rm p}$ by the Love number of the planet $k_{2, \rm p}$ as a function of the modified tidal quality factor $Q^{\prime}_{\rm p}$ of the planet itself as $ k_{2, \rm p} \Delta t_{\rm p} =  3/(2 Q^{\prime}_{\rm p} n)$, where $n=2\pi/P$ is the orbit mean motion with $P$ being the orbital period. 

        The rheology of mini$-$Neptune planets is unknown, thus we investigate two extreme cases. 
        The former assumes $Q^{\prime}_{\rm p} = 10^{5}$ that is comparable with the modified tidal quality factor of Uranus or Neptune as estimated by, e.g., \citet{TittemoreWisdom90} and \citet{Ogilvie14}, and is appropriate for a body where the dissipation of the tidal energy occurs in a fluid interior. 
        The latter assumes that tides are mainly dissipated inside a rocky core encompassing the whole mass of the planet $4.4$~M$_{\oplus}$, but with a radius of 1.3 ~R$_{\oplus}$ that corresponds to the peak of the smaller radius component of the distribution of the transiting planets as observed by Kepler \citep{fulton2017rvalley}. 
        In the latter case, we adopt $Q^{\prime}_{\rm p} =300$ as in the case of our Earth \citep{henning2009tidalmodel} and neglect the small mass and tidal dissipation in the fluid outer envelope of the planet in comparison with those in its rocky core. 
        The present eccentricity tidal decay timescale is found to be $\tau_{\rm e} \equiv |e/(de/dt)| \sim 1040$~Gyr in the former case, or $\tau_{\rm e} \sim 90$~Gyr in the latter case, much longer than the Hubble time in both the cases. 
        Therefore, we conclude that the present eccentricity may well be the result of processes that happened during the formation of the planet and/or its migration to the present distance from its host star. 
        We find that tidal dissipation inside the star has a negligible effect on the evolution of the eccentricity because of the relatively large separation of the planet and its small mass. 

        An interesting consequence of the eccentric orbit of TOI$-$757 b is its expected pseudosynchronization with the orbital motion, that is, its rotation is predicted to be faster than the orbital mean motion $n$ because tides tend to synchronize planet  rotation with the orbital velocity at periastron where they are stronger. 
        Using the formalism of \citet{Leconteetal10}, we predict a rotation period of the planet of 8.8~days, that is, 1.99 times shorter than its orbital period. 
        Such a state of pseudosynchronization is reached after a timescale of $\sim 6$~Myr for $Q^{\prime}_{\rm p} =10^{5}$ or $\sim 0.14$~Myr for $Q^{\prime}_{\rm p} = 300$, either of which is significantly shorter than the estimated age of the host star. 

        We expect tides to dissipate energy inside the planet due to its eccentric orbit and pseudosynchronous rotation. The maximum dissipated power is predicted for $Q^{\prime}_{\rm p} =300$ and is of $\sim 2.3 \times 10^{16}$~W giving a heat flux of $\sim 7.1$~W~m$^{-2}$ at the top of its atmosphere with a radius of $2.5$~R$_{\oplus}$. 
        Such a flux is larger than that measured in the case of Jupiter where the heat flux from the interior of the planet is of $\sim 5.4$~W~m$^{-2}$ \citep{Guillotetal04}. 
        Therefore, it may play a relevant role in the atmospheric dynamics of the planet. 
        On the other hand, adopting $Q^{\prime}_{\rm p} = 10^{5}$, we find a dissipated power of $1.9 \times 10^{15}$~W and a surface flux of only $\sim 0.6$~W~m$^{-2}$.

	\subsection{High$-$eccentricity and Formation Mechanisms}\label{subsec:eccformation} 
 
\begin{figure}
        \centering
            \includegraphics[height=0.9\textheight]{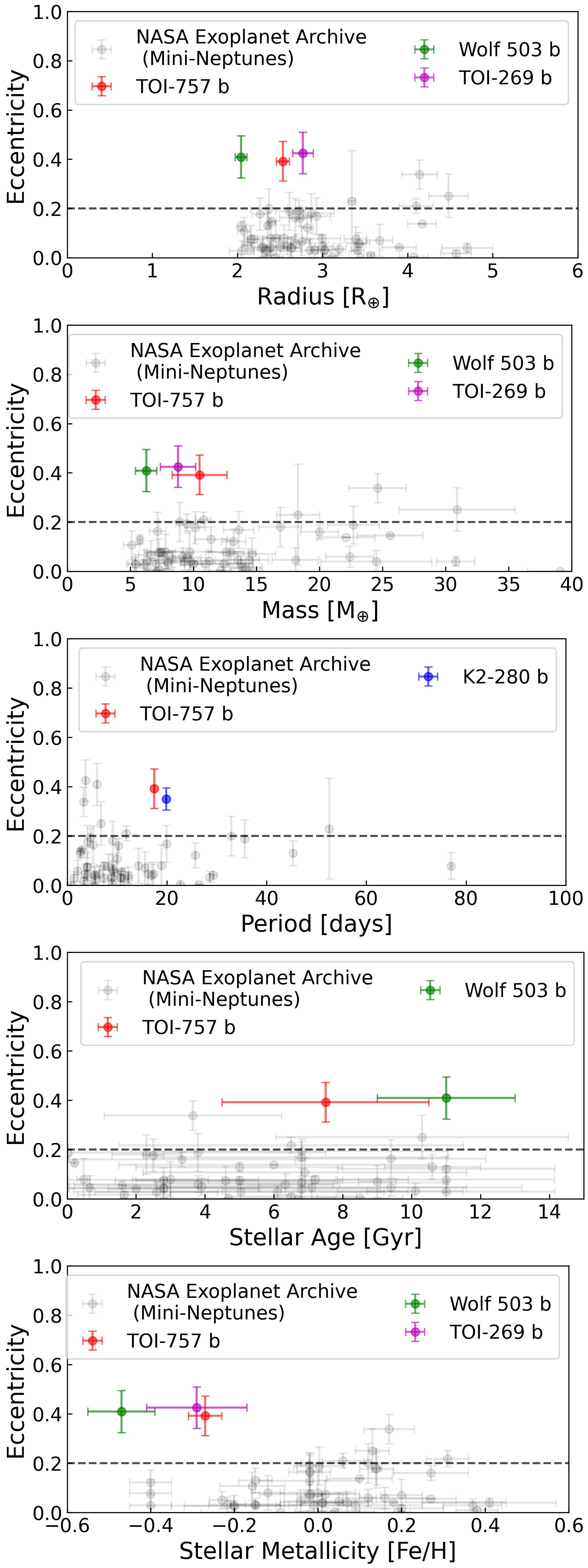}
        \caption{From top to bottom (with Eccentricity on the y-axis), we plot the Radius, Mass, Period, Stellar Age, and Stellar Metallicity [Fe/H] for precisely characterized mini$-$Neptunes. The red dot represents TOI$-$757 b, the blue dot represents K2$-$280 b, the green dot represents Wolf 503 b, and the magenta dot represents TOI$-$269 b. The gray dots show confirmed exoplanets taken from the NASA Exoplanet Archive with mass and radius uncertainties < 20$\%$, filtered for the mini$-$Neptune population ($P$ < 100 days, 5 < $M_{\text{p}} [M_{\oplus}]$ < 40, 2 < $R_{\text{p}} [R_{\oplus}]$ < 5).}
        \label{fig:pop_comparison}
        \end{figure}

        Our findings reveal that TOI$-$757 b is a transiting single$-$planet system on a highly eccentric orbit of $e$ = 0.39$^{+0.08}_{-0.07}$. 
        It has one of the highest eccentricities among all precisely known exoplanets (with mass and radius uncertainties <20$\%$) in the period, radius and mass ranges of the mini$-$Neptune population ($P$ < 100 days, 5 < $M_{\text{p}} [M_{\oplus}]$ < 40, 2 < $R_{\text{p}} [R_{\oplus}]$ < 5), as can be seen in Figure \ref{fig:pop_comparison}.
        \cite{vaneylen2019eccentricity} report that, on average, single$-$planet systems appear to exhibit higher eccentricities than multi$-$planet systems for transiting planets.
        Using transit durations to estimate orbital eccentricities, \cite{vaneylen2019eccentricity} find that the typical eccentricity of single$-$planet systems is $e$ = 0.32$\pm$0.06.
        \cite{xie2016orbitalecc} also find an average eccentricity of $\sim$0.3 for single transiting planet systems.
        So perhaps within this context, it is not too surprising that TOI$-$757 b is highly eccentric.
        While TOI$-$757 b sits among the abundant population of mini$-$Neptunes in relation to the radius valley, it is uniquely one of the very few highly eccentric mini$-$Neptunes, as can be seen in Fig. \ref{fig:pop_comparison}. 
        This makes it an interesting target within the context of the eccentricity distribution of small planets, and poses important questions regarding its formation history.
        \cite{petigura2017subsaturns} studied a sample of four sub$-$Saturn sized planets showing a wide range of masses measured using Keck/HIRES RV data, which implies that their core and envelope masses are diverse in nature.
        Looking at Fig. 11 from their study, we can see that the majority of planets from the sub$-$Saturn population within a similar planet mass and equilibrium temperature range to TOI$-$757 b are all found to be non$-$eccentric ($e$ < 0.1).
        So, even looking at planets larger in size in comparison to the mini$-$Neptune population, TOI$-$757 b stands out as a unique case.

        The high eccentricity of TOI$-$757 b suggests that the history of the planet might be different in comparison with other mini$-$Neptunes, owing to an evolution path involving eccentricity excitation induced by the gravitational interactions with a companion.
        As such, we might expect different atmospheric properties in such scenarios.
        Additionally, when we modeled the tidal dissipation timescale for the system (see Sect. {\ref{subsec:tidaldis}}), even across extreme cases, the circularization timescale was found to be much longer than the age of the system. 
        As such, the planet's present-day eccentricity is consistent with a number of plausible scenarios.
        One possibility is that, regardless of whether the eccentricity developed intrinsically or whether it was induced by a perturber, the planet could have attained its eccentricity close to the formation period of the system.
        Given this scenario, it is also possible that a potential perturber would have been ejected from the system earlier on and can no longer be detected.
        However, we would like to point out that a long circularization timescale does not necessarily prove that the eccentricity arose in the distant past.
        
\begin{figure}
            \centering
            \includegraphics[width=\linewidth, trim={0.2cm 0 1cm 1.4cm}, clip]{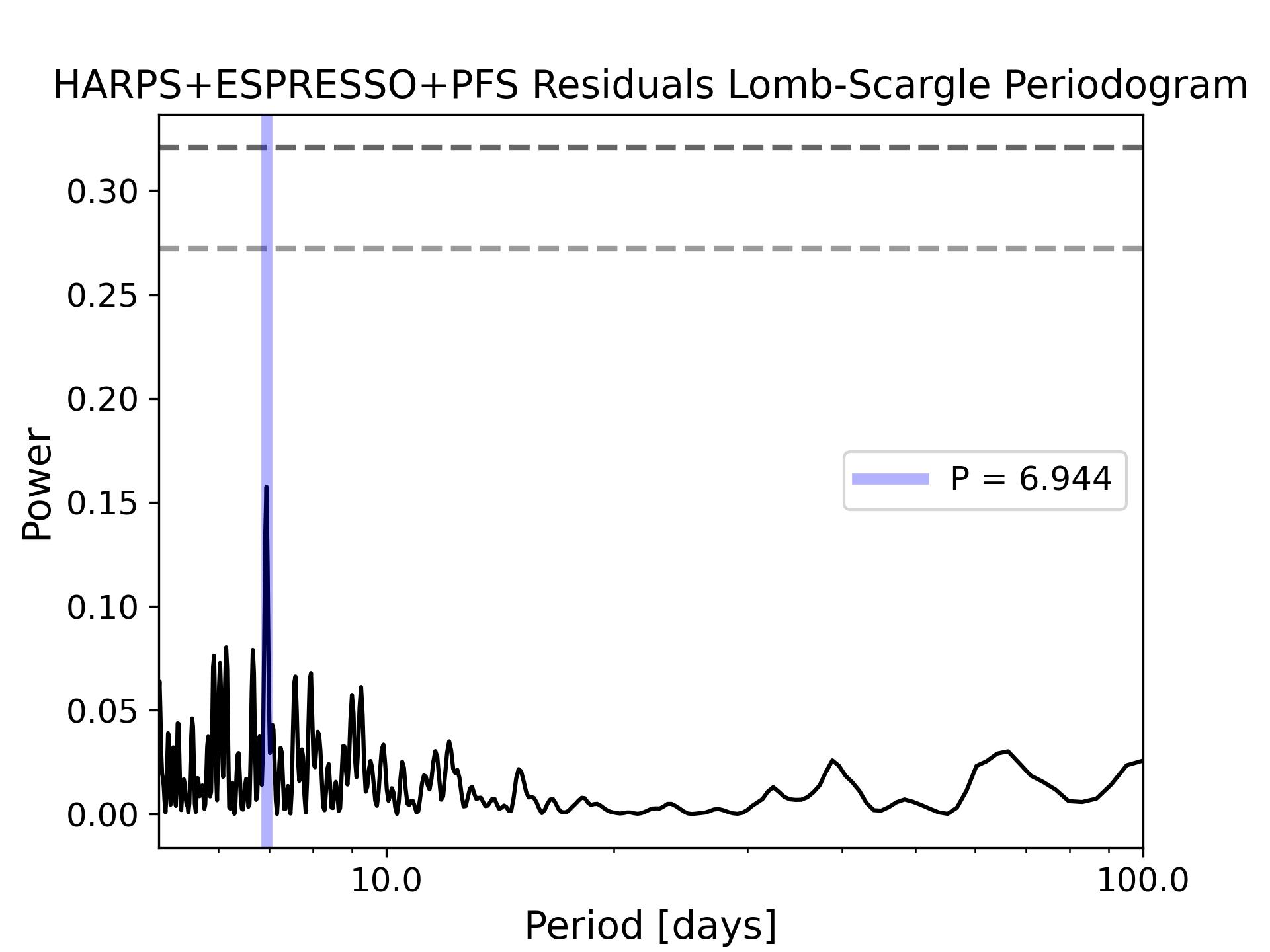}
            \caption{GLS periodogram of the RV residuals (after subtracting the orbit of TOI$-$757 b as well as the GP noise model). The dashed horizontal lines represent the FAP significance at the 0.1\% (dark gray) and 1\% (light gray) levels.}
            \label{fig:rvresid}
        \end{figure}

        To further investigate this scenario, we searched the residuals of the RV data for any signs of periodicity, which could indicate further stellar activity or possibly an outer companion.
        We found no significant peaks in the GLS periodograms (see Fig. \ref{fig:rvresid}), even out to a FAP level of 10$\%$, as well as a lack of evidence for any companions in the high$-$contrast imaging and the \textit{Gaia} data.
        The absence of the peak likely suggests that the outer planet (assuming it is responsible for the large eccentricity of TOI$-$757 b) is very small in mass or it has a much longer period (or it has been ejected from the system already).

        However, it is important to note that TOI$-$757 b’s properties are in agreement with the scenario presented by \cite{huang2017ecc}, which showed that planetary systems that were able to survive scattering due to a giant outer companion exhibited higher eccentricity and lower multiplicity.
        This would put TOI$-$757 b amongst an exciting population of exoplanets for further study and would make it an interesting system to follow$-$up.
        In particular, to verify whether high eccentricity tidal migration due to planet$-$planet interactions is the correct scenario for TOI$-$757 b's evolutionary history, further RV measurements over a longer baseline would aid in searching for the existence of a possible distant giant perturber.
        Another possible mechanism for the high eccentricity of the planet is self$-$excitation, where the planet migrates via planet$-$disk interactions.
        However, planets are expected to have lower eccentricities in this scenario ($e$\,<\,0.3), so this is unlikely to be the mechanism for TOI$-$757 b's formation history given its high eccentricity \citep[e.g.,][]{petrovich2014scattering, matsumoto2015eccentricity, vaneylen2019eccentricity}.

        To attempt to further understand the planet's formation history and evolution given its high eccentricity, we compare TOI$-$757 b to other similar eccentric mini$-$Neptunes that have been previously studied and well$-$characterized. 
        In Fig.~\ref{fig:pop_comparison}, we plot TOI$-$757 b against the confirmed mini$-$Neptune population ($P$\,<\,100\,days, 2\,< $R_{\text{p}}$\,<\,5\,$R_\oplus$, 5\,<\,$M_{\text{p}}$\,<\,40\,\,$M_\oplus$), taken from the NASA Exoplanet Archive with mass and radius uncertainties <\,20$\%$.
        When looking at relatively small exoplanets with longer periods, the closest planet to TOI$-$757 b is K2$-$280 b.
        From the $M_{\text{p}}-e$, $R_{\text{p}}-e$ and [Fe/H]$-e$ plots, we jointly find two planets with similar properties to TOI$-$757 b, namely, TOI$-$269 b and Wolf 503 b.
    
        K2$-$280 b was analyzed by \cite{nowak2020k2280b}, who characterized the planet as a warm sub$-$Saturn with a radius of 7.5 $\pm$ 0.4 $R_{\oplus}$ and a mass of  37.1 $\pm$ 5.6 $M_{\oplus}$, yielding a low planet density of 0.5$^{+0.1}_{-0.1}$ g\,cm$^{-3}$.
        While the planet sits in a different parameter space on the mass$-$radius diagram, it is a transiting single$-$planet system with a similar long$-$period orbit ($\sim$19.9 days) and eccentricity (0.35$^{+0.05}_{-0.04}$) to TOI$-$757 b, which could indicate a similar formation mechanism that resulted in the high eccentricity of our system (eg. migration).
        The study attributes the moderate eccentricity of K2$-$280 b to a formation pathway due to planet–planet gravitational interactions.

        TOI$-$269 b was characterized by \cite{cointepas2021toi269b}, who report a radius of 2.8\,$\pm$\,0.1 $R_{\oplus}$, and a mass of 8.8\,$\pm$\,1.4 $M_{\oplus}$, resulting in a planet density of 2.3$^{+0.5}_{-0.4}$ g\,cm$^{-3}$.
        The host star is reported to have a low metallicity of [Fe/H] = $-$0.3 $\pm$ 0.1, which is very similar to TOI$-$757.
        This study discusses the possibility of planet$-$planet migration as the source of the planet's high eccentricity ($e$ = 0.4 $\pm$ 0.1).
        While it is a short$-$period planet (P\,$\approx$\,3.7 days) orbiting an M2 dwarf star, and as such the host star conditions would affect the planet formation slightly differently, it is interesting that it might possibly share a similar formation history to TOI$-$757 b that led it to obtain an eccentric orbit.
        
        Lastly, we note the properties of Wolf 503 b, a mini$-$Neptune that was studied by \cite{peterson2018wolf503b}, \cite{polanski2021wolf503b} and \cite{bonomo2023wolf503b}.
        This is possibly the most similar system to TOI$-$757 b, orbiting a K$-$dwarf host star. 
        The planet is eccentric (0.4\,$\pm$\,0.1), with a radius of 2.04\,$\pm$\,0.07 $R_{\oplus}$ and a mass of 6.3 $\pm$ 0.7 $M_{\oplus}$, giving a bulk density of  2.9$^{+0.5}_{-0.4}$\,g\,cm$^{-3}$.
        Wolf 503 is similar to TOI$-$757 in that it is an old system (11 $\pm$ 2 Gyr) that is also metal poor ([Fe/H] = $-$0.5 $\pm$ 0.1), which indicates it might share a similar evolution history to our star.
        Despite initial predictions of the existence of a long$-$period companion as the main driving source of the planet's high eccentricity, the system was later followed up over a long RV campaign, where no outer planet companions were found \citep{bonomo2023wolf503b}. 
        The interior composition modeling done in \cite{polanski2021wolf503b} shows that their reported density is consistent with an Earth$-$like core combined with either a significant water mass fraction (45$^{+19}_{-16}$\%), or a 0.5\,$\pm$\,0.3\% H/He atmosphere. 
        The study also suggests that Wolf 503 b would be a good target to test XUV$-$driven mass loss mechanisms due to the predicted low H/He mass fraction, as well as the old age of the host star.
        The system's strikingly similar properties to our planet suggest that TOI$-$757 b might also be favorable for testing theories of XUV$-$driven mass loss.

        Based on the plots in Figure \ref{fig:pop_comparison}, we can see that TOI$-$757 b stands out as part of a sparsely populated group of eccentric mini$-$Neptune planets.
        The $M_{\text{p}}-e$ plot in particular highlights that low$-$mass planets do not often appear to be eccentric. 
        This could in part be because systems like TOI$-$757 b are hard to observe. 
        Eccentric planet systems are most likely at longer periods, and such systems often lack precise mass measurements and are also less likely to transit.
        The Age$-e$ plot does not show any visible trends, but stellar ages are also harder to estimate and have much larger uncertainties, so we cannot make any reliable conclusions from this plot.
        The [Fe/H]$-e$ plot on the other hand is intriguing because it shows two separate sub$-$groups of planets. 
        There is a large group of planets (but not highly eccentric) between $\sim$ 0 < [Fe/H] < 0.2. 
        Then, we observe a very small subset of planets (3 planets, including TOI$-$757 b) that are completely isolated from this large population, and all of them have much higher eccentricities ($e$ > 0.4). 
        2 out of the 3 planets in this subset (TOI$-$757 b and Wolf 503 b) are part of systems that are old in age and low in metallicity, while the third system (TOI$-$269 b) is low in metallicity but does not have a well$-$constrained age measurement.
        All 3 of these cases are single$-$planet systems, with the source of their eccentricity still unknown. 

        While TOI$-$757 b's high eccentricity is unlikely due to self$-$excitation, and studies of eccentric systems focus on scenarios involving gravitational interactions with an outer companion, we propose the possibility of a more intrinsic explanation for the high eccentricity due to star$-$star interactions during the earlier epoch of the Galactic disk formation, given the low metallicity and the older age of TOI$-$757. 
        More specifically, the high eccentricity of TOI$-$757 b could be due to star$-$star interactions during the earlier epoch of the Galactic disk formation. 
        Our star is old in age (7.5 $\pm$ 3.0 Gyr) and is also placed during the transition period between thick and thin disks based on its $\alpha$/Fe (0.08 $\pm$ 0.05) and Fe/H ($-$0.27 $\pm$ 0.04) measurements \citep{ciuca2021mwdiscs}.
        During this epoch, the Milky Way disk has a higher gas fraction, and the star formation rate is higher as well.
        Stars in this period may be forming in the denser molecular clouds, which could increase the probability of star$-$star interactions \citep{zwart2015fragility}, and this alternative effect may become important for these old stars.
        This might be another, more intrinsic explanation for the high eccentricity of TOI$-$757 b coming from the initial conditions of the environment in which the planet was formed, given the older age of the system. 
        As such, the discovery of high$-$eccentricity mini$-$Neptunes around old metal poor stars could open up the new possibility of the Milky Way environment affecting the formation of the exoplanet, which has not been previously well$-$studied as a potential scenario for the evolution path of eccentric systems.

\section{Summary and Conclusions}\label{sec:summary}

In this paper, we present the confirmation and the planet properties of TOI$-$757 b, a transiting mini$-$Neptune that was discovered by the \textit{TESS} mission.
We clearly detect TOI$-$757 b in the \textit{TESS} light curves, and we also make use of the \textit{CHEOPS} photometric follow$-$up observations of the system, allowing us to derive a more precise radius measurement of the planet, along with ground$-$based photometry from WASP, KELT, and LCOGT.
We present multiple RV monitoring campaigns that have been carried out using the HARPS, ESPRESSO, and PFS high$-$resolution spectrographs to measure the mass of TOI$-$757 b, where we find a clear confirmation of the planet.

Through the photometric and RV analyses, we find that TOI$-$757 b has a radius of $R_{\text{p}}$ = 2.5 $\pm$ 0.1 $R_{\oplus}$ and a mass of $M_{\text{p}}$ = 10.5$^{+2.2}_{-2.1}$ $M_{\oplus}$, giving us an implied density of $\rho_{\text{p}} = 3.6 \pm 0.8$ g cm$^{-3}$.
It is unclear whether TOI-757 b harbours any water/ice components or whether it has a fraction of a H/He envelope.
Our internal composition modeling reveals that, while the existence of some quantity of H/He was found to be required, the actual quantity is degenerate with the amount of water, and neither scenario is well constrained.
We highlight the importance of acquiring atmospheric observations to break the degeneracy between planet composition models for mini$-$Neptunes and water$-$rich worlds, which would particularly be useful for TOI$-$757 b given its location on the M$-$R diagram. 

We find the planet to be highly eccentric with $e$ = 0.39$^{+0.08}_{-0.07}$, making it one of the very few highly eccentric mini$-$Neptunes among all precisely known exoplanets from the mini$-$Neptune population.
Additionally, the tidal circularization timescale of the planet was found to be much longer than the age of the system, indicating that the planet likely attained its eccentricity very early on during the formation stages of the system. 
Based on comparisons to other similar eccentric systems, we find a likely scenario for TOI$-$757 b's formation to be high eccentricity excitation due to a distant outer companion.
We additionally propose the possibility of a more intrinsic explanation for the high eccentricity due to star$-$star interactions during the earlier epoch of the Galactic disk formation, given the low metallicity and the older age of TOI$-$757.

\section*{Acknowledgements}

We thank the referee for their thoughtful and helpful feedback, which helped improve the paper.
AAQ acknowledges the funding bodies, the UAE Ministry of Presidential Affairs and the UAE Space Agency, for their support through the PhD scholarship. 
This work was supported by the KESPRINT collaboration, an international consortium devoted to the follow-up and characterization of exoplanets discovered from space-based missions (\protect\url{http://www.kesprint.science/}).
We are extremely grateful to the ESO staff members for their unique and superb support during the observations.
Based on observations performed with the \textit{CHEOPS} space telescope. \textit{CHEOPS} is an ESA mission in partnership with Switzerland with important contributions to the payload and the ground segment from Austria, Belgium, France, Germany, Hungary, Italy, Portugal, Spain, Sweden, and the United Kingdom. The \textit{CHEOPS} Consortium would like to gratefully acknowledge the support received by all the agencies, offices, universities, and industries involved. Their flexibility and willingness to explore new approaches were essential to the success of this mission. \textit{CHEOPS} data analysed in this article will be made available in the \textit{CHEOPS} mission archive (\url{https://cheops.unige.ch/archive_browser/}). 
We  thank  the  Swiss  National  Science  Foundation  (SNSF) and the Geneva University for their continuous support to our planet search programs. This work has been in particular carried out in the frame of the National Centre for Competence in Research {\it PlanetS} supported by the SNSF. This publication makes use of The Data \& Analysis Center for Exoplanets (DACE), which is a facility based at the University of Geneva dedicated to extrasolar planets data visualisation, exchange and analysis.
We acknowledge the use of public \textit{TESS} data from pipelines at the \textit{TESS} Science Office and at the \textit{TESS} Science Processing Operations Center (SPOC). Resources supporting this work were provided by the NASA High-End Computing (HEC) Program through the NASA Advanced Supercomputing (NAS) Division at Ames Research Center for the production of the SPOC data products. Funding for the \textit{TESS} mission is provided by NASA's Science Mission Directorate. KAC acknowledges support from the \textit{TESS} mission via subaward s3449 from MIT.
\textit{TESS} data presented in this paper were obtained from the Milkulski Archive for Space Telescopes (MAST) at the Space Telescope Science Institute.
Some of the observations in this paper made use of the High-Resolution Imaging instrument ‘Alopeke and were obtained under Gemini LLP Proposal Number: GN/S-2021A-LP-105. ‘Alopeke was funded by the NASA Exoplanet Exploration Program and built at the NASA Ames Research Center by Steve B. Howell, Nic Scott, Elliott P. Horch, and Emmett Quigley. Alopeke was mounted on the Gemini North telescope of the international Gemini Observatory, a program of NSF’s OIR Lab, which is managed by the Association of Universities for Research in Astronomy (AURA) under a cooperative agreement with the National Science Foundation. on behalf of the Gemini partnership: the National Science Foundation (United States), National Research Council (Canada), Agencia Nacional de Investigación y Desarrollo (Chile), Ministerio de Ciencia, Tecnología e Innovación (Argentina), Ministério da Ciência, Tecnologia, Inovações e Comunicações (Brazil), and Korea Astronomy and Space Science Institute (Republic of Korea).
This work was enabled by observations made from the Subaru, Gemini North, and Keck telescopes, located within the Maunakea Science Reserve and adjacent to the summit of Maunakea. We are grateful for the privilege of observing the Universe from a place that is unique in both its astronomical quality and its cultural significance.
This work makes use of observations from the LCOGT network. Part of the LCOGT telescope time was granted by NOIRLab through the Mid-Scale Innovations Program (MSIP). MSIP is funded by NSF.
This research has made use of the Exoplanet Follow-up Observation Program (ExoFOP; DOI: 10.26134/ExoFOP5) website, which is operated by the California Institute of Technology, under contract with the National Aeronautics and Space Administration under the Exoplanet Exploration Program.
This paper made use of data from the KELT survey \citep{pepper2007kelt}.

AAQ thanks Daisuke Kawata for his valuable comments and input on the discussion.
DG gratefully acknowledges financial support from the CRT foundation under Grant No. 2018.2323 “Gaseousor rocky? Unveiling the nature of small worlds”. 
SGS acknowledge support from FCT through FCT contract nr. CEECIND/00826/2018 and POPH/FSE (EC). 
The Portuguese team thanks the Portuguese Space Agency for the provision of financial support in the framework of the PRODEX Programme of the European Space Agency (ESA) under contract number 4000142255. 
This work has been carried out within the framework of the NCCR PlanetS supported by the Swiss National Science Foundation under grants 51NF40\_182901 and 51NF40\_205606. 
MNG is the ESA CHEOPS Project Scientist and Mission Representative, and as such also responsible for the Guest Observers (GO) Programme. MNG does not relay proprietary information between the GO and Guaranteed Time Observation (GTO) Programmes, and does not decide on the definition and target selection of the GTO Programme. 
TWi acknowledges support from the UKSA and the University of Warwick. 
ABr was supported by the SNSA. 
YAl acknowledges support from the Swiss National Science Foundation (SNSF) under grant 200020\_192038. 
The Belgian participation to CHEOPS has been supported by the Belgian Federal Science Policy Office (BELSPO) in the framework of the PRODEX Program, and by the University of Liège through an ARC grant for Concerted Research Actions financed by the Wallonia-Brussels Federation. 
LD thanks the Belgian Federal Science Policy Office (BELSPO) for the provision of financial support in the framework of the PRODEX Programme of the European Space Agency (ESA) under contract number 4000142531. 
SCCB acknowledges support from FCT through FCT contracts nr. IF/01312/2014/CP1215/CT0004. 
This work was supported by FCT - Funda\c{c}\~{a}o para a Ci\^{e}ncia e a Tecnologia through national funds and by FEDER through COMPETE2020 through the research grants UIDB/04434/2020, UIDP/04434/2020, 2022.06962.PTDC. 
ODSD is supported in the form of work contract (DL 57/2016/CP1364/CT0004) funded by national funds through FCT. 
We acknowledge financial support from the Agencia Estatal de Investigación of the Ministerio de Ciencia e Innovación MCIN/AEI/10.13039/501100011033 and the ERDF “A way of making Europe” through projects PID2019-107061GB-C61, PID2019-107061GB-C66, PID2021-125627OB-C31, and PID2021-125627OB-C32, from the Centre of Excellence “Severo Ochoa” award to the Instituto de Astrofísica de Canarias (CEX2019-000920-S), from the Centre of Excellence “María de Maeztu” award to the Institut de Ciències de l’Espai (CEX2020-001058-M), and from the Generalitat de Catalunya/CERCA programme. 
DBa, EPa, and IRi acknowledge financial support from the Agencia Estatal de Investigación of the Ministerio de Ciencia e Innovación MCIN/AEI/10.13039/501100011033 and the ERDF “A way of making Europe” through projects PID2019-107061GB-C61, PID2019-107061GB-C66, PID2021-125627OB-C31, and PID2021-125627OB-C32, from the Centre of Excellence “Severo Ochoa'' award to the Instituto de Astrofísica de Canarias (CEX2019-000920-S), from the Centre of Excellence “María de Maeztu” award to the Institut de Ciències de l’Espai (CEX2020-001058-M), and from the Generalitat de Catalunya/CERCA programme. 
LBo, VNa, IPa, GPi, RRa, and GSc acknowledge support from CHEOPS ASI-INAF agreement n. 2019-29-HH.0. 
CBr and ASi acknowledge support from the Swiss Space Office through the ESA PRODEX program. 
ACC acknowledges support from STFC consolidated grant numbers ST/R000824/1 and ST/V000861/1, and UKSA grant number ST/R003203/1. 
PEC. is funded by the Austrian Science Fund (FWF) Erwin Schroedinger Fellowship, program J4595-N. 
This project was supported by the CNES. 
B.-O. D. acknowledges support from the Swiss State Secretariat for Education, Research and Innovation (SERI) under contract number MB22.00046. 
This project has received funding from the Swiss National Science Foundation for project 200021\_200726. It has also been carried out within the framework of the National Centre of Competence in Research PlanetS supported by the Swiss National Science Foundation under grant 51NF40\_205606. The authors acknowledge the financial support of the SNSF. 
MF and CMP gratefully acknowledge the support of the Swedish National Space Agency (DNR 65/19, 174/18). 
MG is an F.R.S.-FNRS Senior Research Associate. 
CHe acknowledges support from the European Union H2020-MSCA-ITN-2019 under Grant Agreement no. 860470 (CHAMELEON). 
SH gratefully acknowledges CNES funding through the grant 837319. 
KWFL was supported by Deutsche Forschungsgemeinschaft grants RA714/14-1 within the DFG Schwerpunkt SPP 1992, Exploring the Diversity of Extrasolar Planets. 
This work was granted access to the HPC resources of MesoPSL financed by the Region Ile de France and the project Equip@Meso (reference ANR-10-EQPX-29-01) of the programme Investissements d'Avenir supervised by the Agence Nationale pour la Recherche. 
ML acknowledges support of the Swiss National Science Foundation under grant number PCEFP2\_194576. 
PM acknowledges support from STFC research grant number ST/R000638/1. 
This work was also partially supported by a grant from the Simons Foundation (PI Queloz, grant number 327127). 
NCSa acknowledges funding by the European Union (ERC, FIERCE, 101052347). Views and opinions expressed are however those of the author(s) only and do not necessarily reflect those of the European Union or the European Research Council. Neither the European Union nor the granting authority can be held responsible for them. 
GyMSz acknowledges the support of the Hungarian National Research, Development and Innovation Office (NKFIH) grant K-125015, a PRODEX Experiment Agreement No. 4000137122, the Lend\"ulet LP2018-7/2021 grant of the Hungarian Academy of Science and the support of the city of Szombathely. 
VVG is an F.R.S-FNRS Research Associate. 
JV acknowledges support from the Swiss National Science Foundation (SNSF) under grant PZ00P2\_208945. 
NAW acknowledges UKSA grant ST/R004838/1. 
RL acknowledges funding from University of La Laguna through the Margarita Salas Fellowship from the Spanish Ministry of Universities ref. UNI/551/2021-May 26, and under the EU Next Generation funds.
GN thanks the research funding from the Ministry of Education and Science programme the "Excellence Initiative - Research University" conducted at the Centre of Excellence in Astrophysics and Astrochemistry of the Nicolaus Copernicus University in Toru\'n, Poland.
This work is partly supported by JSPS KAKENHI Grant Number JP18H05439, JST CREST Grant Number JPMJCR1761.
MR acknowledges support from Heising-Simons grant \#2023-4478.
We acknowledge financial support from the Agencia Estatal de Investigaci\'on of the Ministerio de Ciencia e Innovaci\'on MCIN/AEI/10.13039/501100011033 and the ERDF “A way of making Europe” through project PID2021-125627OB-C32, and from the Centre of Excellence "Severo Ochoa" award to the Instituto de Astrofisica de Canarias.
The contributions at the Mullard Space Science Laboratory by E.M.B. have been supported by STFC through the consolidated grant ST/W001136/1.
T.D. acknowledges support by the McDonnell Center for the Space Sciences at Washington University in St. Louis.
N.M.R. acknowledges support from FCT through grant DFA/BD/5472/2020.
ACMC ackowledges support from the FCT, Portugal, through the CFisUC projects UIDB/04564/2020 and UIDP/04564/2020, with DOI identifiers 10.54499/UIDB/04564/2020 and 10.54499/UIDP/04564/2020, respectively. 
J.~K. gratefully acknowledges the support of the Swedish National Space Agency (SNSA; DNR 2020-00104) and of the Swedish Research Council  (VR: Etableringsbidrag 2017-04945)
This project has received funding from the European Research Council (ERC) under the European Union’s Horizon 2020 research and innovation programme (project {\sc Four Aces}; grant agreement No 724427).  It has also been carried out in the frame of the National Centre for Competence in Research PlanetS supported by the Swiss National Science Foundation (SNSF). D.Eh. and A.De. acknowledge financial support from the SNSF for project 200021\_200726.

\section*{Data Availability}
The data and coding scripts used in this work can be made available upon reasonable request.



\bibliographystyle{mnras} 
\bibliography{bibliography} 

\begin{thebibliography}{}
\makeatletter
\relax
\def\mn@urlcharsother{\let\do\@makeother \do\$\do\&\do\#\do\^\do\_\do\%\do\~}
\def\mn@doi{\begingroup\mn@urlcharsother \@ifnextchar [ {\mn@doi@}
  {\mn@doi@[]}}
\def\mn@doi@[#1]#2{\def\@tempa{#1}\ifx\@tempa\@empty \href
  {http://dx.doi.org/#2} {doi:#2}\else \href {http://dx.doi.org/#2} {#1}\fi
  \endgroup}
\def\mn@eprint#1#2{\mn@eprint@#1:#2::\@nil}
\def\mn@eprint@arXiv#1{\href {http://arxiv.org/abs/#1} {{\tt arXiv:#1}}}
\def\mn@eprint@dblp#1{\href {http://dblp.uni-trier.de/rec/bibtex/#1.xml}
  {dblp:#1}}
\def\mn@eprint@#1:#2:#3:#4\@nil{\def\@tempa {#1}\def\@tempb {#2}\def\@tempc
  {#3}\ifx \@tempc \@empty \let \@tempc \@tempb \let \@tempb \@tempa \fi \ifx
  \@tempb \@empty \def\@tempb {arXiv}\fi \@ifundefined
  {mn@eprint@\@tempb}{\@tempb:\@tempc}{\expandafter \expandafter \csname
  mn@eprint@\@tempb\endcsname \expandafter{\@tempc}}}

\bibitem[\protect\citeauthoryear{{Adibekyan}, {Sousa}, {Santos}, {Delgado
  Mena}, {Gonz{\'a}lez Hern{\'a}ndez}, {Israelian}, {Mayor}  \&
  {Khachatryan}}{{Adibekyan} et~al.}{2012}]{Adibekyan-12}
{Adibekyan} V.~Z.,  {Sousa} S.~G.,  {Santos} N.~C.,  {Delgado Mena} E.,
  {Gonz{\'a}lez Hern{\'a}ndez} J.~I.,  {Israelian} G.,  {Mayor} M.,
  {Khachatryan} G.,  2012, \mn@doi [\aap] {10.1051/0004-6361/201219401}, \href
  {http://adsabs.harvard.edu/abs/2012A%26A...545A..32A} {545, A32}

\bibitem[\protect\citeauthoryear{{Adibekyan} et~al.,}{{Adibekyan}
  et~al.}{2015}]{Adibekyan-15}
{Adibekyan} V.,  et~al., 2015, \mn@doi [\aap] {10.1051/0004-6361/201527120},
  \href {http://adsabs.harvard.edu/abs/2015A%26A...583A..94A} {583, A94}

\bibitem[\protect\citeauthoryear{{Agol}, {Luger}  \& {Foreman-Mackey}}{{Agol}
  et~al.}{2020}]{exoplanet:agol20}
{Agol} E.,  {Luger} R.,   {Foreman-Mackey} D.,  2020, \mn@doi [\aj]
  {10.3847/1538-3881/ab4fee}, \href
  {https://ui.adsabs.harvard.edu/abs/2020AJ....159..123A} {159, 123}

\bibitem[\protect\citeauthoryear{{Akeson} et~al.,}{{Akeson}
  et~al.}{2013}]{akeson2013nexi}
{Akeson} R.~L.,  et~al., 2013, \mn@doi [\pasp] {10.1086/672273}, \href
  {https://ui.adsabs.harvard.edu/abs/2013PASP..125..989A} {125, 989}

\bibitem[\protect\citeauthoryear{{Anderson} et~al.,}{{Anderson}
  et~al.}{2011}]{anderson2011wasp30b}
{Anderson} D.~R.,  et~al., 2011, \mn@doi [\apjl] {10.1088/2041-8205/726/2/L19},
  \href {https://ui.adsabs.harvard.edu/abs/2011ApJ...726L..19A} {726, L19}

\bibitem[\protect\citeauthoryear{{Anglada-Escud{\'e}} \&
  {Butler}}{{Anglada-Escud{\'e}} \& {Butler}}{2012}]{Anglada2012}
{Anglada-Escud{\'e}} G.,  {Butler} R.~P.,  2012, \mn@doi [\apjs]
  {10.1088/0067-0049/200/2/15}, \href
  {https://ui.adsabs.harvard.edu/abs/2012ApJS..200...15A} {200, 15}

\bibitem[\protect\citeauthoryear{{Armitage}, {Eisner}  \& {Simon}}{{Armitage}
  et~al.}{2016}]{armitage2016snowline}
{Armitage} P.~J.,  {Eisner} J.~A.,   {Simon} J.~B.,  2016, \mn@doi [\apjl]
  {10.3847/2041-8205/828/1/L2}, \href
  {https://ui.adsabs.harvard.edu/abs/2016ApJ...828L...2A} {828, L2}

\bibitem[\protect\citeauthoryear{{Astropy Collaboration} et~al.,}{{Astropy
  Collaboration} et~al.}{2013}]{astropy:2013}
{Astropy Collaboration} et~al., 2013, \mn@doi [\aap]
  {10.1051/0004-6361/201322068}, \href
  {http://adsabs.harvard.edu/abs/2013A%26A...558A..33A} {558, A33}

\bibitem[\protect\citeauthoryear{{Astropy Collaboration} et~al.,}{{Astropy
  Collaboration} et~al.}{2018}]{astropy:2018}
{Astropy Collaboration} et~al., 2018, \mn@doi [\aj] {10.3847/1538-3881/aabc4f},
  \href {https://ui.adsabs.harvard.edu/abs/2018AJ....156..123A} {156, 123}

\bibitem[\protect\citeauthoryear{{Astropy Collaboration} et~al.,}{{Astropy
  Collaboration} et~al.}{2022}]{astropy:2022}
{Astropy Collaboration} et~al., 2022, \mn@doi [\apj]
  {10.3847/1538-4357/ac7c74}, \href
  {https://ui.adsabs.harvard.edu/abs/2022ApJ...935..167A} {935, 167}

\bibitem[\protect\citeauthoryear{{Baranne} et~al.,}{{Baranne}
  et~al.}{1996}]{Baranne1996}
{Baranne} A.,  et~al., 1996, \aaps, \href
  {https://ui.adsabs.harvard.edu/abs/1996A&AS..119..373B} {119, 373}

\bibitem[\protect\citeauthoryear{{Barnes}, {Jackson}, {Raymond}, {West}  \&
  {Greenberg}}{{Barnes} et~al.}{2009}]{barnes2009ecc}
{Barnes} R.,  {Jackson} B.,  {Raymond} S.~N.,  {West} A.~A.,   {Greenberg} R.,
  2009, \mn@doi [\apj] {10.1088/0004-637X/695/2/1006}, \href
  {https://ui.adsabs.harvard.edu/abs/2009ApJ...695.1006B} {695, 1006}

\bibitem[\protect\citeauthoryear{{Barrag{\'a}n}, {Gandolfi}  \&
  {Antoniciello}}{{Barrag{\'a}n} et~al.}{2019}]{barragan2019pyaneti}
{Barrag{\'a}n} O.,  {Gandolfi} D.,   {Antoniciello} G.,  2019, \mn@doi [\mnras]
  {10.1093/mnras/sty2472}, \href
  {https://ui.adsabs.harvard.edu/abs/2019MNRAS.482.1017B} {482, 1017}

\bibitem[\protect\citeauthoryear{{Bean}, {Raymond}  \& {Owen}}{{Bean}
  et~al.}{2021}]{bean2021subneptunes}
{Bean} J.~L.,  {Raymond} S.~N.,   {Owen} J.~E.,  2021, \mn@doi [Journal of
  Geophysical Research (Planets)] {10.1029/2020JE006639}, \href
  {https://ui.adsabs.harvard.edu/abs/2021JGRE..12606639B} {126, e06639}

\bibitem[\protect\citeauthoryear{{Benneke} \& {Seager}}{{Benneke} \&
  {Seager}}{2013}]{benneke2013cloudy}
{Benneke} B.,  {Seager} S.,  2013, \mn@doi [\apj]
  {10.1088/0004-637X/778/2/153}, \href
  {https://ui.adsabs.harvard.edu/abs/2013ApJ...778..153B} {778, 153}

\bibitem[\protect\citeauthoryear{{Benz} et~al.,}{{Benz}
  et~al.}{2021}]{2021ExA....51..109B}
{Benz} W.,  et~al., 2021, \mn@doi [Experimental Astronomy]
  {10.1007/s10686-020-09679-4}, \href
  {https://ui.adsabs.harvard.edu/abs/2021ExA....51..109B} {51, 109}

\bibitem[\protect\citeauthoryear{{Bitsch}, {Raymond}  \& {Izidoro}}{{Bitsch}
  et~al.}{2019}]{bitsch2019rockywater}
{Bitsch} B.,  {Raymond} S.~N.,   {Izidoro} A.,  2019, \mn@doi [\aap]
  {10.1051/0004-6361/201935007}, \href
  {https://ui.adsabs.harvard.edu/abs/2019A&A...624A.109B} {624, A109}

\bibitem[\protect\citeauthoryear{{Blackwell} \& {Shallis}}{{Blackwell} \&
  {Shallis}}{1977}]{Blackwell1977}
{Blackwell} D.~E.,  {Shallis} M.~J.,  1977, \mn@doi [\mnras]
  {10.1093/mnras/180.2.177}, \href
  {https://ui.adsabs.harvard.edu/abs/1977MNRAS.180..177B} {180, 177}

\bibitem[\protect\citeauthoryear{{Bonfanti}, {Ortolani}, {Piotto}  \&
  {Nascimbeni}}{{Bonfanti} et~al.}{2015}]{bonfanti2015}
{Bonfanti} A.,  {Ortolani} S.,  {Piotto} G.,   {Nascimbeni} V.,  2015, \mn@doi
  [\aap] {10.1051/0004-6361/201424951}, \href
  {http://adsabs.harvard.edu/abs/2015A%26A...575A..18B} {575, A18}

\bibitem[\protect\citeauthoryear{{Bonfanti}, {Ortolani}  \&
  {Nascimbeni}}{{Bonfanti} et~al.}{2016}]{bonfanti2016}
{Bonfanti} A.,  {Ortolani} S.,   {Nascimbeni} V.,  2016, \mn@doi [\aap]
  {10.1051/0004-6361/201527297}, \href
  {http://adsabs.harvard.edu/abs/2016A%26A...585A...5B} {585, A5}

\bibitem[\protect\citeauthoryear{{Bonfanti} et~al.,}{{Bonfanti}
  et~al.}{2021}]{bonfanti2021}
{Bonfanti} A.,  et~al., 2021, \mn@doi [\aap] {10.1051/0004-6361/202039608},
  \href {https://ui.adsabs.harvard.edu/abs/2021A&A...646A.157B} {646, A157}

\bibitem[\protect\citeauthoryear{{Bonomo} et~al.,}{{Bonomo}
  et~al.}{2023}]{bonomo2023wolf503b}
{Bonomo} A.~S.,  et~al., 2023, \mn@doi [arXiv e-prints]
  {10.48550/arXiv.2304.05773}, \href
  {https://ui.adsabs.harvard.edu/abs/2023arXiv230405773B} {p. arXiv:2304.05773}

\bibitem[\protect\citeauthoryear{{Brown} et~al.,}{{Brown}
  et~al.}{2013}]{Brown:2013}
{Brown} T.~M.,  et~al., 2013, \mn@doi [\pasp] {10.1086/673168}, \href
  {https://ui.adsabs.harvard.edu/abs/2013PASP..125.1031B} {125, 1031}

\bibitem[\protect\citeauthoryear{{Butler}, {Marcy}, {Williams}, {McCarthy},
  {Dosanjh}  \& {Vogt}}{{Butler} et~al.}{1996}]{butler1996}
{Butler} R.~P.,  {Marcy} G.~W.,  {Williams} E.,  {McCarthy} C.,  {Dosanjh} P.,
   {Vogt} S.~S.,  1996, \mn@doi [\pasp] {10.1086/133755}, \href
  {http://adsabs.harvard.edu/abs/1996PASP..108..500B} {108, 500}

\bibitem[\protect\citeauthoryear{{Cabrera}, {Csizmadia}, {Erikson}, {Rauer}  \&
  {Kirste}}{{Cabrera} et~al.}{2012}]{2012A&A...548A..44C}
{Cabrera} J.,  {Csizmadia} S.,  {Erikson} A.,  {Rauer} H.,   {Kirste} S.,
  2012, \mn@doi [\aap] {10.1051/0004-6361/201219337}, \href
  {https://ui.adsabs.harvard.edu/abs/2012A&A...548A..44C} {548, A44}

\bibitem[\protect\citeauthoryear{{Castelli} \& {Kurucz}}{{Castelli} \&
  {Kurucz}}{2003}]{Castelli2003}
{Castelli} F.,  {Kurucz} R.~L.,  2003, in {Piskunov} N.,  {Weiss} W.~W.,
  {Gray} D.~F.,  eds,  IAU Symposium Vol. 210, Modelling of Stellar
  Atmospheres. p.~A20 (\mn@eprint {arXiv} {astro-ph/0405087})

\bibitem[\protect\citeauthoryear{{Ciuc{\u{a}}}, {Kawata}, {Miglio}, {Davies}
  \& {Grand}}{{Ciuc{\u{a}}} et~al.}{2021}]{ciuca2021mwdiscs}
{Ciuc{\u{a}}} I.,  {Kawata} D.,  {Miglio} A.,  {Davies} G.~R.,   {Grand} R.
  J.~J.,  2021, \mn@doi [\mnras] {10.1093/mnras/stab639}, \href
  {https://ui.adsabs.harvard.edu/abs/2021MNRAS.503.2814C} {503, 2814}

\bibitem[\protect\citeauthoryear{{Cointepas} et~al.,}{{Cointepas}
  et~al.}{2021}]{cointepas2021toi269b}
{Cointepas} M.,  et~al., 2021, \mn@doi [\aap] {10.1051/0004-6361/202140328},
  \href {https://ui.adsabs.harvard.edu/abs/2021A&A...650A.145C} {650, A145}

\bibitem[\protect\citeauthoryear{{Collins}}{{Collins}}{2019}]{collins:2019}
{Collins} K.,  2019, in American Astronomical Society Meeting Abstracts \#233.
  p. 140.05

\bibitem[\protect\citeauthoryear{{Collins}, {Kielkopf}, {Stassun}  \&
  {Hessman}}{{Collins} et~al.}{2017}]{Collins:2017}
{Collins} K.~A.,  {Kielkopf} J.~F.,  {Stassun} K.~G.,   {Hessman} F.~V.,  2017,
  \mn@doi [\aj] {10.3847/1538-3881/153/2/77}, \href
  {http://adsabs.harvard.edu/abs/2017AJ....153...77C} {153, 77}

\bibitem[\protect\citeauthoryear{{Crane}, {Shectman}  \& {Butler}}{{Crane}
  et~al.}{2006}]{crane2006}
{Crane} J.~D.,  {Shectman} S.~A.,   {Butler} R.~P.,  2006, in Society of
  Photo-Optical Instrumentation Engineers (SPIE) Conference Series. p. 626931,
  \mn@doi{10.1117/12.672339}

\bibitem[\protect\citeauthoryear{{Crane}, {Shectman}, {Butler}, {Thompson}  \&
  {Burley}}{{Crane} et~al.}{2008}]{crane2008}
{Crane} J.~D.,  {Shectman} S.~A.,  {Butler} R.~P.,  {Thompson} I.~B.,
  {Burley} G.~S.,  2008, in Ground-based and Airborne Instrumentation for
  Astronomy II. p. 701479, \mn@doi{10.1117/12.789637}

\bibitem[\protect\citeauthoryear{{Crane}, {Shectman}, {Butler}, {Thompson},
  {Birk}, {Jones}  \& {Burley}}{{Crane} et~al.}{2010}]{crane2010}
{Crane} J.~D.,  {Shectman} S.~A.,  {Butler} R.~P.,  {Thompson} I.~B.,  {Birk}
  C.,  {Jones} P.,   {Burley} G.~S.,  2010, in Ground-based and Airborne
  Instrumentation for Astronomy III. p. 773553, \mn@doi{10.1117/12.857792}

\bibitem[\protect\citeauthoryear{{Cutri} et~al.,}{{Cutri}
  et~al.}{2003}]{cutri2003sloan}
{Cutri} R.~M.,  et~al., 2003, VizieR Online Data Catalog, \href
  {https://ui.adsabs.harvard.edu/abs/2003yCat.2246....0C} {p. II/246}

\bibitem[\protect\citeauthoryear{{Cutri} et~al.,}{{Cutri}
  et~al.}{2021}]{cutri2014wise}
{Cutri} R.~M.,  et~al., 2021, VizieR Online Data Catalog, \href
  {https://ui.adsabs.harvard.edu/abs/2014yCat.2328....0C} {p. II/328}

\bibitem[\protect\citeauthoryear{{Dorn}, {Venturini}, {Khan}, {Heng},
  {Alibert}, {Helled}, {Rivoldini}  \& {Benz}}{{Dorn} et~al.}{2017}]{Dorn+2017}
{Dorn} C.,  {Venturini} J.,  {Khan} A.,  {Heng} K.,  {Alibert} Y.,  {Helled}
  R.,  {Rivoldini} A.,   {Benz} W.,  2017, \mn@doi [\aap]
  {10.1051/0004-6361/201628708}, \href
  {https://ui.adsabs.harvard.edu/abs/2017A&A...597A..37D} {597, A37}

\bibitem[\protect\citeauthoryear{{Eastman}, {Siverd}  \& {Gaudi}}{{Eastman}
  et~al.}{2010}]{eastman2010timeconversion}
{Eastman} J.,  {Siverd} R.,   {Gaudi} B.~S.,  2010, \mn@doi [\pasp]
  {10.1086/655938}, \href
  {https://ui.adsabs.harvard.edu/abs/2010PASP..122..935E} {122, 935}

\bibitem[\protect\citeauthoryear{{Eastman}, {Gaudi}  \& {Agol}}{{Eastman}
  et~al.}{2013}]{eastman2013exofast}
{Eastman} J.,  {Gaudi} B.~S.,   {Agol} E.,  2013, \mn@doi [\pasp]
  {10.1086/669497}, \href
  {https://ui.adsabs.harvard.edu/abs/2013PASP..125...83E} {125, 83}

\bibitem[\protect\citeauthoryear{{El-Badry}, {Rix}  \& {Heintz}}{{El-Badry}
  et~al.}{2021}]{elbadry2021gaiabinaries}
{El-Badry} K.,  {Rix} H.-W.,   {Heintz} T.~M.,  2021, \mn@doi [\mnras]
  {10.1093/mnras/stab323}, \href
  {https://ui.adsabs.harvard.edu/abs/2021MNRAS.506.2269E} {506, 2269}

\bibitem[\protect\citeauthoryear{{Foreman-Mackey}}{{Foreman-Mackey}}{2018}]{exoplanet:foremanmackey18}
{Foreman-Mackey} D.,  2018, \mn@doi [Research Notes of the American
  Astronomical Society] {10.3847/2515-5172/aaaf6c}, \href
  {http://adsabs.harvard.edu/abs/2018RNAAS...2a..31F} {2, 31}

\bibitem[\protect\citeauthoryear{{Foreman-Mackey}, {Agol}, {Ambikasaran}  \&
  {Angus}}{{Foreman-Mackey} et~al.}{2017}]{exoplanet:foremanmackey17}
{Foreman-Mackey} D.,  {Agol} E.,  {Ambikasaran} S.,   {Angus} R.,  2017,
  \mn@doi [\aj] {10.3847/1538-3881/aa9332}, \href
  {http://adsabs.harvard.edu/abs/2017AJ....154..220F} {154, 220}

\bibitem[\protect\citeauthoryear{Foreman-Mackey et~al.,}{Foreman-Mackey
  et~al.}{2021a}]{exoplanet:zenodo}
Foreman-Mackey D.,  et~al., 2021a, exoplanet-dev/exoplanet v0.5.1,
  \mn@doi{10.5281/zenodo.1998447}, \url
  {https://doi.org/10.5281/zenodo.1998447}

\bibitem[\protect\citeauthoryear{{Foreman-Mackey} et~al.,}{{Foreman-Mackey}
  et~al.}{2021b}]{exoplanet:joss}
{Foreman-Mackey} D.,  et~al., 2021b, arXiv e-prints, \href
  {https://ui.adsabs.harvard.edu/abs/2021arXiv210501994F} {p. arXiv:2105.01994}

\bibitem[\protect\citeauthoryear{{Fragoso} \& {Louzada Neto}}{{Fragoso} \&
  {Louzada Neto}}{2015}]{fragoso2015bma}
{Fragoso} T.~M.,  {Louzada Neto} F.,  2015, \mn@doi [arXiv e-prints]
  {10.48550/arXiv.1509.08864}, \href
  {https://ui.adsabs.harvard.edu/abs/2015arXiv150908864F} {p. arXiv:1509.08864}

\bibitem[\protect\citeauthoryear{{Fulton} et~al.,}{{Fulton}
  et~al.}{2017}]{fulton2017rvalley}
{Fulton} B.~J.,  et~al., 2017, \mn@doi [\aj] {10.3847/1538-3881/aa80eb}, \href
  {https://ui.adsabs.harvard.edu/abs/2017AJ....154..109F} {154, 109}

\bibitem[\protect\citeauthoryear{{Gaia Collaboration} et~al.,}{{Gaia
  Collaboration} et~al.}{2022}]{GaiaCollaboration2022}
{Gaia Collaboration} et~al., 2022, arXiv e-prints, \href
  {https://ui.adsabs.harvard.edu/abs/2022arXiv220800211G} {p. arXiv:2208.00211}

\bibitem[\protect\citeauthoryear{{Gaia Collaboration} et~al.,}{{Gaia
  Collaboration} et~al.}{2023}]{gaia2023dr3}
{Gaia Collaboration} et~al., 2023, \mn@doi [\aap]
  {10.1051/0004-6361/202243940}, \href
  {https://ui.adsabs.harvard.edu/abs/2023A&A...674A...1G} {674, A1}

\bibitem[\protect\citeauthoryear{{Giacalone} et~al.,}{{Giacalone}
  et~al.}{2021}]{giacalone2021vetting}
{Giacalone} S.,  et~al., 2021, \mn@doi [\aj] {10.3847/1538-3881/abc6af}, \href
  {https://ui.adsabs.harvard.edu/abs/2021AJ....161...24G} {161, 24}

\bibitem[\protect\citeauthoryear{{Ginzburg}, {Schlichting}  \&
  {Sari}}{{Ginzburg} et~al.}{2018}]{ginzburg2018corepower}
{Ginzburg} S.,  {Schlichting} H.~E.,   {Sari} R.,  2018, \mn@doi [\mnras]
  {10.1093/mnras/sty290}, \href
  {https://ui.adsabs.harvard.edu/abs/2018MNRAS.476..759G} {476, 759}

\bibitem[\protect\citeauthoryear{{Guerrero} et~al.,}{{Guerrero}
  et~al.}{2021}]{guerrero2021toi}
{Guerrero} N.~M.,  et~al., 2021, \mn@doi [\apjs] {10.3847/1538-4365/abefe1},
  \href {https://ui.adsabs.harvard.edu/abs/2021ApJS..254...39G} {254, 39}

\bibitem[\protect\citeauthoryear{{Guillot}, {Stevenson}, {Hubbard}  \&
  {Saumon}}{{Guillot} et~al.}{2004}]{Guillotetal04}
{Guillot} T.,  {Stevenson} D.~J.,  {Hubbard} W.~B.,   {Saumon} D.,  2004, in
  {Bagenal} F.,  {Dowling} T.~E.,   {McKinnon} W.~B.,  eds, , Vol.~1, Jupiter.
  The Planet, Satellites and Magnetosphere.
pp 35--57

\bibitem[\protect\citeauthoryear{{Gupta} \& {Schlichting}}{{Gupta} \&
  {Schlichting}}{2019}]{gupta2019corepower}
{Gupta} A.,  {Schlichting} H.~E.,  2019, \mn@doi [\mnras]
  {10.1093/mnras/stz1230}, \href
  {https://ui.adsabs.harvard.edu/abs/2019MNRAS.487...24G} {487, 24}

\bibitem[\protect\citeauthoryear{{Gupta} \& {Schlichting}}{{Gupta} \&
  {Schlichting}}{2020}]{gupta2020corepower}
{Gupta} A.,  {Schlichting} H.~E.,  2020, \mn@doi [\mnras]
  {10.1093/mnras/staa315}, \href
  {https://ui.adsabs.harvard.edu/abs/2020MNRAS.493..792G} {493, 792}

\bibitem[\protect\citeauthoryear{{Hakim}, {Rivoldini}, {Van Hoolst},
  {Cottenier}, {Jaeken}, {Chust}  \& {Steinle-Neumann}}{{Hakim}
  et~al.}{2018}]{Hakim+2018}
{Hakim} K.,  {Rivoldini} A.,  {Van Hoolst} T.,  {Cottenier} S.,  {Jaeken} J.,
  {Chust} T.,   {Steinle-Neumann} G.,  2018, \mn@doi [\icarus]
  {10.1016/j.icarus.2018.05.005}, \href
  {https://ui.adsabs.harvard.edu/abs/2018Icar..313...61H} {313, 61}

\bibitem[\protect\citeauthoryear{{Haldemann}, {Alibert}, {Mordasini}  \&
  {Benz}}{{Haldemann} et~al.}{2020}]{Haldemann+2020}
{Haldemann} J.,  {Alibert} Y.,  {Mordasini} C.,   {Benz} W.,  2020, \mn@doi
  [\aap] {10.1051/0004-6361/202038367}, \href
  {https://ui.adsabs.harvard.edu/abs/2020A&A...643A.105H} {643, A105}

\bibitem[\protect\citeauthoryear{{Hatzes}}{{Hatzes}}{2019}]{Hatzes2019}
{Hatzes} A.~P.,  2019, {The Doppler Method for the Detection of Exoplanets},
  \mn@doi{10.1088/2514-3433/ab46a3.
}

\bibitem[\protect\citeauthoryear{{Henning}, {O'Connell}  \&
  {Sasselov}}{{Henning} et~al.}{2009}]{henning2009tidalmodel}
{Henning} W.~G.,  {O'Connell} R.~J.,   {Sasselov} D.~D.,  2009, \mn@doi [\apj]
  {10.1088/0004-637X/707/2/1000}, \href
  {https://ui.adsabs.harvard.edu/abs/2009ApJ...707.1000H} {707, 1000}

\bibitem[\protect\citeauthoryear{{Ho} \& {Van Eylen}}{{Ho} \& {Van
  Eylen}}{2023}]{ho2023rvalley}
{Ho} C. S.~K.,  {Van Eylen} V.,  2023, \mn@doi [\mnras]
  {10.1093/mnras/stac3802}, \href
  {https://ui.adsabs.harvard.edu/abs/2023MNRAS.519.4056H} {519, 4056}

\bibitem[\protect\citeauthoryear{Hoffman, Gelman  et~al.}{Hoffman
  et~al.}{2014}]{hoffman2014nuts}
Hoffman M.~D.,  Gelman A.,   et~al., 2014, J. Mach. Learn. Res., 15, 1593

\bibitem[\protect\citeauthoryear{{H{\o}g} et~al.,}{{H{\o}g}
  et~al.}{2000}]{hog2000tycho}
{H{\o}g} E.,  et~al., 2000, \aap, \href
  {https://ui.adsabs.harvard.edu/abs/2000A&A...355L..27H} {355, L27}

\bibitem[\protect\citeauthoryear{Howell, Everett, Sherry, Horch  \&
  Ciardi}{Howell et~al.}{2011}]{howell2011speckle}
Howell S.~B.,  Everett M.~E.,  Sherry W.,  Horch E.,   Ciardi D.~R.,  2011, The
  Astronomical Journal, 142, 19

\bibitem[\protect\citeauthoryear{Hoyer, Guterman, Demangeon, Sousa, Deleuil,
  Meunier  \& Benz}{Hoyer et~al.}{2020}]{hoyer2020expected}
Hoyer S.,  Guterman P.,  Demangeon O.,  Sousa S.,  Deleuil M.,  Meunier J.,
  Benz W.,  2020, Astronomy \& Astrophysics, 635, A24

\bibitem[\protect\citeauthoryear{{Huang}, {Petrovich}  \& {Deibert}}{{Huang}
  et~al.}{2017}]{huang2017ecc}
{Huang} C.~X.,  {Petrovich} C.,   {Deibert} E.,  2017, \mn@doi [\aj]
  {10.3847/1538-3881/aa67fb}, \href
  {https://ui.adsabs.harvard.edu/abs/2017AJ....153..210H} {153, 210}

\bibitem[\protect\citeauthoryear{{Hut}}{{Hut}}{1981}]{hut1981tidalevo}
{Hut} P.,  1981, \aap, \href
  {https://ui.adsabs.harvard.edu/abs/1981A&A....99..126H} {99, 126}

\bibitem[\protect\citeauthoryear{{Jenkins}}{{Jenkins}}{2002}]{jenkins2002solarvar}
{Jenkins} J.~M.,  2002, \mn@doi [\apj] {10.1086/341136}, \href
  {https://ui.adsabs.harvard.edu/abs/2002ApJ...575..493J} {575, 493}

\bibitem[\protect\citeauthoryear{{Jenkins} et~al.,}{{Jenkins}
  et~al.}{2010}]{jenkins2010kepler}
{Jenkins} J.~M.,  et~al., 2010, in {Radziwill} N.~M.,  {Bridger} A.,  eds,
  Society of Photo-Optical Instrumentation Engineers (SPIE) Conference Series
  Vol. 7740, Software and Cyberinfrastructure for Astronomy. p. 77400D,
  \mn@doi{10.1117/12.856764}

\bibitem[\protect\citeauthoryear{{Jenkins} et~al.,}{{Jenkins}
  et~al.}{2016}]{jenkins2016tess}
{Jenkins} J.~M.,  et~al., 2016, in {Chiozzi} G.,  {Guzman} J.~C.,  eds,
  Society of Photo-Optical Instrumentation Engineers (SPIE) Conference Series
  Vol. 9913, Software and Cyberinfrastructure for Astronomy IV. p. 99133E,
  \mn@doi{10.1117/12.2233418}

\bibitem[\protect\citeauthoryear{{Jenkins}, {Tenenbaum}, {Seader}, {Burke},
  {McCauliff}, {Smith}, {Twicken}  \& {Chandrasekaran}}{{Jenkins}
  et~al.}{2020}]{jenkins2020keplerhandbook}
{Jenkins} J.~M.,  {Tenenbaum} P.,  {Seader} S.,  {Burke} C.~J.,  {McCauliff}
  S.~D.,  {Smith} J.~C.,  {Twicken} J.~D.,   {Chandrasekaran} H.,  2020,
  {Kepler Data Processing Handbook: Transiting Planet Search}, Kepler Science
  Document KSCI-19081-003, id. 9. Edited by Jon M. Jenkins.

\bibitem[\protect\citeauthoryear{{Jensen}}{{Jensen}}{2013}]{Jensen:2013}
{Jensen} E.,  2013, {Tapir: A web interface for transit/eclipse observability},
  Astrophysics Source Code Library (\mn@eprint {ascl} {1306.007})

\bibitem[\protect\citeauthoryear{{Jin} \& {Mordasini}}{{Jin} \&
  {Mordasini}}{2018}]{jin2018rvalleyice}
{Jin} S.,  {Mordasini} C.,  2018, \mn@doi [\apj] {10.3847/1538-4357/aa9f1e},
  \href {https://ui.adsabs.harvard.edu/abs/2018ApJ...853..163J} {853, 163}

\bibitem[\protect\citeauthoryear{{Kempton} et~al.,}{{Kempton}
  et~al.}{2018}]{kempton2018esmtsm}
{Kempton} E. M.~R.,  et~al., 2018, \mn@doi [\pasp] {10.1088/1538-3873/aadf6f},
  \href {https://ui.adsabs.harvard.edu/abs/2018PASP..130k4401K} {130, 114401}

\bibitem[\protect\citeauthoryear{{Kipping}}{{Kipping}}{2013a}]{exoplanet:kipping13}
{Kipping} D.~M.,  2013a, \mn@doi [\mnras] {10.1093/mnras/stt1435}, \href
  {http://adsabs.harvard.edu/abs/2013MNRAS.435.2152K} {435, 2152}

\bibitem[\protect\citeauthoryear{{Kipping}}{{Kipping}}{2013b}]{kipping2013quadlimb}
{Kipping} D.~M.,  2013b, \mn@doi [\mnras] {10.1093/mnras/stt1435}, \href
  {https://ui.adsabs.harvard.edu/abs/2013MNRAS.435.2152K} {435, 2152}

\bibitem[\protect\citeauthoryear{{Kov{\'a}cs}, {Bakos}  \&
  {Noyes}}{{Kov{\'a}cs} et~al.}{2005}]{Kovacs2005}
{Kov{\'a}cs} G.,  {Bakos} G.,   {Noyes} R.~W.,  2005, \mn@doi [\mnras]
  {10.1111/j.1365-2966.2004.08479.x}, \href
  {http://adsabs.harvard.edu/abs/2005MNRAS.356..557K} {356, 557}

\bibitem[\protect\citeauthoryear{{Kuerster}, {Schmitt}, {Cutispoto}  \&
  {Dennerl}}{{Kuerster} et~al.}{1997}]{Kuester1997}
{Kuerster} M.,  {Schmitt} J.~H.~M.~M.,  {Cutispoto} G.,   {Dennerl} K.,  1997,
  \aap, \href {https://ui.adsabs.harvard.edu/abs/1997A&A...320..831K} {320,
  831}

\bibitem[\protect\citeauthoryear{Kumar, Carroll, Hartikainen  \& Martin}{Kumar
  et~al.}{2019}]{exoplanet:arviz}
Kumar R.,  Carroll C.,  Hartikainen A.,   Martin O.~A.,  2019, \mn@doi [The
  Journal of Open Source Software] {10.21105/joss.01143}

\bibitem[\protect\citeauthoryear{{Kurucz}}{{Kurucz}}{1993a}]{Kurucz-93}
{Kurucz} R.~L.,  1993a, {SYNTHE spectrum synthesis programs and line data}

\bibitem[\protect\citeauthoryear{{Kurucz}}{{Kurucz}}{1993b}]{Kurucz1993}
{Kurucz} R.~L.,  1993b, {SYNTHE spectrum synthesis programs and line data}.
Astrophysics Source Code Library

\bibitem[\protect\citeauthoryear{{Leconte}, {Chabrier}, {Baraffe}  \&
  {Levrard}}{{Leconte} et~al.}{2010}]{Leconteetal10}
{Leconte} J.,  {Chabrier} G.,  {Baraffe} I.,   {Levrard} B.,  2010, \mn@doi
  [\aap] {10.1051/0004-6361/201014337}, \href
  {https://ui.adsabs.harvard.edu/abs/2010A&A...516A..64L} {516, A64}

\bibitem[\protect\citeauthoryear{{Leleu} et~al.,}{{Leleu}
  et~al.}{2021}]{Leleu+2021}
{Leleu} A.,  et~al., 2021, \mn@doi [\aap] {10.1051/0004-6361/202039767}, \href
  {https://ui.adsabs.harvard.edu/abs/2021A&A...649A..26L} {649, A26}

\bibitem[\protect\citeauthoryear{{Li}, {Tenenbaum}, {Twicken}, {Burke},
  {Jenkins}, {Quintana}, {Rowe}  \& {Seader}}{{Li}
  et~al.}{2019}]{li2019keplervalid}
{Li} J.,  {Tenenbaum} P.,  {Twicken} J.~D.,  {Burke} C.~J.,  {Jenkins} J.~M.,
  {Quintana} E.~V.,  {Rowe} J.~F.,   {Seader} S.~E.,  2019, \mn@doi [\pasp]
  {10.1088/1538-3873/aaf44d}, \href
  {https://ui.adsabs.harvard.edu/abs/2019PASP..131b4506L} {131, 024506}

\bibitem[\protect\citeauthoryear{Lindegren et~al.}{Lindegren
  et~al.}{2018}]{lindegren2018ruwe}
Lindegren L.,  et~al., 2018, Gaia Technical Note: GAIA-C3-TN-LU-LL-124-01

\bibitem[\protect\citeauthoryear{{Lindegren} et~al.,}{{Lindegren}
  et~al.}{2021}]{Lindegren2021}
{Lindegren} L.,  et~al., 2021, \mn@doi [\aap] {10.1051/0004-6361/202039653},
  \href {https://ui.adsabs.harvard.edu/abs/2021A&A...649A...4L} {649, A4}

\bibitem[\protect\citeauthoryear{{Lopez} \& {Fortney}}{{Lopez} \&
  {Fortney}}{2013}]{lopez2013photoevap}
{Lopez} E.~D.,  {Fortney} J.~J.,  2013, \mn@doi [\apj]
  {10.1088/0004-637X/776/1/2}, \href
  {https://ui.adsabs.harvard.edu/abs/2013ApJ...776....2L} {776, 2}

\bibitem[\protect\citeauthoryear{{Lopez} \& {Fortney}}{{Lopez} \&
  {Fortney}}{2014}]{lopez2014massradius}
{Lopez} E.~D.,  {Fortney} J.~J.,  2014, \mn@doi [\apj]
  {10.1088/0004-637X/792/1/1}, \href
  {https://ui.adsabs.harvard.edu/abs/2014ApJ...792....1L} {792, 1}

\bibitem[\protect\citeauthoryear{{Lovis} \& {Pepe}}{{Lovis} \&
  {Pepe}}{2007}]{Lovis2007}
{Lovis} C.,  {Pepe} F.,  2007, \mn@doi [\aap] {10.1051/0004-6361:20077249},
  \href {https://ui.adsabs.harvard.edu/abs/2007A&A...468.1115L} {468, 1115}

\bibitem[\protect\citeauthoryear{{Luger}, {Agol}, {Foreman-Mackey}, {Fleming},
  {Lustig-Yaeger}  \& {Deitrick}}{{Luger} et~al.}{2019}]{exoplanet:luger18}
{Luger} R.,  {Agol} E.,  {Foreman-Mackey} D.,  {Fleming} D.~P.,
  {Lustig-Yaeger} J.,   {Deitrick} R.,  2019, \mn@doi [\aj]
  {10.3847/1538-3881/aae8e5}, \href
  {http://adsabs.harvard.edu/abs/2019AJ....157...64L} {157, 64}

\bibitem[\protect\citeauthoryear{{Marboeuf}, {Thiabaud}, {Alibert}, {Cabral}
  \& {Benz}}{{Marboeuf} et~al.}{2014}]{Marboeuf+2014}
{Marboeuf} U.,  {Thiabaud} A.,  {Alibert} Y.,  {Cabral} N.,   {Benz} W.,  2014,
  \mn@doi [\aap] {10.1051/0004-6361/201423431}, \href
  {https://ui.adsabs.harvard.edu/abs/2014A&A...570A..36M} {570, A36}

\bibitem[\protect\citeauthoryear{{Marigo} et~al.,}{{Marigo}
  et~al.}{2017}]{marigo2017}
{Marigo} P.,  et~al., 2017, \mn@doi [\apj] {10.3847/1538-4357/835/1/77}, \href
  {http://adsabs.harvard.edu/abs/2017ApJ...835...77M} {835, 77}

\bibitem[\protect\citeauthoryear{{Matsumoto}, {Nagasawa}  \& {Ida}}{{Matsumoto}
  et~al.}{2015}]{matsumoto2015eccentricity}
{Matsumoto} Y.,  {Nagasawa} M.,   {Ida} S.,  2015, \mn@doi [\apj]
  {10.1088/0004-637X/810/2/106}, \href
  {https://ui.adsabs.harvard.edu/abs/2015ApJ...810..106M} {810, 106}

\bibitem[\protect\citeauthoryear{{Maxted} et~al.,}{{Maxted}
  et~al.}{2011}]{maxted2011wasp41b}
{Maxted} P.~F.~L.,  et~al., 2011, \mn@doi [\pasp] {10.1086/660007}, \href
  {https://eur01.safelinks.protection.outlook.com/?url=https%3A%2F%2Fui.adsabs.harvard.edu%2Fabs%2F2011PASP..123..547M&data=05%7C01%7Cahlam.alqasim.17%40ucl.ac.uk%7C8437c8c27a894b825a4f08db9829fc3a%7C1faf88fea9984c5b93c9210a11d9a5c2%7C0%7C0%7C638271078319991143%7CUnknown%7CTWFpbGZsb3d8eyJWIjoiMC4wLjAwMDAiLCJQIjoiV2luMzIiLCJBTiI6Ik1haWwiLCJXVCI6Mn0%3D%7C3000%7C%7C%7C&sdata=S136Jq8T7JyuAz9S8B7NP%2F%2FQ0QtOwttnv%2BCPt1smIXM%3D&reserved=0}
  {123, 547}

\bibitem[\protect\citeauthoryear{{Maxted} et~al.,}{{Maxted}
  et~al.}{2022}]{maxted2022pycheops}
{Maxted} P.~F.~L.,  et~al., 2022, \mn@doi [\mnras] {10.1093/mnras/stab3371},
  \href {https://ui.adsabs.harvard.edu/abs/2022MNRAS.514...77M} {514, 77}

\bibitem[\protect\citeauthoryear{{Mayor} et~al.,}{{Mayor}
  et~al.}{2003}]{Mayor2003}
{Mayor} M.,  et~al., 2003, The Messenger, \href
  {https://ui.adsabs.harvard.edu/abs/2003Msngr.114...20M} {114, 20}

\bibitem[\protect\citeauthoryear{{McCully}, {Volgenau}, {Harbeck}, {Lister},
  {Saunders}, {Turner}, {Siiverd}  \& {Bowman}}{{McCully}
  et~al.}{2018}]{McCully:2018}
{McCully} C.,  {Volgenau} N.~H.,  {Harbeck} D.-R.,  {Lister} T.~A.,  {Saunders}
  E.~S.,  {Turner} M.~L.,  {Siiverd} R.~J.,   {Bowman} M.,  2018, in \procspie.
  p. 107070K (\mn@eprint {arXiv} {1811.04163}), \mn@doi{10.1117/12.2314340}

\bibitem[\protect\citeauthoryear{{Mordasini}}{{Mordasini}}{2020}]{mordasini2020rvalleyice}
{Mordasini} C.,  2020, \mn@doi [\aap] {10.1051/0004-6361/201935541}, \href
  {https://ui.adsabs.harvard.edu/abs/2020A&A...638A..52M} {638, A52}

\bibitem[\protect\citeauthoryear{{Nowak} et~al.,}{{Nowak}
  et~al.}{2020}]{nowak2020k2280b}
{Nowak} G.,  et~al., 2020, \mn@doi [\mnras] {10.1093/mnras/staa2077}, \href
  {https://ui.adsabs.harvard.edu/abs/2020MNRAS.497.4423N} {497, 4423}

\bibitem[\protect\citeauthoryear{{Oelkers} et~al.,}{{Oelkers}
  et~al.}{2018}]{oelkers2018kelt}
{Oelkers} R.~J.,  et~al., 2018, \mn@doi [\aj] {10.3847/1538-3881/aa9bf4}, \href
  {https://ui.adsabs.harvard.edu/abs/2018AJ....155...39O} {155, 39}

\bibitem[\protect\citeauthoryear{{Ogilvie}}{{Ogilvie}}{2014}]{Ogilvie14}
{Ogilvie} G.~I.,  2014, \mn@doi [\araa] {10.1146/annurev-astro-081913-035941},
  \href {https://ui.adsabs.harvard.edu/abs/2014ARA&A..52..171O} {52, 171}

\bibitem[\protect\citeauthoryear{{Owen} \& {Wu}}{{Owen} \&
  {Wu}}{2013}]{owenwu2013photoevap}
{Owen} J.~E.,  {Wu} Y.,  2013, \mn@doi [\apj] {10.1088/0004-637X/775/2/105},
  \href {https://ui.adsabs.harvard.edu/abs/2013ApJ...775..105O} {775, 105}

\bibitem[\protect\citeauthoryear{{Owen} \& {Wu}}{{Owen} \&
  {Wu}}{2017}]{owenwu2017rvalley}
{Owen} J.~E.,  {Wu} Y.,  2017, \mn@doi [\apj] {10.3847/1538-4357/aa890a}, \href
  {https://ui.adsabs.harvard.edu/abs/2017ApJ...847...29O} {847, 29}

\bibitem[\protect\citeauthoryear{{Paegert}, {Stassun}, {Collins}, {Pepper},
  {Torres}, {Jenkins}, {Twicken}  \& {Latham}}{{Paegert}
  et~al.}{2022}]{paegert2022tess}
{Paegert} M.,  {Stassun} K.~G.,  {Collins} K.~A.,  {Pepper} J.,  {Torres} G.,
  {Jenkins} J.,  {Twicken} J.~D.,   {Latham} D.~W.,  2022, VizieR Online Data
  Catalog, \href {https://ui.adsabs.harvard.edu/abs/2022yCat.4039....0P} {p.
  IV/39}

\bibitem[\protect\citeauthoryear{{Pecaut} \& {Mamajek}}{{Pecaut} \&
  {Mamajek}}{2013}]{mamajek2013intrinsic}
{Pecaut} M.~J.,  {Mamajek} E.~E.,  2013, \mn@doi [\apjs]
  {10.1088/0067-0049/208/1/9}, \href
  {https://ui.adsabs.harvard.edu/abs/2013ApJS..208....9P} {208, 9}

\bibitem[\protect\citeauthoryear{{Pepe}, {Mayor}, {Galland}, {Naef}, {Queloz},
  {Santos}, {Udry}  \& {Burnet}}{{Pepe} et~al.}{2002}]{Pepe2002}
{Pepe} F.,  {Mayor} M.,  {Galland} F.,  {Naef} D.,  {Queloz} D.,  {Santos}
  N.~C.,  {Udry} S.,   {Burnet} M.,  2002, \mn@doi [\aap]
  {10.1051/0004-6361:20020433}, \href
  {https://ui.adsabs.harvard.edu/abs/2002A&A...388..632P} {388, 632}

\bibitem[\protect\citeauthoryear{{Pepe} et~al.,}{{Pepe}
  et~al.}{2021}]{pepe2021}
{Pepe} F.,  et~al., 2021, \mn@doi [\aap] {10.1051/0004-6361/202038306}, \href
  {https://ui.adsabs.harvard.edu/abs/2021A&A...645A..96P} {645, A96}

\bibitem[\protect\citeauthoryear{{Pepper} et~al.,}{{Pepper}
  et~al.}{2007a}]{Pepper2007}
{Pepper} J.,  et~al., 2007a, \mn@doi [\pasp] {10.1086/521836}, \href
  {http://adsabs.harvard.edu/abs/2007PASP..119..923P} {119, 923}

\bibitem[\protect\citeauthoryear{{Pepper} et~al.,}{{Pepper}
  et~al.}{2007b}]{pepper2007kelt}
{Pepper} J.,  et~al., 2007b, \mn@doi [\pasp] {10.1086/521836}, \href
  {https://ui.adsabs.harvard.edu/abs/2007PASP..119..923P} {119, 923}

\bibitem[\protect\citeauthoryear{{Pepper}, {Kuhn}, {Siverd}, {James}  \&
  {Stassun}}{{Pepper} et~al.}{2012}]{Pepper2012}
{Pepper} J.,  {Kuhn} R.~B.,  {Siverd} R.,  {James} D.,   {Stassun} K.,  2012,
  \mn@doi [\pasp] {10.1086/665044}, \href
  {http://adsabs.harvard.edu/abs/2012PASP..124..230P} {124, 230}

\bibitem[\protect\citeauthoryear{{Peterson} et~al.,}{{Peterson}
  et~al.}{2018}]{peterson2018wolf503b}
{Peterson} M.~S.,  et~al., 2018, \mn@doi [\aj] {10.3847/1538-3881/aaddfe},
  \href {https://ui.adsabs.harvard.edu/abs/2018AJ....156..188P} {156, 188}

\bibitem[\protect\citeauthoryear{{Petigura} et~al.,}{{Petigura}
  et~al.}{2017}]{petigura2017subsaturns}
{Petigura} E.~A.,  et~al., 2017, \mn@doi [\aj] {10.3847/1538-3881/aa5ea5},
  \href {https://ui.adsabs.harvard.edu/abs/2017AJ....153..142P} {153, 142}

\bibitem[\protect\citeauthoryear{{Petrovich}, {Tremaine}  \&
  {Rafikov}}{{Petrovich} et~al.}{2014}]{petrovich2014scattering}
{Petrovich} C.,  {Tremaine} S.,   {Rafikov} R.,  2014, \mn@doi [\apj]
  {10.1088/0004-637X/786/2/101}, \href
  {https://ui.adsabs.harvard.edu/abs/2014ApJ...786..101P} {786, 101}

\bibitem[\protect\citeauthoryear{{Polanski} et~al.,}{{Polanski}
  et~al.}{2021}]{polanski2021wolf503b}
{Polanski} A.~S.,  et~al., 2021, \mn@doi [\aj] {10.3847/1538-3881/ac1590},
  \href {https://ui.adsabs.harvard.edu/abs/2021AJ....162..238P} {162, 238}

\bibitem[\protect\citeauthoryear{{Pollacco} et~al.,}{{Pollacco}
  et~al.}{2006}]{pollaco2006superwasp}
{Pollacco} D.~L.,  et~al., 2006, \mn@doi [\pasp] {10.1086/508556}, \href
  {https://eur01.safelinks.protection.outlook.com/?url=http%3A%2F%2Fadsabs.harvard.edu%2Fabs%2F2006PASP..118.1407P&data=05%7C01%7Cahlam.alqasim.17%40ucl.ac.uk%7C8437c8c27a894b825a4f08db9829fc3a%7C1faf88fea9984c5b93c9210a11d9a5c2%7C0%7C0%7C638271078319991143%7CUnknown%7CTWFpbGZsb3d8eyJWIjoiMC4wLjAwMDAiLCJQIjoiV2luMzIiLCJBTiI6Ik1haWwiLCJXVCI6Mn0%3D%7C3000%7C%7C%7C&sdata=xrq6moZH41XkHfnp5pICAKA%2B0nQL1PylunUXHPJeRL8%3D&reserved=0}
  {118, 1407}

\bibitem[\protect\citeauthoryear{{Portegies Zwart} \&
  {J{\'\i}lkov{\'a}}}{{Portegies Zwart} \&
  {J{\'\i}lkov{\'a}}}{2015}]{zwart2015fragility}
{Portegies Zwart} S.~F.,  {J{\'\i}lkov{\'a}} L.,  2015, \mn@doi [\mnras]
  {10.1093/mnras/stv877}, \href
  {https://ui.adsabs.harvard.edu/abs/2015MNRAS.451..144P} {451, 144}

\bibitem[\protect\citeauthoryear{{Ricker} et~al.,}{{Ricker}
  et~al.}{2015}]{ricker2015tess}
{Ricker} G.~R.,  et~al., 2015, \mn@doi [Journal of Astronomical Telescopes,
  Instruments, and Systems] {10.1117/1.JATIS.1.1.014003}, \href
  {https://ui.adsabs.harvard.edu/abs/2015JATIS...1a4003R} {1, 014003}

\bibitem[\protect\citeauthoryear{{Rogers}, {Schlichting}  \& {Owen}}{{Rogers}
  et~al.}{2023}]{rogers2023waterworlds}
{Rogers} J.~G.,  {Schlichting} H.~E.,   {Owen} J.~E.,  2023, \mn@doi [\apjl]
  {10.3847/2041-8213/acc86f}, \href
  {https://ui.adsabs.harvard.edu/abs/2023ApJ...947L..19R} {947, L19}

\bibitem[\protect\citeauthoryear{Salvatier, Wiecki  \& Fonnesbeck}{Salvatier
  et~al.}{2016}]{exoplanet:pymc3}
Salvatier J.,  Wiecki T.~V.,   Fonnesbeck C.,  2016, PeerJ Computer Science, 2,
  e55

\bibitem[\protect\citeauthoryear{{Santos} et~al.,}{{Santos}
  et~al.}{2013}]{Santos-13}
{Santos} N.~C.,  et~al., 2013, \mn@doi [\aap] {10.1051/0004-6361/201321286},
  \href {http://adsabs.harvard.edu/abs/2013A%26A...556A.150S} {556, A150}

\bibitem[\protect\citeauthoryear{{Schanche} et~al.,}{{Schanche}
  et~al.}{2020}]{Schanche2020}
{Schanche} N.,  et~al., 2020, \mn@doi [\mnras] {10.1093/mnras/staa2848}, \href
  {https://ui.adsabs.harvard.edu/abs/2020MNRAS.499..428S} {499, 428}

\bibitem[\protect\citeauthoryear{Scott et~al.,}{Scott
  et~al.}{2021}]{scott2021gemini}
Scott N.~J.,  et~al., 2021, Frontiers in Astronomy and Space Sciences, 8,
  716560

\bibitem[\protect\citeauthoryear{{Sebring}, {Krabbendam}  \&
  {Heathcote}}{{Sebring} et~al.}{2003}]{sebring2003soar}
{Sebring} T.~A.,  {Krabbendam} V.~L.,   {Heathcote} S.~R.,  2003, in {Oschmann}
  J.~M.,  {Stepp} L.~M.,  eds,  Society of Photo-Optical Instrumentation
  Engineers (SPIE) Conference Series Vol. 4837, Large Ground-based Telescopes.
  pp 71--81, \mn@doi{10.1117/12.456692}

\bibitem[\protect\citeauthoryear{{Sim{\'o}n-D{\'\i}az} \&
  {Herrero}}{{Sim{\'o}n-D{\'\i}az} \& {Herrero}}{2014}]{Simon2014}
{Sim{\'o}n-D{\'\i}az} S.,  {Herrero} A.,  2014, \mn@doi [\aap]
  {10.1051/0004-6361/201322758}, \href
  {https://ui.adsabs.harvard.edu/abs/2014A&A...562A.135S} {562, A135}

\bibitem[\protect\citeauthoryear{{Skrutskie} et~al.,}{{Skrutskie}
  et~al.}{2006}]{Skrutskie2006}
{Skrutskie} M.~F.,  et~al., 2006, \mn@doi [\aj] {10.1086/498708}, \href
  {https://ui.adsabs.harvard.edu/abs/2006AJ....131.1163S} {131, 1163}

\bibitem[\protect\citeauthoryear{Smith et~al.,}{Smith
  et~al.}{2012}]{smith2012kepler}
Smith J.~C.,  et~al., 2012, Publications of the Astronomical Society of the
  Pacific, 124, 1000

\bibitem[\protect\citeauthoryear{{Sneden}}{{Sneden}}{1973}]{Sneden-73}
{Sneden} C.~A.,  1973, PhD thesis, THE UNIVERSITY OF TEXAS AT AUSTIN.

\bibitem[\protect\citeauthoryear{{Sotin}, {Grasset}  \& {Mocquet}}{{Sotin}
  et~al.}{2007}]{Sotin+2007}
{Sotin} C.,  {Grasset} O.,   {Mocquet} A.,  2007, \mn@doi [\icarus]
  {10.1016/j.icarus.2007.04.006}, \href
  {https://ui.adsabs.harvard.edu/abs/2007Icar..191..337S} {191, 337}

\bibitem[\protect\citeauthoryear{{Sousa}}{{Sousa}}{2014}]{Sousa-14}
{Sousa} S.~G.,  2014, [arXiv:1407.5817], \href
  {http://adsabs.harvard.edu/abs/2014arXiv1407.5817S} {}

\bibitem[\protect\citeauthoryear{{Sousa}, {Santos}, {Israelian}, {Mayor}  \&
  {Monteiro}}{{Sousa} et~al.}{2007}]{Sousa-07}
{Sousa} S.~G.,  {Santos} N.~C.,  {Israelian} G.,  {Mayor} M.,   {Monteiro}
  M.~J.~P.~F.~G.,  2007, \mn@doi [A\&A] {10.1051/0004-6361:20077288}, \href
  {http://adsabs.harvard.edu/abs/2007A%26A...469..783S} {469, 783}

\bibitem[\protect\citeauthoryear{{Sousa} et~al.,}{{Sousa}
  et~al.}{2008}]{Sousa-08}
{Sousa} S.~G.,  et~al., 2008, \mn@doi [\aap] {10.1051/0004-6361:200809698},
  \href {https://ui.adsabs.harvard.edu/abs/2008A&A...487..373S} {487, 373}

\bibitem[\protect\citeauthoryear{{Sousa}, {Santos}, {Adibekyan}, {Delgado-Mena}
   \& {Israelian}}{{Sousa} et~al.}{2015}]{Sousa-15}
{Sousa} S.~G.,  {Santos} N.~C.,  {Adibekyan} V.,  {Delgado-Mena} E.,
  {Israelian} G.,  2015, \mn@doi [\aap] {10.1051/0004-6361/201425463}, \href
  {http://adsabs.harvard.edu/abs/2015A%26A...577A..67S} {577, A67}

\bibitem[\protect\citeauthoryear{{Sousa} et~al.,}{{Sousa}
  et~al.}{2021}]{Sousa-21}
{Sousa} S.~G.,  et~al., 2021, arXiv e-prints, \href
  {https://ui.adsabs.harvard.edu/abs/2021arXiv210904781S} {p. arXiv:2109.04781}

\bibitem[\protect\citeauthoryear{{Stassun} \& {Torres}}{{Stassun} \&
  {Torres}}{2021}]{stassun2021ruwe}
{Stassun} K.~G.,  {Torres} G.,  2021, \mn@doi [\apjl]
  {10.3847/2041-8213/abdaad}, \href
  {https://ui.adsabs.harvard.edu/abs/2021ApJ...907L..33S} {907, L33}

\bibitem[\protect\citeauthoryear{Stumpe et~al.,}{Stumpe
  et~al.}{2012}]{stumpe2012kepler}
Stumpe M.~C.,  et~al., 2012, Publications of the Astronomical Society of the
  Pacific, 124, 985

\bibitem[\protect\citeauthoryear{Stumpe, Smith, Catanzarite, Van~Cleve,
  Jenkins, Twicken  \& Girouard}{Stumpe et~al.}{2014}]{stumpe2014multiscale}
Stumpe M.~C.,  Smith J.~C.,  Catanzarite J.~H.,  Van~Cleve J.~E.,  Jenkins
  J.~M.,  Twicken J.~D.,   Girouard F.~R.,  2014, Publications of the
  Astronomical Society of the Pacific, 126, 100

\bibitem[\protect\citeauthoryear{{Su{\'a}rez Mascare{\~n}o}, {Rebolo},
  {Gonz{\'a}lez Hern{\'a}ndez}  \& {Esposito}}{{Su{\'a}rez Mascare{\~n}o}
  et~al.}{2015}]{suarez2015}
{Su{\'a}rez Mascare{\~n}o} A.,  {Rebolo} R.,  {Gonz{\'a}lez Hern{\'a}ndez}
  J.~I.,   {Esposito} M.,  2015, \mn@doi [\mnras] {10.1093/mnras/stv1441},
  \href {https://ui.adsabs.harvard.edu/abs/2015MNRAS.452.2745S} {452, 2745}

\bibitem[\protect\citeauthoryear{{Sundqvist}, {Sim{\'o}n-D{\'\i}az}, {Puls}  \&
  {Markova}}{{Sundqvist} et~al.}{2013}]{Sundqvist2013}
{Sundqvist} J.~O.,  {Sim{\'o}n-D{\'\i}az} S.,  {Puls} J.,   {Markova} N.,
  2013, \mn@doi [\aap] {10.1051/0004-6361/201322761}, \href
  {https://ui.adsabs.harvard.edu/abs/2013A&A...559L..10S} {559, L10}

\bibitem[\protect\citeauthoryear{{Tayar}, {Claytor}, {Huber}  \& {van
  Saders}}{{Tayar} et~al.}{2022}]{tayar2022realistic}
{Tayar} J.,  {Claytor} Z.~R.,  {Huber} D.,   {van Saders} J.,  2022, \mn@doi
  [\apj] {10.3847/1538-4357/ac4bbc}, \href
  {https://ui.adsabs.harvard.edu/abs/2022ApJ...927...31T} {927, 31}

\bibitem[\protect\citeauthoryear{{Teske} et~al.,}{{Teske}
  et~al.}{2021}]{Teske2021}
{Teske} J.,  et~al., 2021, \mn@doi [\apjs] {10.3847/1538-4365/ac0f0a}, \href
  {https://ui.adsabs.harvard.edu/abs/2021ApJS..256...33T} {256, 33}

\bibitem[\protect\citeauthoryear{{Theano Development Team}}{{Theano Development
  Team}}{2016}]{exoplanet:theano}
{Theano Development Team} 2016, arXiv e-prints, abs/1605.02688

\bibitem[\protect\citeauthoryear{{Thiabaud}, {Marboeuf}, {Alibert}, {Cabral},
  {Leya}  \& {Mezger}}{{Thiabaud} et~al.}{2014}]{Thiabaud+2014}
{Thiabaud} A.,  {Marboeuf} U.,  {Alibert} Y.,  {Cabral} N.,  {Leya} I.,
  {Mezger} K.,  2014, \mn@doi [\aap] {10.1051/0004-6361/201322208}, \href
  {https://ui.adsabs.harvard.edu/abs/2014A&A...562A..27T} {562, A27}

\bibitem[\protect\citeauthoryear{{Thiabaud}, {Marboeuf}, {Alibert}, {Leya}  \&
  {Mezger}}{{Thiabaud} et~al.}{2015}]{Thiabaud+2015}
{Thiabaud} A.,  {Marboeuf} U.,  {Alibert} Y.,  {Leya} I.,   {Mezger} K.,  2015,
  \mn@doi [\aap] {10.1051/0004-6361/201424868}, \href
  {https://ui.adsabs.harvard.edu/abs/2015A&A...574A.138T} {574, A138}

\bibitem[\protect\citeauthoryear{{Tittemore} \& {Wisdom}}{{Tittemore} \&
  {Wisdom}}{1990}]{TittemoreWisdom90}
{Tittemore} W.~C.,  {Wisdom} J.,  1990, \mn@doi [\icarus]
  {10.1016/0019-1035(90)90125-S}, \href
  {https://ui.adsabs.harvard.edu/abs/1990Icar...85..394T} {85, 394}

\bibitem[\protect\citeauthoryear{{Tokovinin}}{{Tokovinin}}{2018}]{tokovinin2018soar}
{Tokovinin} A.,  2018, \mn@doi [\pasp] {10.1088/1538-3873/aaa7d9}, \href
  {https://ui.adsabs.harvard.edu/abs/2018PASP..130c5002T} {130, 035002}

\bibitem[\protect\citeauthoryear{{Tokovinin} \& {Cantarutti}}{{Tokovinin} \&
  {Cantarutti}}{2008}]{tokovinin2008soar}
{Tokovinin} A.,  {Cantarutti} R.,  2008, \mn@doi [\pasp] {10.1086/528809},
  \href {https://ui.adsabs.harvard.edu/abs/2008PASP..120..170T} {120, 170}

\bibitem[\protect\citeauthoryear{{Tokovinin}, {Mason}  \&
  {Hartkopf}}{{Tokovinin} et~al.}{2010}]{tokovinin2010soar}
{Tokovinin} A.,  {Mason} B.~D.,   {Hartkopf} W.~I.,  2010, \mn@doi [\aj]
  {10.1088/0004-6256/139/2/743}, \href
  {https://ui.adsabs.harvard.edu/abs/2010AJ....139..743T} {139, 743}

\bibitem[\protect\citeauthoryear{{Twicken} et~al.,}{{Twicken}
  et~al.}{2018}]{twicken2018keplervalid}
{Twicken} J.~D.,  et~al., 2018, \mn@doi [\pasp] {10.1088/1538-3873/aab694},
  \href {https://ui.adsabs.harvard.edu/abs/2018PASP..130f4502T} {130, 064502}

\bibitem[\protect\citeauthoryear{{Van Eylen} et~al.,}{{Van Eylen}
  et~al.}{2019}]{vaneylen2019eccentricity}
{Van Eylen} V.,  et~al., 2019, \mn@doi [\aj] {10.3847/1538-3881/aaf22f}, \href
  {https://ui.adsabs.harvard.edu/abs/2019AJ....157...61V} {157, 61}

\bibitem[\protect\citeauthoryear{{Winn}}{{Winn}}{2010}]{winn2010transitdur}
{Winn} J.~N.,  2010, \mn@doi [arXiv e-prints] {10.48550/arXiv.1001.2010}, \href
  {https://ui.adsabs.harvard.edu/abs/2010arXiv1001.2010W} {p. arXiv:1001.2010}

\bibitem[\protect\citeauthoryear{{Wright} et~al.,}{{Wright}
  et~al.}{2010}]{Wright2010}
{Wright} E.~L.,  et~al., 2010, \mn@doi [\aj] {10.1088/0004-6256/140/6/1868},
  \href {https://ui.adsabs.harvard.edu/abs/2010AJ....140.1868W} {140, 1868}

\bibitem[\protect\citeauthoryear{{Xie} et~al.,}{{Xie}
  et~al.}{2016}]{xie2016orbitalecc}
{Xie} J.-W.,  et~al., 2016, \mn@doi [Proceedings of the National Academy of
  Science] {10.1073/pnas.1604692113}, \href
  {https://ui.adsabs.harvard.edu/abs/2016PNAS..11311431X} {113, 11431}

\bibitem[\protect\citeauthoryear{{Yi}, {Demarque}, {Kim}, {Lee}, {Ree},
  {Lejeune}  \& {Barnes}}{{Yi} et~al.}{2001}]{Yi-01}
{Yi} S.,  {Demarque} P.,  {Kim} Y.-C.,  {Lee} Y.-W.,  {Ree} C.~H.,  {Lejeune}
  T.,   {Barnes} S.,  2001, \mn@doi [\apjs] {10.1086/321795}, \href
  {https://ui.adsabs.harvard.edu/abs/2001ApJS..136..417Y} {136, 417}

\bibitem[\protect\citeauthoryear{{Zeng} et~al.,}{{Zeng}
  et~al.}{2019}]{zeng2019massradius}
{Zeng} L.,  et~al., 2019, \mn@doi [Proceedings of the National Academy of
  Science] {10.1073/pnas.1812905116}, \href
  {https://ui.adsabs.harvard.edu/abs/2019PNAS..116.9723Z} {116, 9723}

\makeatother
\end{thebibliography}



\appendix

\section{Affiliations}\label{sec:affiliations}

\textsuperscript{\hypertarget{inst:1}{1}} UCL Mullard Space Science Laboratory, Holmbury Hill Road, Dorking, Surrey, RH5 6NT, UK \\
\textsuperscript{\hypertarget{inst:2}{2}} Observatoire de Gen{\`e}ve, Universit{\'e} de Gen{\`e}ve, 51 Chemin Pegasi, 1290 Versoix, Switzerland \\
\textsuperscript{\hypertarget{inst:3}{3}} Instituto de Astrofisica e Ciencias do Espaco, Universidade do Porto, CAUP, Rua das Estrelas, 4150-762 Porto, Portugal \\
\textsuperscript{\hypertarget{inst:4}{4}} Departamento de Fisica e Astronomia, Faculdade de Ciencias, Universidade do Porto, Rua do Campo Alegre, 4169-007 Porto, Portugal \\
\textsuperscript{\hypertarget{inst:5}{5}} Dipartimento di Fisica, Universit\'a degli Studi di Torino, Via Pietro Giuria 1, I-10125, Torino, Italy \\
\textsuperscript{\hypertarget{inst:6}{6}} Department of Astronomy, The University of Tokyo, 7-3-1 Hongo, Bunkyo-ku, Tokyo 113-0033, Japan \\
\textsuperscript{\hypertarget{inst:7}{7}} Center for Astrophysics \textbar \ Harvard \& Smithsonian, 60 Garden Street, Cambridge, MA 02138, USA \\
\textsuperscript{\hypertarget{inst:8}{8}} Earth and Planets Laboratory, Carnegie Institution for Science, 5241 Broad Branch Rd NW, Washington, DC 20015, USA \\
\textsuperscript{\hypertarget{inst:9}{9}} Chalmers University of Technology, Department of Space, Earth and Environment, Onsala Space Observatory,  SE-439 92 Onsala, Sweden \\
\textsuperscript{\hypertarget{inst:10}{10}} Leiden Observatory, University of Leiden, PO Box 9513, 2300 RA, Leiden, The Netherlands \\
\textsuperscript{\hypertarget{inst:11}{11}} Weltraumforschung und Planetologie, Physikalisches Institut, University of Bern, Gesellschaftsstrasse 6, 3012 Bern, Switzerland \\
\textsuperscript{\hypertarget{inst:12}{12}} Institute of Planetary Research, German Aerospace Center (DLR), Rutherfordstrasse 2, D-12489 Berlin, Germany \\
\textsuperscript{\hypertarget{inst:13}{13}} Astrophysics Group, Lennard Jones Building, Keele University, Staffordshire, ST5 5BG, UK \\
\textsuperscript{\hypertarget{inst:14}{14}} INAF-Osservatorio Astrofisico di Catania, Via S. Sofia, 78 - 95123 Catania, Italy \\
\textsuperscript{\hypertarget{inst:15}{15}} Department of Physics and Astronomy, Texas A\&M University, College Station, TX 77843-4242, USA \\
\textsuperscript{\hypertarget{inst:16}{16}} Center for Gravitational Wave Astronomy, The University of Texas, Rio Grande Valley, Brownsville, TX 78520, USA \\
\textsuperscript{\hypertarget{inst:17}{17}} Kotizarovci Observatory, Sarsoni 90, 51216 Viskovo, Croatia \\
\textsuperscript{\hypertarget{inst:18}{18}} The Observatories of the Carnegie Institution for Science, 813 Santa Barbara Street, Pasadena, CA 91101, USA \\
\textsuperscript{\hypertarget{inst:19}{19}} European Space Agency (ESA), European Space Research and Technology Centre (ESTEC), Keplerlaan 1, 2201 AZ Noordwijk, The Netherlands \\
\textsuperscript{\hypertarget{inst:20}{20}} Department of Physics, University of Warwick, Gibbet Hill Road, Coventry CV4 7AL, UK \\
\textsuperscript{\hypertarget{inst:21}{21}} Department of Astronomy, Stockholm University, AlbaNova University Center, 10691 Stockholm, Sweden \\
\textsuperscript{\hypertarget{inst:22}{22}} Department of Astronomy, Tsinghua University, Beijing 100084, People's Republic of China \\
\textsuperscript{\hypertarget{inst:23}{23}} Observatoire astronomique de l'Université de Genève, Chemin Pegasi 51, 1290 Versoix, Switzerland \\
\textsuperscript{\hypertarget{inst:24}{24}} Space Research Institute, Austrian Academy of Sciences, Schmiedlstrasse 6, A-8042 Graz, Austria \\
\textsuperscript{\hypertarget{inst:25}{25}} Center for Space and Habitability, University of Bern, Gesellschaftsstrasse 6, 3012 Bern, Switzerland \\
\textsuperscript{\hypertarget{inst:26}{26}} Astrobiology Research Unit, Université de Liège, Allée du 6 Août 19C, B-4000 Liège, Belgium \\
\textsuperscript{\hypertarget{inst:27}{27}} Space sciences, Technologies and Astrophysics Research (STAR) Institute, Université de Liège, Allée du 6 Août 19C, 4000 Liège, Belgium \\
\textsuperscript{\hypertarget{inst:28}{28}} Institute of Astronomy, KU Leuven, Celestijnenlaan 200D, 3001 Leuven, Belgium \\
\textsuperscript{\hypertarget{inst:29}{29}} South African Astronomical Observatory, P.O. Box 9, Observatory, Cape Town 7935, South Africa \\
\textsuperscript{\hypertarget{inst:30}{30}} George Mason University, 4400 University Drive, Fairfax, VA 22030, USA \\
\textsuperscript{\hypertarget{inst:31}{31}} Stellar Astrophysics Centre, Department of Physics and Astronomy, Aarhus University, Ny Munkegade 120, DK-8000 Aarhus C, Denmark \\
\textsuperscript{\hypertarget{inst:32}{32}} Instituto de Astrofísica de Canarias, Vía Láctea s/n, 38200 La Laguna, Tenerife, Spain \\
\textsuperscript{\hypertarget{inst:33}{33}} Departamento de Astrof\'isica, Universidad de La Laguna (ULL), 38206 La Laguna, Tenerife, Spain \\
\textsuperscript{\hypertarget{inst:34}{34}} Admatis, 5. Kandó Kálmán Street, 3534 Miskolc, Hungary \\
\textsuperscript{\hypertarget{inst:35}{35}} Depto. de Astrofísica, Centro de Astrobiología (CSIC-INTA), ESAC campus, 28692 Villanueva de la Cañada (Madrid), Spain \\
\textsuperscript{\hypertarget{inst:36}{36}} INAF, Osservatorio Astronomico di Padova, Vicolo dell'Osservatorio 5, 35122 Padova, Italy \\
\textsuperscript{\hypertarget{inst:37}{37}} McDonald Observatory, The University of Texas, Austin Texas, USA \\
\textsuperscript{\hypertarget{inst:38}{38}} Center for Planetary Systems Habitability, The University of Texas, Austin Texas, USA \\
\textsuperscript{\hypertarget{inst:39}{39}} Centre for Exoplanet Science, SUPA School of Physics and Astronomy, University of St Andrews, North Haugh, St Andrews KY16 9SS, UK \\
\textsuperscript{\hypertarget{inst:40}{40}} CFisUC, Department of Physics, University of Coimbra, 3004-516 Coimbra, Portugal \\
\textsuperscript{\hypertarget{inst:41}{41}} INAF, Osservatorio Astrofisico di Torino, Via Osservatorio, 20, I-10025 Pino Torinese To, Italy \\
\textsuperscript{\hypertarget{inst:42}{42}} Centre for Mathematical Sciences, Lund University, Box 118, 221 00 Lund, Sweden \\
\textsuperscript{\hypertarget{inst:43}{43}} Department of Physics and McDonnell Center for the Space Sciences, Washington University, St. Louis, MO 63130, USA \\ 
\textsuperscript{\hypertarget{inst:44}{44}} Aix Marseille Univ, CNRS, CNES, LAM, 38 rue Frédéric Joliot-Curie, 13388 Marseille, France \\
\textsuperscript{\hypertarget{inst:45}{45}} ELTE E\"otv\"os Lor\'and University, Gothard Astrophysical Observatory, 9700 Szombathely, Szent Imre h. u. 112, Hungary \\
\textsuperscript{\hypertarget{inst:46}{46}} SRON Netherlands Institute for Space Research, Niels Bohrweg 4, 2333 CA Leiden, Netherlands \\
\textsuperscript{\hypertarget{inst:47}{47}} Centre Vie dans l’Univers, Faculté des sciences, Université de Genève, Quai Ernest-Ansermet 30, 1211 Genève 4, Switzerland \\
\textsuperscript{\hypertarget{inst:48}{48}} Department of Physics and Astronomy, The University of New Mexico, 210 Yale Blvd NE, Albuquerque, NM 87106, USA \\
\textsuperscript{\hypertarget{inst:49}{49}} National and Kapodistrian University of Athens, Department of Physics, University Campus, Zografos GR-157 84, Athens, Greece \\
\textsuperscript{\hypertarget{inst:50}{50}} Department of Astrophysics, University of Vienna, Türkenschanzstrasse 17, 1180 Vienna, Austria \\
\textsuperscript{\hypertarget{inst:51}{51}} Thüringer Landessternwarte Tautenburg, Sternwarte 5, D-07778 Tautenburg, Germany \\
\textsuperscript{\hypertarget{inst:51}{51}} Institute for Theoretical Physics and Computational Physics, Graz University of Technology, Petersgasse 16, 8010 Graz, Austria \\
\textsuperscript{\hypertarget{inst:52}{52}} NASA Ames Research Center, Moffett Field, CA 94035, USA \\
\textsuperscript{\hypertarget{inst:53}{53}} Konkoly Observatory, Research Centre for Astronomy and Earth Sciences, 1121 Budapest, Konkoly Thege Miklós út 15-17, Hungary \\
\textsuperscript{\hypertarget{inst:54}{54}} ELTE E\"otv\"os Lor\'and University, Institute of Physics, P\'azm\'any P\'eter s\'et\'any 1/A, 1117 Budapest, Hungary \\
\textsuperscript{\hypertarget{inst:55}{55}} Lund Observatory, Division of Astrophysics, Department of Physics, Lund University, Box 43, 22100 Lund, Sweden \\
\textsuperscript{\hypertarget{inst:56}{56}} IMCCE, UMR8028 CNRS, Observatoire de Paris, PSL Univ., Sorbonne Univ., 77 av. Denfert-Rochereau, 75014 Paris, France \\
\textsuperscript{\hypertarget{inst:57}{57}} Institut d'astrophysique de Paris, UMR7095 CNRS, Université Pierre \& Marie Curie, 98bis blvd. Arago, 75014 Paris, France \\
\textsuperscript{\hypertarget{inst:58}{58}} Department of Astronomy \& Astrophysics, University of Chicago, Chicago, IL 60637, USA \\
\textsuperscript{\hypertarget{inst:59}{59}} Department of Physics and Astronomy, The University of North Carolina at Chapel Hill, Chapel Hill, NC 27599, USA \\
\textsuperscript{\hypertarget{inst:60}{60}} Komaba Institute for Science, The University of Tokyo, 3-8-1 Komaba, Meguro, Tokyo 153-8902, Japan \\
\textsuperscript{\hypertarget{inst:61}{61}} Astrobiology Center, 2-21-1 Osawa, Mitaka, Tokyo 181-8588, Japan \\
\textsuperscript{\hypertarget{inst:62}{62}} Instituto de Astrof\'isica de Canarias (IAC), c/ Via Lactea, s/n, 38205 La Laguna, Tenerife, Spain \\
\textsuperscript{\hypertarget{inst:63}{63}} Institute of Astronomy, Faculty of Physics, Astronomy and Informatics, Nicolaus Copernicus University, Grudzi\c{a}dzka 5, 87-100, Toru\'n, Poland \\
\textsuperscript{\hypertarget{inst:64}{64}} European Southern Observatory, Karl-Schwarzschild-Straße 2, 85748 Garching bei München, Germany \\
\textsuperscript{\hypertarget{inst:65}{65}} Las Campanas Observatory, Carnegie Institution for Science, Colina el Pino, Casilla 601 La Serena, Chile \\
\textsuperscript{\hypertarget{inst:66}{66}} Institute of Optical Sensor Systems, German Aerospace Center (DLR), Rutherfordstrasse 2, 12489 Berlin, Germany \\
\textsuperscript{\hypertarget{inst:67}{67}} Dipartimento di Fisica e Astronomia "Galileo Galilei", Università degli Studi di Padova, Vicolo dell'Osservatorio 3, 35122 Padova, Italy \\
\textsuperscript{\hypertarget{inst:68}{68}} ETH Zurich, Department of Physics, Wolfgang-Pauli-Strasse 2, CH-8093 Zurich, Switzerland \\
\textsuperscript{\hypertarget{inst:69}{69}} Cavendish Laboratory, JJ Thomson Avenue, Cambridge CB3 0HE, UK \\
\textsuperscript{\hypertarget{inst:70}{70}} Zentrum für Astronomie und Astrophysik, Technische Universität Berlin, Hardenbergstr. 36, D-10623 Berlin, Germany \\
\textsuperscript{\hypertarget{inst:71}{71}} Institut fuer Geologische Wissenschaften, Freie Universitaet Berlin, Maltheserstrasse 74-100, 12249 Berlin, Germany \\
\textsuperscript{\hypertarget{inst:72}{72}} Astronomy Department and Van Vleck Observatory, Wesleyan University, Middletown, CT 06459, USA \\
\textsuperscript{\hypertarget{inst:73}{73}} Institut de Ciencies de l'Espai (ICE, CSIC), Campus UAB, Can Magrans s/n, 08193 Bellaterra, Spain \\
\textsuperscript{\hypertarget{inst:74}{74}} Institut d’Estudis Espacials de Catalunya (IEEC), Gran Capità 2-4, 08034 Barcelona, Spain \\
\textsuperscript{\hypertarget{inst:75}{75}} Department of Astronomy, Yale University, New Haven, CT 06511, USA \\
\textsuperscript{\hypertarget{inst:76}{76}} Department of Physics and Kavli Institute for Astrophysics and Space Research, Massachusetts Institute of Technology, Cambridge, MA 02139, USA \\
\textsuperscript{\hypertarget{inst:77}{77}} HUN-REN--ELTE Exoplanet Research Group, Szent Imre h. u. 112., Szombathely, H-9700, Hungary \\
\textsuperscript{\hypertarget{inst:78}{78}} Institute of Astronomy, University of Cambridge, Madingley Road, Cambridge, CB3 0HA, UK \\
\textsuperscript{\hypertarget{inst:79}{79}} Department of Astrophysical Sciences, Princeton University, 4 Ivy Lane, Princeton, NJ 08544, USA \\
\textsuperscript{\hypertarget{inst:80}{80}} Department of Earth, Atmospheric and Planetary Sciences, Massachusetts Institute of Technology, Cambridge, MA 02139, USA \\
\textsuperscript{\hypertarget{inst:81}{81}} Department of Aeronautics and Astronautics, MIT, 77 Massachusetts Avenue, Cambridge, MA 02139, USA \\
\textsuperscript{\hypertarget{inst:82}{82}} National Astronomical Observatory of Japan, 2-21-1 Osawa, Mitaka, Tokyo 181-8588, Japan \\
\textsuperscript{\hypertarget{inst:83}{83}} SETI Institute, Mountain View, CA 94043, USA \\
\textsuperscript{\hypertarget{inst:84}{84}} NASA Exoplanet Science Institute, IPAC, California Institute of Technology, Pasadena, CA 91125, USA \\

\section{Radial Velocity Datasets}

\begin{table*}
\centering
\caption{Radial velocities, spectral activity indicators and exposure times for the HARPS, ESPRESSO and PFS measurements. The times are given in \textit{TESS} BJD time (BTJD), which is defined as BJD - 2457000.}
\label{table:rvdata}
\resizebox{\textwidth}{!}{
\begin{tabular}{lrlrllllrrllllr}
\toprule
Telescope &       Time &           RV & $\sigma_{\text{RV}}$ &          BIS & $\sigma_{\text{BIS}}$ &         FWHM & $\sigma_{\text{FWHM}}$ & S-Index & $\sigma_{\text{S-Index}}$ & $\mathrm{H_{\alpha}}$ & $\sigma_{\mathrm{H_{\alpha}}}$ & $\mathrm{\log\,R ^\prime_\mathrm{HK}}$ & $\sigma_{\mathrm{\log\,R ^\prime_\mathrm{HK}}}$ & T$_{\text{exp}}$ \\
          &     [BTJD] & [m s$^{-1}$] &         [m s$^{-1}$] & [m s$^{-1}$] &          [m s$^{-1}$] & [m s$^{-1}$] &           [m s$^{-1}$] &         &                           &                       &                                &                               &                                        &              [s] \\
\midrule
    HARPS & 2236.80 &         5.78 &                 1.00 &        -20.5 &                   2.0 &      6595.88 &                    2.0 &    0.22 &                     0.001 &                   0.9 &                             $--$ &                          -4.8 &                                  0.006 &           1800.0 \\
    HARPS & 2238.75 &          5.6 &                 0.91 &       -16.36 &                  1.81 &      6599.89 &                   1.81 &    0.23 &                     0.001 &                  0.89 &                             $--$ &                         -4.81 &                                  0.005 &           1800.0 \\
    HARPS & 2239.73 &          4.0 &                 0.85 &        -21.2 &                  1.69 &      6598.55 &                   1.69 &    0.23 &                     0.001 &                  0.89 &                             $--$ &                          -4.8 &                                  0.005 &           1800.0 \\
    HARPS & 2239.79 &         2.47 &                 0.77 &       -17.76 &                  1.55 &      6593.97 &                   1.55 &    0.23 &                     0.001 &                  0.89 &                             $--$ &                          -4.8 &                                  0.004 &           1800.0 \\
    HARPS & 2240.85 &         1.77 &                 0.88 &       -19.32 &                  1.75 &      6597.62 &                   1.75 &    0.23 &                     0.001 &                  0.89 &                             $--$ &                          -4.8 &                                  0.004 &           1500.0 \\
    HARPS & 2242.81 &        -0.95 &                 0.84 &       -17.01 &                  1.68 &      6594.76 &                   1.68 &    0.23 &                     0.001 &                   0.9 &                             $--$ &                          -4.8 &                                  0.004 &           1800.0 \\
    HARPS & 2243.86 &        -1.05 &                 0.80 &       -17.79 &                   1.6 &      6590.69 &                    1.6 &    0.23 &                     0.001 &                   0.9 &                             $--$ &                         -4.81 &                                  0.004 &           1600.0 \\
    HARPS & 2244.86 &        -2.82 &                 0.96 &       -20.25 &                  1.93 &       6587.7 &                   1.93 &    0.23 &                     0.001 &                   0.9 &                             $--$ &                         -4.81 &                                  0.004 &           1800.0 \\
    HARPS & 2245.81 &        -3.29 &                 1.06 &        -24.3 &                  2.12 &      6592.22 &                   2.12 &    0.23 &                     0.002 &                   0.9 &                             $--$ &                         -4.81 &                                  0.008 &           1500.0 \\
    HARPS & 2246.83 &        -0.71 &                 1.09 &       -17.21 &                  2.18 &      6578.91 &                   2.18 &    0.23 &                     0.001 &                   0.9 &                             $--$ &                         -4.81 &                                  0.006 &           1500.0 \\
    HARPS & 2250.86 &         5.27 &                 1.32 &       -26.68 &                  2.64 &      6577.18 &                   2.64 &    0.22 &                     0.001 &                   0.9 &                             $--$ &                         -4.83 &                                  0.006 &           1800.0 \\
    HARPS & 2261.84 &        -3.25 &                 1.02 &       -24.61 &                  2.04 &      6591.91 &                   2.04 &    0.23 &                     0.002 &                   0.9 &                             $--$ &                         -4.79 &                                  0.006 &           1500.0 \\
    HARPS & 2263.79 &          0.0 &                 1.09 &        -23.9 &                  2.17 &      6587.85 &                   2.17 &    0.23 &                     0.002 &                   0.9 &                             $--$ &                         -4.79 &                                  0.006 &           1500.0 \\
    HARPS & 2264.77 &          2.1 &                 0.98 &       -24.25 &                  1.97 &      6582.24 &                   1.97 &    0.22 &                     0.001 &                   0.9 &                             $--$ &                         -4.81 &                                  0.005 &           1500.0 \\
    HARPS & 2266.77 &         8.44 &                 0.93 &       -19.67 &                  1.86 &      6582.97 &                   1.86 &    0.22 &                     0.001 &                   0.9 &                             $--$ &                         -4.82 &                                  0.005 &           1500.0 \\
    HARPS & 2267.83 &         7.52 &                 0.85 &       -22.06 &                  1.71 &      6585.02 &                   1.71 &    0.22 &                     0.001 &                  0.91 &                             $--$ &                         -4.84 &                                  0.005 &           1500.0 \\
    HARPS & 2268.86 &         7.21 &                 1.01 &       -24.65 &                  2.03 &      6587.25 &                   2.03 &    0.22 &                     0.001 &                  0.91 &                             $--$ &                         -4.83 &                                  0.006 &           1500.0 \\
    HARPS & 2269.74 &          5.2 &                 0.86 &        -25.0 &                  1.71 &      6592.71 &                   1.71 &    0.22 &                     0.001 &                   0.9 &                             $--$ &                         -4.83 &                                  0.004 &           1500.0 \\
    HARPS & 2273.73 &         2.03 &                 1.06 &        -19.3 &                  2.13 &      6592.05 &                   2.13 &    0.23 &                     0.002 &                   0.9 &                             $--$ &                          -4.8 &                                  0.007 &           1500.0 \\
    HARPS & 2274.69 &         -0.7 &                 0.97 &       -15.44 &                  1.94 &      6599.29 &                   1.94 &    0.23 &                     0.002 &                   0.9 &                             $--$ &                         -4.81 &                                  0.006 &           1500.0 \\
    HARPS & 2275.82 &        -4.16 &                 1.08 &       -17.52 &                  2.16 &      6593.96 &                   2.16 &    0.22 &                     0.002 &                  0.89 &                             $--$ &                         -4.82 &                                  0.009 &           1500.0 \\
    HARPS & 2276.82 &        -3.63 &                 0.98 &       -16.25 &                  1.96 &      6593.83 &                   1.96 &    0.23 &                     0.002 &                   0.9 &                             $--$ &                          -4.8 &                                  0.008 &           1500.0 \\
    HARPS & 2277.78 &        -5.45 &                 0.99 &       -17.14 &                  1.98 &      6580.53 &                   1.98 &    0.22 &                     0.001 &                   0.9 &                             $--$ &                         -4.82 &                                  0.004 &           1500.0 \\
    HARPS & 2286.84 &         2.87 &                 1.05 &       -25.27 &                  2.11 &      6634.04 &                   2.11 &    0.21 &                     0.002 &                  0.89 &                             $--$ &                         -4.85 &                                  0.007 &           1500.0 \\
    HARPS & 2288.79 &        -2.93 &                 1.50 &       -21.62 &                  3.01 &      6586.59 &                   3.01 &    0.21 &                     0.002 &                   0.9 &                             $--$ &                         -4.85 &                                   0.01 &           1500.0 \\
    HARPS & 2291.75 &        -3.02 &                 0.96 &       -21.93 &                  1.91 &      6571.49 &                   1.91 &    0.22 &                     0.001 &                  0.91 &                             $--$ &                         -4.83 &                                  0.006 &           1500.0 \\
    HARPS & 2364.67 &        -0.29 &                 1.35 &       -12.28 &                   2.7 &      6584.38 &                    2.7 &    0.21 &                     0.002 &                  0.91 &                             $--$ &                         -4.85 &                                   0.01 &           1500.0 \\
 ESPRESSO & 2308.65 &        -0.94 &                 0.61 &       -73.03 &                  1.21 &      6896.45 &                   1.21 &    0.17 &                    0.0002 &                  0.24 &                        0.00007 &                         -5.11 &                                  0.001 &            600.0 \\
 ESPRESSO & 2310.82 &        -0.22 &                 0.76 &       -70.83 &                  1.51 &      6887.61 &                   1.51 &    0.16 &                    0.0004 &                  0.24 &                        0.00008 &                         -5.19 &                                  0.002 &            600.0 \\
 ESPRESSO & 2313.75 &        -0.49 &                 0.40 &       -71.49 &                   0.8 &      6894.54 &                    0.8 &    0.20 &                    0.0001 &                  0.24 &                        0.00004 &                         -4.96 &                                 0.0005 &            600.0 \\
 ESPRESSO & 2316.61 &        -0.66 &                 0.64 &       -71.03 &                  1.28 &      6879.84 &                   1.28 &    0.18 &                    0.0003 &                  0.24 &                        0.00007 &                         -5.08 &                                  0.001 &            600.0 \\
 ESPRESSO & 2318.64 &         2.95 &                 0.43 &       -68.77 &                  0.86 &      6883.65 &                   0.86 &    0.20 &                    0.0001 &                  0.24 &                        0.00005 &                         -4.99 &                                 0.0006 &            600.0 \\
 ESPRESSO & 2321.73 &         4.02 &                 0.65 &        -72.8 &                  1.31 &      6884.01 &                   1.31 &    0.17 &                    0.0003 &                  0.24 &                        0.00007 &                         -5.12 &                                  0.002 &            600.0 \\
 ESPRESSO & 2324.57 &        -1.15 &                 0.83 &       -76.38 &                  1.66 &      6882.39 &                   1.66 &    0.16 &                    0.0004 &                  0.24 &                        0.00009 &                         -5.19 &                                  0.003 &            600.0 \\
 ESPRESSO & 2327.54 &        0.008 &                 0.44 &        -76.2 &                  0.87 &      6884.69 &                   0.87 &    0.20 &                    0.0002 &                  0.24 &                        0.00005 &                         -4.96 &                                 0.0006 &            600.0 \\
 ESPRESSO & 2329.58 &         0.49 &                 0.69 &       -76.38 &                  1.38 &      6896.57 &                   1.38 &    0.18 &                    0.0003 &                  0.24 &                        0.00007 &                         -5.07 &                                  0.001 &            600.0 \\
 ESPRESSO & 2332.58 &          0.4 &                 0.54 &       -68.78 &                  1.07 &      6907.25 &                   1.07 &    0.20 &                    0.0002 &                  0.24 &                        0.00006 &                         -4.98 &                                 0.0008 &            600.0 \\
 ESPRESSO & 2335.77 &         1.89 &                 0.53 &       -67.36 &                  1.07 &      6904.64 &                   1.07 &    0.20 &                    0.0002 &                  0.24 &                        0.00006 &                         -4.98 &                                 0.0008 &            600.0 \\
 ESPRESSO & 2341.54 &        -0.36 &                 0.80 &       -67.41 &                   1.6 &      6890.32 &                    1.6 &    0.16 &                    0.0004 &                  0.23 &                        0.00009 &                         -5.19 &                                  0.002 &            600.0 \\
 ESPRESSO & 2344.68 &        -0.77 &                 0.41 &       -73.07 &                  0.81 &      6883.64 &                   0.81 &    0.20 &                    0.0001 &                  0.24 &                        0.00004 &                         -4.99 &                                 0.0005 &            600.0 \\
 ESPRESSO & 2349.73 &        -1.93 &                 0.60 &       -74.55 &                   1.2 &       6875.9 &                    1.2 &    0.15 &                    0.0003 &                  0.23 &                        0.00006 &                         -5.24 &                                  0.002 &            600.0 \\
 ESPRESSO & 2371.66 &         4.38 &                 0.52 &        -63.3 &                  1.03 &      6903.33 &                   1.03 &    0.22 &                    0.0002 &                  0.24 &                        0.00006 &                          -4.9 &                                 0.0006 &            600.0 \\
 ESPRESSO & 2374.61 &        -0.61 &                 0.45 &       -68.73 &                   0.9 &      6890.44 &                    0.9 &    0.22 &                    0.0002 &                  0.24 &                        0.00005 &                         -4.91 &                                 0.0005 &            600.0 \\
 ESPRESSO & 2406.57 &         3.12 &                 0.53 &       -72.97 &                  1.06 &      6880.69 &                   1.06 &    0.20 &                    0.0002 &                  0.24 &                        0.00006 &                         -4.97 &                                 0.0008 &            600.0 \\
 ESPRESSO & 2410.53 &        -1.89 &                 0.64 &       -69.47 &                  1.28 &      6870.54 &                   1.28 &    0.16 &                    0.0003 &                  0.23 &                        0.00007 &                         -5.18 &                                  0.002 &            600.0 \\
 ESPRESSO & 2413.52 &        -4.56 &                 0.48 &       -74.56 &                  0.96 &      6870.99 &                   0.96 &    0.20 &                    0.0002 &                  0.23 &                        0.00005 &                         -4.97 &                                 0.0007 &            600.0 \\
 ESPRESSO & 2415.52 &        -5.82 &                 0.51 &       -78.36 &                  1.02 &      6874.03 &                   1.02 &    0.19 &                    0.0002 &                  0.23 &                        0.00005 &                          -5.0 &                                 0.0008 &            600.0 \\
 ESPRESSO & 2430.46 &        -1.44 &                 0.47 &       -73.78 &                  0.94 &       6890.0 &                   0.94 &    0.23 &                    0.0002 &                  0.24 &                        0.00005 &                         -4.88 &                                 0.0005 &            600.0 \\
 ESPRESSO & 2432.51 &         3.58 &                 1.55 &       -75.34 &                   3.1 &      6914.51 &                    3.1 &    0.18 &                     0.001 &                  0.24 &                         0.0002 &                         -5.06 &                                  0.005 &            600.0 \\
      PFS & 2207.84 &        -2.49 &                 0.89 &           $--$ &                    $--$ &           $--$ &                     $--$ &    0.23 &                      0.01 &                    $--$ &                             $--$ &                            $--$ &                                     $--$ &            751.0 \\
      PFS & 2207.85 &        -3.64 &                 0.81 &           $--$ &                    $--$ &           $--$ &                     $--$ &    0.22 &                      0.01 &                    $--$ &                             $--$ &                            $--$ &                                     $--$ &            750.0 \\
      PFS & 2211.86 &        -1.58 &                 0.62 &           $--$ &                    $--$ &           $--$ &                     $--$ &    0.21 &                      0.01 &                    $--$ &                             $--$ &                            $--$ &                                     $--$ &           1200.0 \\
      PFS & 2212.84 &        -1.15 &                 0.89 &           $--$ &                    $--$ &           $--$ &                     $--$ &    0.24 &                      0.01 &                    $--$ &                             $--$ &                            $--$ &                                     $--$ &            600.0 \\
      PFS & 2212.85 &         0.43 &                 0.84 &           $--$ &                    $--$ &           $--$ &                     $--$ &    0.24 &                      0.01 &                    $--$ &                             $--$ &                            $--$ &                                     $--$ &            600.0 \\
      PFS & 2213.84 &         0.73 &                 0.82 &           $--$ &                    $--$ &           $--$ &                     $--$ &    0.25 &                      0.01 &                    $--$ &                             $--$ &                            $--$ &                                     $--$ &            901.0 \\
      PFS & 2213.85 &        -0.63 &                 0.90 &           $--$ &                    $--$ &           $--$ &                     $--$ &    0.24 &                      0.01 &                    $--$ &                             $--$ &                            $--$ &                                     $--$ &            900.0 \\
      PFS & 2214.84 &         3.68 &                 1.03 &           $--$ &                    $--$ &           $--$ &                     $--$ &    0.25 &                      0.01 &                    $--$ &                             $--$ &                            $--$ &                                     $--$ &            600.0 \\
      PFS & 2214.84 &        -0.17 &                 1.05 &           $--$ &                    $--$ &           $--$ &                     $--$ &    0.26 &                      0.01 &                    $--$ &                             $--$ &                            $--$ &                                     $--$ &            600.0 \\
      PFS & 2215.86 &        -1.26 &                 0.85 &           $--$ &                    $--$ &           $--$ &                     $--$ &    0.24 &                      0.01 &                    $--$ &                             $--$ &                            $--$ &                                     $--$ &            600.0 \\
      PFS & 2215.87 &        -0.48 &                 0.88 &           $--$ &                    $--$ &           $--$ &                     $--$ &    0.25 &                      0.01 &                    $--$ &                             $--$ &                            $--$ &                                     $--$ &            600.0 \\
      PFS & 2239.80 &          0.0 &                 0.88 &           $--$ &                    $--$ &           $--$ &                     $--$ &    0.24 &                      0.01 &                    $--$ &                             $--$ &                            $--$ &                                     $--$ &           1200.0 \\
      PFS & 2356.59 &         7.15 &                 1.32 &           $--$ &                    $--$ &           $--$ &                     $--$ &    0.22 &                      0.01 &                    $--$ &                             $--$ &                            $--$ &                                     $--$ &           1001.0 \\
      PFS & 2356.69 &         5.36 &                 1.36 &           $--$ &                    $--$ &           $--$ &                     $--$ &    0.24 &                      0.01 &                    $--$ &                             $--$ &                            $--$ &                                     $--$ &           1501.0 \\
      PFS & 2363.57 &        -0.34 &                 1.14 &           $--$ &                    $--$ &           $--$ &                     $--$ &    0.21 &                      0.01 &                    $--$ &                             $--$ &                            $--$ &                                     $--$ &            900.0 \\
      PFS & 2363.65 &        -0.83 &                 1.15 &           $--$ &                    $--$ &           $--$ &                     $--$ &    0.22 &                      0.01 &                    $--$ &                             $--$ &                            $--$ &                                     $--$ &           1201.0 \\
      PFS & 2596.85 &        -0.08 &                 0.81 &           $--$ &                    $--$ &           $--$ &                     $--$ &    0.22 &                      0.01 &                    $--$ &                             $--$ &                            $--$ &                                     $--$ &           1001.0 \\
      PFS & 2597.85 &         5.38 &                 0.79 &           $--$ &                    $--$ &           $--$ &                     $--$ &    0.22 &                      0.01 &                    $--$ &                             $--$ &                            $--$ &                                     $--$ &           1201.0 \\
      PFS & 2597.86 &         3.94 &                 0.80 &           $--$ &                    $--$ &           $--$ &                     $--$ &    0.21 &                      0.01 &                    $--$ &                             $--$ &                            $--$ &                                     $--$ &           1001.0 \\
      PFS & 2598.81 &         8.38 &                 0.98 &           $--$ &                    $--$ &           $--$ &                     $--$ &    0.24 &                      0.01 &                    $--$ &                             $--$ &                            $--$ &                                     $--$ &           1201.0 \\
      PFS & 2598.82 &         7.82 &                 1.19 &           $--$ &                    $--$ &           $--$ &                     $--$ &    0.27 &                      0.01 &                    $--$ &                             $--$ &                            $--$ &                                     $--$ &           1200.0 \\
      PFS & 2599.84 &         6.22 &                 0.98 &           $--$ &                    $--$ &           $--$ &                     $--$ &    0.25 &                      0.01 &                    $--$ &                             $--$ &                            $--$ &                                     $--$ &            900.0 \\
      PFS & 2599.85 &         6.47 &                 0.88 &           $--$ &                    $--$ &           $--$ &                     $--$ &    0.24 &                      0.01 &                    $--$ &                             $--$ &                            $--$ &                                     $--$ &            900.0 \\
      PFS & 2601.86 &         0.99 &                 0.80 &           $--$ &                    $--$ &           $--$ &                     $--$ &    0.22 &                      0.01 &                    $--$ &                             $--$ &                            $--$ &                                     $--$ &            901.0 \\
\bottomrule
\end{tabular}
}
\end{table*}


\bsp	
\label{lastpage}
\end{document}